\documentclass[preprint,
preprintnumbers,superscriptaddress,supberscriptaddress,nofootinbib,longbibliography]{revtex4-1}

\usepackage[bookmarks=true,colorlinks,linkcolor=blue,urlcolor=cyan,citecolor=red]{hyperref}

\usepackage{graphicx}
\usepackage{latexsym}
\usepackage{amsfonts}
\usepackage{amssymb}
\usepackage{amsmath}
\usepackage{bm}
\usepackage{mathrsfs}
\usepackage{slashed}
\usepackage{comment}

\usepackage{dcolumn}% Align table columns on decimal point

\usepackage[%dvipdfm,
bookmarks=true,colorlinks,linkcolor=blue,urlcolor=cyan,citecolor=red]{hyperref}

\renewcommand{\varXi}{\Xi}

\newcommand{\with}{\quad{\rm with}\quad}

\newcommand{\cout}[1]{ \if 0 {#1} \fi }

\newcommand{\beq}{\begin{eqnarray}}
\newcommand{\eeq}{\end{eqnarray}}
\newcommand{\bseq}{\begin{subequations}}
\newcommand{\eseq}{\end{subequations}}
\newcommand{\nn}{\nonumber}
\renewcommand{\=}{&=&}
\newcommand{\nnb}{\nonumber \\}
\newcommand{\pd}{\partial}

\newcommand{\LR}[1]{\left\langle {#1} \right\rangle}
\newcommand{\lan}{\langle}
\newcommand{\ran}{\rangle}
\renewcommand{\l}{\left}
\renewcommand{\r}{\right}

\renewcommand{\a}{\alpha}
\renewcommand{\b}{\beta}
\renewcommand{\d}{{\delta}}

\renewcommand{\l}{\lambda}
\newcommand{\m}{\mu}
\newcommand{\n}{\nu}
\renewcommand{\r}{\rho}
\newcommand{\s}{\sigma}

\newcommand{\ve}{\varepsilon}

\newcommand{\e}{ \epsilon }
\newcommand{\vep}{ \varepsilon }

\newcommand{\bx}{{\bm{x}}}

\newcommand{\bk}{{\bm{k}}}

\newcommand{\bq}{{\bm{q}}}

\newcommand{\bB}{{\bm B}}
\newcommand{\bE}{{\bm E}}

\newcommand{\order}{{O}}

\newcommand{\R}{ {\mathcal R} }

\newcommand{\para}{ \parallel}

\newcommand{\Tr}{ {\rm Tr} }
\newcommand{\tr}{ {\rm tr} }

\newcommand{\Hall}{H}

\newcommand{\vk}{{\bm k}}

\newcommand{\vv}{\bm{v}}

\newcommand{\zero}{{(0)}}
\newcommand{\one}{{(1)}}

\newcommand{\tilA}{\widetilde{A}}
\newcommand{\tilC}{\widetilde{C}}
\newcommand{\tilD}{\widetilde{D}}

\newcommand{\hSigma}{\hat{\Sigma}}

\newcommand{\hpi}{\hat{\pi}}

\newcommand{\hp}{\hat{p}}

\newcommand{\hE}{\hat{E}}
\newcommand{\hD}{\hat{D}}
\newcommand{\Hcal}{\mathcal{H}}

\newcommand{\tile}{\tilde{e}}
\newcommand{\tilg}{\tilde{g}}
\newcommand{\tilt}{\tilde{t}}
\newcommand{\tilb}{\tilde{b}}
\newcommand{\tilx}{\tilde{x}}

\newcommand{\tils}{\tilde{s}}

\newcommand{\tildelta}{\tilde{\delta}}
\newcommand{\tilpartial}{\tilde{\partial}}
\newcommand{\tilF}{\widetilde{F}}

\newcommand{\diff}{\mathrm{d}}
\newcommand{\average}[1]{\langle#1\rangle}

\newcommand{\baverageLG}[1]{\Big\langle#1\Big\rangle^{\mathrm{LG}}}

\newcommand{\hT}{\hat{T}}

\newcommand{\hc}{\hat{c}}
\newcommand{\hS}{\hat{S}}

\newcommand{\hA}{\hat{A}}
\newcommand{\hB}{\hat{B}}

\newcommand{\pt}{\partial}
\newcommand{\ie}{{\rm i.e.}}
\newcommand{\eg}{{\rm e.g.}}
\newcommand{\h}{\theta}
\newcommand{\eq}{{\rm eq}}

\renewcommand{\equiv}{:=}

\newcommand{\rmi}{\mathrm{i}} 
\newcommand{\rme}{\mathrm{e}} 

\newcommand\Lcal{\mathcal{L}}
\newcommand\Dcal{\mathcal{D}}
\newcommand{\tilLcal}{\widetilde{\mathcal{L}}}
\newcommand{\tilbx}{\widetilde{\bm{x}}}
\newcommand{\tilomega}{\widetilde{\omega}}

\newcommand{\ptc}[1]{{\bar{#1}}}
\newcommand{\ppsi}{\ptc{\psi}}
\newcommand{\qed}{\mathrm{QED}}

\newcommand{\htau}{\hat{\tau}}
\newcommand{\hrho}{\hat{\rho}}
\newcommand{\hrhoLG}{\hat{\rho}_{\mathrm{LG}}}
\newcommand{\hOcal}{\hat{\mathcal{O}}}

\newcommand{\averageLG}[1]{\langle#1\rangle^\text{LG}}
\newcommand{\ip}[1]{\bm{(} {#1}\bm{)} }

\usepackage{color} % For blue in-text comments and additions
\usepackage[normalem]{ulem}  % \sout{old text} for strikeout
\newcommand{\com}[1]{ [{\color[rgb]{0,0,1}{#1}}] }

\graphicspath{{./figs/}}

\begin{document}

\title{New developments in relativistic magnetohydrodynamics}

\author{Koichi Hattori}
\email{koichi.hattori@zju.edu.cn}
\affiliation{Zhejiang Institute of Modern Physics, Department of Physics, 
Zhejiang University, Hangzhou, 310027, China} 
\affiliation{Research Center for Nuclear Physics, Osaka University, Osaka 567-0047 Japan.}

\author{Masaru Hongo}
\email{hongo@phys.sc.niigata-u.ac.jp}
\affiliation{Department of Physics, Niigata University, Niigata 950-2181, Japan}
\affiliation{RIKEN iTHEMS, RIKEN, Wako 351-0198, Japan}

\author{Xu-Guang Huang}
\email{huangxuguang@fudan.edu.cn}
\affiliation{Physics Department and Center for Field Theory and Particle Physics, Fudan University, Shanghai 200438, China}
\affiliation{Key Laboratory of Nuclear Physics and Ion-beam Application (MOE), Fudan University, Shanghai 200433, China}
\affiliation{Shanghai Research Center for Theoretical Nuclear Physics, National Science Foundation of China and Fudan University, Shanghai 200438, China}

%\date{\today}

\begin{abstract}
Relativistic magnetohydrodynamics (RMHD) provides an extremely useful description of the low-energy long-wavelength phenomena in a variety of physical systems from quark-gluon plasma in heavy-ion collisions to matters in supernova, compact stars, and early universe. We review the recent theoretical progresses of RMHD, such as a formulation of RMHD from the perspective of magnetic flux conservation using the entropy-current analysis, the nonequilibrium statistical operator approach applied to quantum electrodynamics, and the relativistic kinetic theory. We discuss how the transport coefficients in RMHD are computed in kinetic theory and perturbative quantum field theories. We also explore the collective modes and instabilities in RMHD with a special emphasis on the role of chirality in a parity-odd plasma. We also give some future prospects of RMHD, including the interaction with spin hydrodynamics and the new kinetic framework with magnetic flux conservation.
\end{abstract}

\maketitle
\tableofcontents

\section{Introduction}\label{sec:intro}

Relativistic hydrodynamics provides an incredibly successful description of the macroscopic dynamics of interacting many-body systems in relativistic arena since its birth \cite{Eckart:1940te,kluitenberg1953relativistic:I,kluitenberg1953relativistic:II,kluitenberg1954relativistic:III,landau1959course:fluid}. Its applicability ranges from very small systems such as the quark gluon plasma (QGP) created in high-energy heavy-ion collisions to very large systems like the expanding universe and the explosive supernovas~\cite{Yagi-Hatsuda-Miake:2005yb,Romatschke:2017ejr}. From a theoretical point of view, relativistic hydrodynamics is a typical example of effective field theories which is valid at low-energy and long-wavelength limit. The dynamical variables in relativistic hydrodynamics are the coarse-grained conserved quantities stemmed from the underlying symmetries. As such, there exist a variety of ways to systematically construct relativistic hydrodynamics when an appropriate derivative expansion is employed. 

Strong magnetic fields exist and play critical roles in a number of systems where the relativistic hydrodynamics can be applied. Examples range from high-energy heavy-ion collisions to supernovas, neutron stars, and the early universe. In heavy-ion collisions, the colliding nuclei induce very strong transient magnetic fields exerting on the produced quark-gluon plasma (QGP)~\cite{Skokov:2009qp,Voronyuk:2011jd,Bzdak:2011yy,Deng:2012pc,Bloczynski:2012en,Tuchin:2013apa,Yan:2021zjc,Wang:2021oqq}. The peak strength of such magnetic fields can reach $10^{19}$ Gauss in Au + Au collisions at Relativistic Heavy Ion Collider (RHIC) and $10^{20}$ Gauss in Pb + Pb collisions at Large Hadron Collider (LHC). Given that the QGP is also an electromagnetic (EM) plasma, such magnetic fields can strongly influence the dynamics of QGP and induce novel transport phenomena in QGP such as the chiral magnetic effect which provides a valuable machinery to access the quantum chromodynamics (QCD) topological sector in an experimentally feasible way~\cite{Kharzeev:2007jp,Fukushima:2008xe}. (For reviews of strong magnetic-field effects in heavy-ion collisions, see Refs.~\cite{Miransky:2015ava,Huang:2015oca,Hattori:2016emy,Kharzeev:2015znc,Kharzeev:2020jxw}.) 

In compact stellar objects such as the neutron stars, the surface magnetic fields can reach the order of $10^{12}–10^{13}$ Gauss~\cite{Shapiro:1983du}, with a subclass of neutron stars (called magnetars) having surface magnetic fields of the order of $10^{14}–10^{15}$ Gauss~\cite{Duncan:1992hi} (Even stronger transient magnetic fields may be created in binary neutron star mergers~\cite{Price:2006fi,Kiuchi:2015sga}). The internal magnetic fields could be by orders of magnitude stronger than the surface magnetic fields. Although the origin of magnetar magnetic fields is not fully understood, one widely accepted theory is that they result from the magnetohydrodynamic dynamo processes in the interior fluids of some newly born neutron stars~\cite{Thompson:1993hn}. Observationally, soft gamma ray repeaters (SGRs) and anomalous X-ray pulsars (AXPs) are commonly identified as magnetars. The strong magnetic fields are crucial for understanding the phenomenology of neutron stars, particularly magnetars, such as radio emission, cooling properties, equations of state, shape deformation and gravitational-wave emission, merging processes and post-merger evolution of binary neutron stars, among other things. (See Refs.~\cite{Lai:2000at,Harding:2006qn,Mereghetti:2015asa,Turolla:2015mwa,  Kaspi:2017fwg,Enoto:2019vcg,Ciolfi:2020cpf} for reviews.) 

Although a conclusive evidence for primordial magnetic fields in the early universe is still lacking, astrophysical studies of the large-scale intergalactic  magnetic fields in the current universe strongly support their existence. The origin of the primordial magnetic fields has been pursued over the last three decades with possible scenarios stemming from the big-bang era  \cite{Turner:1987bw, Carroll:1989vb, Garretson:1992vt} and cosmic phase transitions like the QCD phase transition and the electroweak phase transition \cite{Hogan:1983zz,Quashnock:1988vs,Vachaspati:1991nm,Cheng:1994yr,Baym:1995fk,Son:1998my}. 
%The estimated strength of the magnetic fields during EW phase transition can reach $10^{23}$ Gauss~\cite{Grasso:2000wj}. These primordial magnetic fields might evolve into the seed fields for magnetohydrodynamic turbulent dynamos, which in turn generate the magnetic fields that survive the universe expansion and provide a natural explanation for the large-scale magnetic fields observed in galaxies, galaxy clusters, and intergalactic medium in the current universe. 
Besides, the strong primordial magnetic fields may have played an important role in understanding the cosmological structure formation, the thermal spectrum and polarization anisotropies of the cosmic microwave background, the big-bang nucleosynthesis and baryogenesis, and so on. (For reviews of the primordial magnetic fields, see Refs.~\cite{Grasso:2000wj,Kandus:2010nw,Subramanian:2015lua}.)

The relativistic magnetohydrodynamics (relativistic MHD or RMHD) is often used as a standard tool to analyze the physical processes in the systems mentioned above. It provides a macroscopic framework to self-consistently describe the evolution of the matter coupled with either dynamical electromagnetic (EM) fields or in external EM fields. There is a long history of astrophysical and cosmological applications of RMHD (typically in the presence of gravity), and there are already a number of excellent reviews and textbooks (see, for example, books~\cite{Lichnerowicz:book,Anile:book,Rezzolla:book,Goedbloed:book,Shoj:book}). Over the last decade, due to the realization of strong magnetic fields in heavy ion collisions, various aspects and applications of RMHD in the context of heavy ion collisions have been extensively investigated, including the study of evolution of magnetic fields in the QGP~\cite{Tuchin:2013ie,Li:2016tel}, the computation of RMHD transport coefficients in QCD matter using perturbative field theory~\cite{Hattori:2016lqx,Hattori:2016cnt,Hattori:2016idp,Hattori:2017qih,Li:2017tgi}, kinetic theories~\cite{Denicol:2018rbw,Denicol:2019iyh,Dey:2019vkn,Chen:2019usj,Dash:2020vxk,Singh:2020faa,Panda:2020zhr,Ghosh:2022xtv}, and holographic models~\cite{Critelli:2014kra,Finazzo:2016mhm,Li:2018ufq,Fukushima:2021got}, the anisotropic evolution of the QGP coupled with magnetic fields~\cite{Tuchin:2011jw,Gursoy:2014aka,Roy:2015kma,Pu:2016bxy,Pu:2016ayh,Inghirami:2016iru,Gursoy:2018yai,She:2019wdt,Inghirami:2019mkc,Kord:2019nko,Emamian:2020psw}, the simulation of chiral magnetic effect and other anomalous transport phenomena~\cite{Hongo:2013cqa,Yee:2013cya,Hirono:2014oda,Yin:2015fca,Huang:2015fqj,Jiang:2016wve,Shi:2017cpu,Guo:2017jxs,Shi:2019wzi,Siddique:2019gqh}, among other things. In particular, the presence of the axial charge induced by a topological nature of the system motivated people to extend RMHD to the chiral magnetohydrodynamics (chiral MHD) \cite{Hattori:2017usa}, which gives a potential theoretical tool to investigate the dynamical origin of the primordial large-scale magnetic field~\cite{Tashiro:2012mf,Boyarsky:2015faa, Rogachevskii:2017uyc, Schober:2017cdw,Brandenburg:2017rcb, Boyarsky:2020cyk} (see also Refs.~\cite{Joyce:1997uy, Field:1998hi, Giovannini:2003yn, Semikoz:2004rr, Laine:2005bt, Boyarsky:2011uy} for earlier approaches to this problem based on RMHD).

Instead of exploring into the details of RMHD applications in many subfields of physics, the goal of this article is to present a two-fold overview of the theoretical aspects of the special relativistic MHD: One based on the recent formulation of RMHD motivated by the generalized symmetry viewpoint, and the other with the conventional approach in which the matter components and the EM fields are separated discussed. We will present our view to relate these two formulations with each other.

As a branch of relativistic hydrodynamics, the RMHD can also be considered as an effective theory of low-energy long-wavelength modes of the system. Such modes are usually the conserved modes like the energy and momentum modes. When the EM fields are dynamical, it has been suggested recently that the Bianchi identity for EM fields can be regarded as a conservation law (associated with a one-form magnetic $U(1)$ symmetry)~\cite{Grozdanov:2016tdf,Hattori:2017usa,Glorioso:2018kcp,Armas:2018atq,Armas:2018zbe,Hongo:2020qpv}, allowing a formulation of RMHD based solely on symmetry argument and derivative expansions of conserved quantities (in this case, called hydrodynamic variables)~\footnote{See also Refs.~\cite{Grozdanov:2018ewh,Armas:2018ibg,Gralla:2018kif,Delacretaz:2019brr,Iqbal:2020lrt,Landry:2021kko} for several applications of higher-form (and higher-group) symmetry to hydrodynamics.}. We will give detailed discussion about the construction of RMHD in this manner in Sec.~\ref{sec:entropy} using a phenomenological method based on entropy-current analysis and in Sec.~\ref{sec:mhd:qed} using a nonequilibrium statistical operator method. A crucial observation of such a construction is that the magnetic field, like temperature and fluid velocity, persists at thermal equilibrium and can thus be assigned as a leading-order variable in derivative expansion. This is known as a strong magnetic field, and is the scenario that we are concerned with~\footnote{We will not discuss the scenario with weak magnetic fields at sub-leading orders in derivative expansion because the results are less novel, though it is also very useful in practical applications.}. As a result, the constitutive relations exhibit anisotropies at both the ideal and dissipative levels, which is a distinguished feature of RMHD (also non-relativistic MHD, in fact)~\cite{Braginskii:1965,Landau10:kinetics,Huang:2009ue,Huang:2011dc,Hernandez:2017mch}. 

We will then examine how this new formulation is connected to the conventional approach in Sec.~\ref{sec:comparison}, and proceed with the evaluations of transport coefficients in kinetic theory and in perturbative quantum field theories (QED, as an example) in Sec.~\ref{sec:kine} and Sec.~\ref{sec:transport-coeff}, respectively. More novel features can appear in RMHD when the system allows parity violation, leading to the chiral MHD. This will be explored in Sec.~\ref{sec:CMHD}. We note that there is an significant difference between the treatments of EM fields in Secs.~\ref{sec:entropy}, ~\ref{sec:mhd:qed}, ~\ref{sec:CMHD} and that in Secs.~\ref{sec:kine} and ~\ref{sec:transport-coeff}. In Secs.~\ref{sec:entropy}, \ref{sec:mhd:qed}, and \ref{sec:CMHD}, we treat the EM fields as dynamical variables following a recent formulation of RMHD. On the other hand, the EM fields in Secs.~\ref{sec:kine} and ~\ref{sec:transport-coeff} are considered as backgrounds, and formulations are closer to the conventional approach with the EM fields treated separately from the matter components. Considering this difference, we compare the result of Secs.~\ref{sec:entropy}-\ref{sec:mhd:qed} to the conventional one to clarify these two formulations in the intermediate section~\ref{sec:comparison}. Finally, we discuss the future prospects of RMHD in Sec.~\ref{sec:futur}.

Here is a summary of our notations. We use the natural units, $\hbar=c=k_B=1$. For the electromagnetism, we use the Heaviside-Lorentz convention $\e_0=\mu_0=1$ and $\a=e^2/(4\pi)$=1/137. Other notations are: 
\vspace{-0.5em}
\begin{itemize}
\setlength\itemsep{-0.5em}
\item Minkowski metric: $\eta_{\m\n}=\eta^{\m\n}={\rm diag}(1,-1,-1,-1)$. A  curved-spacetime metric is denoted by $g_{\m\n}$. 
\item Levi-Civita tensor in Minkowski spacetime: $\ve^{\m\n\r\s}$ with $\ve^{0123}=-\ve_{0123}=1$. 
\item Fluid velocity four vector: $u^\m=\gamma\,(1,\vv)^T$ with $\gamma=1/\sqrt{1-\vv^2}$ and $u^2=1$.  
\item Direction of the magnetic field: $b^\m=B^\m/B$ with $B=\sqrt{-B_\m B^\m}$. Note that $b^2=-1$ and $b\cdot u=0$.
\item Projector transverse to $u^\m$: $\Delta^{\m\n}=\eta^{\m\n}-u^\m u^\n$.
\item Projector transverse to both $u^\m$ and $b^\m$: $\Xi^{\m\n}=\Delta^{\m\n}+b^\m b^\n$. 
\item Cross projector: $b_\star^{\m\n}=-b_\star^{\n\m}=\ve^{\m\n\r\s}u_\r b_\s$. Note that $u_\m b_\star^{\m\n}=0=b_\m b_\star^{\m\n}$, $b_\star^{\m\l}b^\n_{\star\l}=\Xi^{\m\n}$.
\item Co-moving derivative (or material derivative or proper-time derivative) of $A$: $D A=\dot{A}=dA/d\tau=u^\m \pd_\m A$.
\item Spatial gradient of $A$: $\nabla_\m A=\Delta_{\m\n}\pt^\n A$. 
\item Symmetrization, anti-symmetrization, and traceless symmetrization of a rank-two tensor $A^{\m\n}$: $A^{(\m\n)}=(A^{\m\n}+A^{\n\m})/2$, $A^{[\m\n]}=(A^{\m\n}-A^{\n\m})/2$, and $A^{\lan\m\n\ran}=(1/2)\Delta^\m_\a \Delta^\n_\b(A^{\a\b}+A^{\a\b})-(1/3)\Delta^{\m\n}\Delta_{\a\b}A^{\a\b}$.
\item Decomposition of velocity gradient: $\pd_\m u_\n=u_\m D u_\n+\omega_{\m\n}+w_{\m\n}$ where $\omega_{\m\n}=\nabla_{[\m}u_{\n]}$ is the vorticity tensor and $w_{\m\n}=\nabla_{(\m}u_{\n)}=\nabla_{\lan\m}u_{\n\ran}+\frac{1}{3}\Delta_{\m\n}\h$ with $\nabla_{\lan\m}u_{\n\ran}$ the shear tensor and $\h=\Delta_{\a\b}w^{\a\b}$ the expansion rate of the fluid.
\end{itemize}

\section{Macroscopic approach: the entropy-current analysis}
\label{sec:entropy}

In this section, we review a phenomenological way to derive the RMHD equations based on the second law of local thermodynamics.
In Sec.~\ref{sec:hydro:macro}, we first give a brief overview of how one can build relativistic hydrodynamics without coupling to the EM fields before going into the discussion of RMHD. This can help us understand some basic ingredients that are essential to the construction of RMHD. In Sec.~\ref{sec:MHD}, we generalize our discussion to RMHD based on the symmetry associated with the Bianchi identity.

As mentioned in Sec.~\ref{sec:intro}, hydrodynamics is regarded as a low-energy effective theory 
that describes the dynamics of gapless modes. 
In other words, hydrodynamics describes the macroscopic behavior %dynamics 
of conserved charges that do not dissipate away.
%long spacetime limit. 
Therefore, it is crucial to identify symmetries of the system 
and associated conservation laws that serve as the equations of motion (EOMs) in hydrodynamics. 
This observation allows us to derive hydrodynamics from the viewpoint of the symmetry and thermodynamics of irreversible processes~\cite{landau1959course:fluid,deGroot1962}
without going into details of the system. 
This macroscopic (phenomenological) approach is often called the entropy-current analysis since the constitutive relations (see definition below Eq.~(\ref{eq:hydro:exp})) are derived by requiring the semi-positivity of the entropy-current divergence, or the second law of local thermodynamics. 

\subsection{Primer to the entropy-current analysis}
\label{sec:hydro:macro}

For the sake of simplicity, let us consider a relativistic system that only enjoys spacetime translational symmetries, and thus, respects the energy-momentum conservation law:
\begin{eqnarray}
\label{eq:hydro:tmn}
\pd_\m T^{\mu\nu} = 0,
\end{eqnarray}
where $T^{\mu\nu}$ denotes the energy-momentum tensor.Throughout this paper, we always consider relativistic systems respecting the Lorentz symmetry, so that one can write down the hydrodynamic equations in a Lorentz covariant manner. 
One may think that the Lorentz symmetry itself leads to the angular momentum conservation law, which should give independent hydrodynamic equations as well.
However, in the strict hydrodynamic limit (\ie, the long wavelength and low frequency limit), the angular momentum conservation law reduces to the constraint equation that forces the energy-momentum tensor to be symmetric under the exchange of its two Lorentz indices.
We thus use the symmetric energy-momentum tensor $T^{\m\n} = T^{\n\m} $ in the following discussion~\footnote{
In the transient time scale, called the spin hydrodynamic regime in Refs.~\cite{Hongo:2021ona,Hongo:2022izs,Cao:2022aku}, the spin density of microscopic constituents may show its intrinsic dynamics, and the energy-momentum tensor has the anti-symmetric components. This is the main topic of relativistic spin hydrodynamics which we will briefly discuss in Sec.~\ref{sec:futur}.}. We also note that the discrete charge-conjugation (C), time reversal (T), and parity (P) symmetries also impose strong constraints on the transport phenomena. As will be seen in Sec.~\ref{sec:CMHD}, 
breaking one or two of them will allow the appearance of new transport terms in the constitutive relations. 
In this section, however, we assume that C, P, T symmetries are not violated. 

Dynamical variables of hydrodynamics are the conserved energy-momentum densities, or the corresponding thermodynamic conjugate variables given by the fluid four-velocity $u^\m$ --- normalized as $\eta_{\m\n} u^\m u^\n = 1$ with the Minkowski metric $\eta_{\m\n} = \mathrm{diag} (+1,-1,-1,-1)$ --- and the local inverse temperature $\beta = 1/T$.
In this paper, we employ the Landau-Lifshitz frame~\cite{landau1959course:fluid} to define the fluid velocity by
\begin{equation}
 \label{eq:def-fluid-velocity}
 T^{\m}_{~\n} u^{\n} = \e u^{\n},
\end{equation}
with the energy density $\e$.
The local inverse temperature is related to the energy density $\e$ through the usual thermodynamic relation.
We will later give a concrete definition of the inverse temperature in Eq.~\eqref{eq:def-beta} when we start to consider the entropy density. 

Then, as a long-wavelength low-frequency effective theory, one can formulate hydrodynamics based on an expansion of $T^{\m\n}$ with respect to the small derivatives of $u^\m$ and $\beta$ as
\begin{eqnarray}
\label{eq:hydro:exp}
T^{\mu\nu} = T^{\mu\nu}_{(0)} + T^{\mu\nu}_{(1)}+ O(\pt^2),
\end{eqnarray}
where the subscripts $(0)$ and $(1)$ denote numbers of derivatives. This equation, which expresses $T^{\mu\nu}$ in terms of the dynamical variable $u^\m$ and $\beta$ (or $\e$) is called the \textit{constitutive relation}. We will construct the constitutive relation on an order-by-order basis.

In the leading-order derivative expansion, one can uniquely decompose the zeroth-order term $T^{\mu\nu}_{(0)}$ as 
\begin{eqnarray}
\label{eq:hydro:zeroth}
T^{\mu\nu}_{(0)} = \e u^\m u^\n - p \Delta^{\m\n},
\end{eqnarray}
with the projection tensor $\Delta_{\m\n}=\eta_{\m\n}-u_\m u_\n$
satisfying $u^\m \Delta_{\m\n} = 0$. 
This is because no other rank-two symmetric tensors can be constructed from algebraic combinations of available building blocks $u^\m, \eta^{\m\n}$, and Levi-Civita tensor $\ve^{\m\n\r\s}$ (normalized to be $\ve^{0123}=-\ve_{0123}=1$).
The scalar coefficient functions $\e$ and $p$ are regarded as the energy density and pressure in the fluid rest frame. This interpretation becomes manifest, $T_{(0)}^{\m\n}={\rm diag}(\e,p,p,p)$, 
when we take $u^\m=(1,\bm{0})$. We note that we have already imposed the matching condition for $\e$, $T^{\m\n}u_\m u_\n=T_{(0)}^{\m\n}u_\m u_\n$, in Eq.~(\ref{eq:hydro:zeroth}), which enforces that $T^{\m\n}_\one u_\n=0$ following Eqs.~(\ref{eq:def-fluid-velocity}). On the other hand, no such matching condition is required for $p$, and there could be $O(\pd^1)$ correction to the physical pressure in the fluid.

At this stage, Eq.~\eqref{eq:hydro:zeroth} is just a parameterization of $T^{\mu\nu}_{(0)}$ using the unknown function $p$.
In order to organize Eq.~\eqref{eq:hydro:tmn} 
in a solvable leading-order hydrodynamic equation with $T^{\mu\nu}_{(0)}$, we have to relate the pressure $p$ to $\e$. If our system is close to the local thermal equilibrium, we can expect that 
such a relation is provided by 
the equation of state (EOS), $p=p(\e)$, so that $p$ is identified as the \textit{thermodynamic} pressure.
As we will see shortly, $p(\e)$ has to satisfy a certain thermodynamic relation in order to respect the second law of thermodynamics. Then, substituting the leading-order constitutive relation~(\ref{eq:hydro:zeroth}) into the conservation law~(\ref{eq:hydro:tmn}), one obtains the relativistic Euler equations:
\begin{subequations}
\begin{eqnarray}
\label{eq:euler:u}&&
(\e+p)D u^\m-\nabla^\m p = 0
\, ,
\\
\label{eq:euler:e}&&
D\e+(\e+p)\theta =0
\, ,
\end{eqnarray}
\end{subequations}
where we defined the material (or co-moving time) derivative $D=u^\m \pt_\m$, the spatial gradient $\nabla_\m=\Delta_{\m\n}\pt^\n$, and the expansion rate of the fluid $\theta=\pt_\m u^\m$.

In the same manner, we can continue to perform 
the tensor decomposition of
the first-order term $T^{\mu\nu}_{(1)}$. 
Noting that $T^{\mu\nu}_{(1)}$ satisfies $T^{\mu\nu}_{(1)} u_\n =0 $ enforced by the Landau-Lifshitz frame condition and the matching condition for $\e$, we use $\Delta^{\m\n}$ to decompose the rank-two symmetric tensor $T^{\mu\nu}_{(1)}$ into the trace part $\Pi$ and symmetric traceless part $\pi^{\m\n}$ (or $\pi^{\m\n} \Delta_{\m\n} = 0$) as
\begin{eqnarray}
\label{eq:hydro:first}
T^{\mu\nu}_{(1)} = 
%h^\m u^\n + h^\n u^\m 
-\Delta^{\m\n} \Pi + \pi^{\m\n}.
\end{eqnarray}
This equation gives the decomposition of the rank-two symmetric tensor into irreducible representations of the $\mathrm{SO}(3)$ rotation, and thus, $\Delta^{\m\n}\Pi$ and $\pi^{\m\n}$ do not mix under the $\mathrm{SO}(3)$ rotation.
%where $h^\m$ and $\pi^{\m\n}$ are transverse to $u^\m$, or $h^\m u_\m = 0 = \pi^{\m\n} u_\m$, because of the condition  $T^{\mu\nu}_{(1)}u_\m u_\n=0$, and $\pi^{\m\n}$ is traceless, $\pi^{\mu\nu}\eta_{\m\n}=0$. 
Here, we can regard $\Pi$ as the derivative correction to the
%viscous 
pressure because it appears at the same place as the thermodynamic pressure $p$ does.
As we will see, 
$\Pi$ describes a viscous correction of the pressure 
proportional to the expansion rate $\theta$, while $\pi^{\m\n}$ 
gives the shear viscous contribution describing a friction-like process induced by the velocity gradient. As being $O(\pt)$-order quantities, $\Pi$, and $\pi^{\m\n}$ should be structured linearly in the gradients of $u^\m$ and $\beta$ (or equivalently, $\e$).

The entropy current analysis gives an elegant way to work out such linear structures based only on the thermodynamic laws. 
First, we write down the entropy current in a similar manner as that for $T^{\m\n}$,
\begin{eqnarray}
\label{eq:hydro:ent}
s^\mu = s u^\m + s_{(1)}^\m + O(\pt^2),
\end{eqnarray}
where $s = s(\e)$ is the \textit{thermodynamic} entropy density in the fluid rest frame and $s_{(1)}^\mu$ is a derivative correction transverse to $u^\m$. Note that the thermodynamic entropy density is a function of the energy density $\e$, whose derivative with respect to $\e$ gives a definition of the local inverse temperature
$\beta (x) = 1/T(x)$ as
\begin{equation}
\label{eq:def-beta}
 \beta (x) \equiv 
 \left. \frac{\partial s (\e)}{\partial \e} \right|_{\e = \e (x)}
 ~\Leftrightarrow~
 T (x) \diff s (x) = \diff \e (x).
\end{equation}
Then, using the chain rule for the material derivative of the entropy density, and the contracted equation of motion 
$u_\n \pt_\m T^{\m\n} = 0 $ with Eqs.~\eqref{eq:hydro:zeroth} and \eqref {eq:hydro:first}, we find the divergence of the entropy current to be
\begin{eqnarray}
\label{eq:hydro:divs}
\pt_\m s^\mu = 
\beta (Ts-\e-p)\h - \beta \Pi\h + \beta 
\pt_{\langle\m} u_{\n\rangle}
%\s_{\m\n}
\pi^{\m\n}
%- \beta \pt_\m h^\m 
%+ \beta h^\m D u_\m 
+ \pt_\m s_{(1)}^\m 
+ O(\pt^3),
%T\pt_\m s^\mu = (Ts-\e-p)\h-\Pi\h+\s_{\m\n}\pi^{\m\n}-\pt_\m h^\m + h^\m D u_\m+ T\pt_\m s_{(1)}^\m +O(\pt^3),
\end{eqnarray}
where we introduced the symmetric traceless projection of the velocity gradient $\pt_{\langle\m} u_{\n\rangle}$ (called shear tensor) as
%shear tensor $\s^{\m\n}$
%the first law of thermodynamics, $Tds=d\e$, is used, and $\s^{\m\n}$ is the shear tensor,
\begin{eqnarray}
\label{eq:hydro:shear}
\pt^{\langle\m} u^{\n\rangle}= \nabla^{\langle\m} u^{\n\rangle} 
= \frac{1}{2}\left(\nabla^\m u^\n + \nabla^\n u^\m\right) -\frac{1}{3}\Delta^{\m\n}\h.
\end{eqnarray}

Since the divergence of the entropy current gives a local entropy production rate, the second law of local thermodynamics requires $\pt_\m s^\m\geq0$ for any configuration of the hydrodynamic variables $\beta$ and $u^\m$. 
%The second law of local thermodynamics requires $T\pt_\m s^\m\geq0$. 
This is achieved by identifying the first-order entropy current as $s_{(1)}^\mu= 0$
%$s_{(1)}^\mu=\b h^\m$ 
and the following conditions:
%with $\b=1/T$ and 
\begin{subequations}
\begin{eqnarray}
\label{eq:hydro:gibbs}
Ts&=&\e+p, \\
\label{eq:hydro:Pi}
\Pi &=&-\zeta\,\h,\\
\label{eq:hydro:pi}
\pi^{\mu\nu} &=& 2\eta\,\pt^{\langle\m} u^{\n\rangle}.
%\s^{\m\n}.
%\\
%\label{eq:hydro:hmu}
%h^\m &=&-\kappa \left( Du^\m-\b\nabla^\m T\right).
\end{eqnarray}
\end{subequations}
Here, Eq.~(\ref{eq:hydro:gibbs}) gives the thermodynamic relation, which restricts $p(\e)$ appearing in the leading-order constitutive relation to be the thermodynamic pressure.
%where, Eq.~(\ref{eq:hydro:gibbs}) is the usual thermodynamic relation and 
Equations~(\ref{eq:hydro:Pi})-(\ref{eq:hydro:pi}) complete the first-order constitutive relations with the phenomenological parameters $\zeta, \eta$ identified as the transport coefficients 
called bulk viscosity and shear viscosity, respectively. The second law of local thermodynamics $\pt_\m s^\m\geq 0$ requires the semi-positivity of those transport coefficients.

With the identification of $\Pi$ and $\pi^{\m\n}$
given in Eqs.~\eqref{eq:hydro:Pi}-\eqref{eq:hydro:pi}, 
we obtain the hydrodynamic equations up to $O(\pt^2)$ by
substituting the derived constitutive relations into the conservation law \eqref{eq:hydro:tmn}.
By projecting the obtained equation into the spatial and temporal directions, we have 
\begin{subequations}
\begin{eqnarray}
\label{eq:NS:u}&&
(\e+p-\zeta\h)D u^\m-\nabla^\m (p-\zeta\h) +2\Delta^\m_\n\pt_\r(\eta 
\pt^{\langle\n} u^{\r\rangle}
%\sigma^{\n\r}
) = 0
\, ,
\\
\label{eq:NS:e}&&
D\e+(\e+p-\zeta\h)\theta-2\eta
\pt_{\langle\m} u_{\n\rangle} \pt^{\langle\m} u^{\n\rangle} 
%\s_{\m\n}\s^{\m\n} 
=0
\, .
\end{eqnarray}
\end{subequations}
These are the relativistic Navier-Stokes equations.

One can continue the above procedure to higher orders in derivative expansion.
The resultant constitutive relations and the hydrodynamic equations become complicated, and we do not report these higher order results here. The readers who are interested in such results can find excellent discussions in, \eg, Refs.~\cite{Israel:1979wp,Baier:2007ix,Romatschke:2017ejr}.

\subsection{Relativistic MHD from the magnetic flux conservation}
\label{sec:MHD}

We now proceed to the discussion of RMHD. Here, what is different from the usual relativistic hydrodynamics is the existence of the $\mathrm{U}(1)$ gauge field, or dynamical electric and magnetic fields. 
We thus first need to examine whether or not those fields deserve to be qualified as hydrodynamic variables 
that persist in the low-energy long-wavelength limit without being damped out in a fluid. 
 
Firstly, one finds that the electric field is not a hydrodynamic variable. To see this, notice that electric fields are subject to the Debye screening, which is a {\it static} screening effect in the long spacetime limit~\cite{Kapusta:2006pm, Bellac:2011kqa}. 
Namely, an electric-charge density is redistributed 
%\footnote{
%Here, we implicitly assume that our fluid has any finite conductivity so that the charge redistribution occurs. 
%However, we do not need to assume that the conductivity is infinitely large 
%as often stated in the conventional formulation of an ``ideal MHD'' 
%(see, e.g., Refs.~\cite{LandauLifshitz_contmedia, lichnerowicz1967relativistic}). 
%One just needs to focus on a sufficiently longer time scale than an inverse of the conductivity 
%which is the time scale necessary for the charge redistribution to occur (see Appendix~\ref{sec:time}). 
%%Note also that the net charge neutrality is essential for the above discussion. 
%%Namely, a charged plasma in an infinite volume will not reach an equilibrium. 
%%A net charge neutrality of commonly discussed systems is guaranteed by
%%an equal number of particles and antiparticles or the presence of positively charged ions for an electron plasma. 
%}  
so that the Coulomb field offsets the electric field in an equilibrium state. Therefore, an electric field is a gapped excitation due to the Debye screening mass, and does not deserve a hydrodynamic variable. 
One also finds the same conclusion from the Maxwell equation $\pt_\m F^{\m\n} = J^\n $,
where $F_{\m\n} \equiv \pt_\m A_\n - \pt_\n A_\n$ is the field strength tensor and $J^\m$ is the electric current.
The Maxwell equation indicates that the electric flux is not conserved due to the presence of the electric current. With a given constitutive relation for $J^\mu$ describing the Ohmic current, as shown in Appendix \ref{sec:time}, the electric field behaves like a decay mode in neutral plasma. %Thus, a non-conserved electric field does not give a hydrodynamic variable.

In contrast, there is no static screening effect on a magnetic field, 
meaning that magnetic field enjoys its intrinsic dynamics
%can persist 
in the macroscopic scale even in a medium.
This qualitative difference from the electric fields stems from the absence of a magnetic monopole. 
Namely, there is no ``magnetic charge'' distribution that can screen magnetic fields. This is a consequence of the Bianchi identity $\pt_\m \tilF^{\m\n} = 0 $, 
where we defined the Hodge dual of $F_{\m\n}$ by
$\tilF^{\mu\nu} = \frac{1}{2} \varepsilon^{\mu\nu\rho\sigma} F_{\rho\sigma}$.

The novel formulation of RMHD reviewed in 
Secs.~\ref{sec:entropy} and \ref{sec:mhd:qed} of this paper is based on this simple observation that 
{\it the magnetic flux is a conserved quantity}~\cite{Schubring:2014iwa, 
Grozdanov:2016tdf, Hattori:2017usa, Glorioso:2018kcp,Armas:2018atq,Armas:2018zbe, Hongo:2020qpv}
due to the Bianchi identity. 
This observation allows us to straightforwardly derive RMHD according to the philosophy of hydrodynamics that motivates us to keep only the conserved quantities resulting from the associate symmetries~\footnote{The corresponding symmetry associated with the conserved magnetic flux is called a one-form magnetic $\mathrm{U}(1)$ symmetry~\cite{Gaiotto:2014kfa}.
This gives a generalization of the usual global symmetry because the conserved magnetic-flux acts on a one-dimensional extended object, or a 't Hooft loop, rather than a conventional local operator having a point-like charge.
}.
%from the concept of hydrodynamics and motivates us to identify the associate symmetry. 
In this section, we formulate RMHD by plugging the magnetic-flux conservation law into the entropy-current analysis. 
In Sec.~\ref{sec:mhd:qed}, we identify the symmetry for the magnetic-flux conservation inherent in QED and provide the derivation of RMHD on the basis of the nonequilibrium statistical method. 

Before diving into the entropy-current analysis, it is useful to pay our attention to a basic difference with the conventional formulation of MHD~\cite{LandauLifshitz_contmedia, kluitenberg1954relativistic:IV, kluitenberg1954relativistic:V, lichnerowicz1967relativistic, Israel:1978up, gedalin1991relativistic}. In the conventional formulation of MHD, the starting point is to couple the Navier-Stokes (or Euler) equation with the Maxwell equation and Bianchi identity. 
This means that we also keep the electric field as a dynamical variable, which shows a non-hydrodynamic relaxational behavior according to our identification.
Thus, the conventional MHD describes the transient dynamics including the electric field. On the other hand, the new formulation only keeps conserved quantities as the dynamical variable, and gives a low-energy effective theory in a strict hydrodynamic limit.

\subsubsection{Entropy production rate with magnetic flux} 

Let us perform the entropy-current analysis to derive the RMHD equation.
This is carried out as a direct extension of the simple analysis in Sec.~\ref{sec:hydro:macro} by additionally considering the magnetic-flux conservation law. 
%once we identify the set of equations of motion and the first law of thermodynamics. 

First of all, the energy-momentum and magnetic-flux conservation laws are given as 
\begin{eqnarray}
\label{eq:EoMs}
\pd_\m T^{\mu\nu} = 0
\, , \quad
\pd_\mu \tilde F^{\mu\nu} =0
\, .
\end{eqnarray} 
Here, we note that $ T^{\mu\nu} $ is the {\it total} energy-momentum tensor of the %translation-invariant 
system 
that includes not only the matter part but also the contribution from the EM fields.
%Hydrodynamic equations should work independently of such a microscopic separation or coupling. 
The second equation, or the Bianchi identity, indicates the conservation of magnetic flux, which corresponds to
%which we denote as $\tilde F^{\mu\nu}$ motivated by the underlying theory, i.e., QED. 
%This is the Bianchi identity that contains 
the Gauss's law for the magnetic flux in the absence of a magnetic monopole. 
Following the philosophy of hydrodynamics discussed above, we do {\it not} include another Maxwell equation into the set of hydrodynamic equations because it describes the time evolution of the gapped electric fields. 

We then introduce the dynamical variables of RMHD.
Associated with the conserved energy-momentum density, we again employ the Landau-Lifshitz frame to define the energy density $\e$ and the normalized fluid four-velocity $u^\m$ by Eq.~\eqref{eq:def-fluid-velocity}.
Besides, we define a magnetic flux density by $B^\mu = \tilde F^{\mu\nu} u_\nu$ in a covariant manner.
This four vector reduces to $ B^\mu = (0, {\bm B})$ in the local rest frame with $u^\mu =(1,0,0,0)$.
Thus, the number of components of it is not four but three, which is manifest by noting that $B^\m$ is transverse to the fluid velocity, i.e., $ u_\mu B^\mu = 0 $.

Let us next introduce the conjugate variables based on the first law of local thermodynamics. Relying on the thermodynamic entropy density $s (\e,B^\m)$ in the rest frame, which is a function of the energy density $\e$ and magnetic flux density $B^\m$, we define the local inverse temperature $\beta$ and (in-medium) magnetic field~\footnote{We note that in Sec.~\ref{sec:entropy}, \ref{sec:mhd:qed}, and \ref{sec:CMHD} we call $H_\m$ the magnetic field and $B^\m$ the magnetic flux density, but in other sections, we call $B^\m$ the magnetic field without confusion.} $H_\m$ as 
\begin{eqnarray}
\label{eq:1st-law-B}
 \beta \equiv 
 \frac{\partial s (\e,B^\m)}{\partial \e}, \quad
 \beta H_\m \equiv 
 - \frac{\partial s (\e,B^\m)}{\partial B^\m} 
 ~\Leftrightarrow~
 T \diff s = \diff \epsilon - H_\mu \diff B^\mu 
% = Tds 
%-  d \left( \frac{1}{2} B_\mu B^\mu - M_\mu B^\mu \right)
 \, .
\end{eqnarray}
%where we make the magnetic field in the local rest frame to be $H_\m = (0,\bm{H})$ .
%One can also introduce the magnetization as $M^\mu = B^\mu + H^\mu$ in a covariant manner.
With the help of these variables, we will derive the constitutive relations for $T^{\m\n}$ and $\tilF^{\m\n}$.

To perform the entropy-current analysis with Eqs.~(\ref{eq:EoMs}) and (\ref{eq:1st-law-B}), it is useful to decompose $T^{\m\n}$ and $\tilF^{\m\n}$ into  possible tensor structures.
%As a result, the resulting constitutive relations show richer structures. 
In the present setup, we have the magnetic flux vector $B^\mu$ in addition to the flow vector $u^\mu$ as available zeroth-order vectors.
It is then convenient to introduce a normalized vector $b^\m=B^\m/B$ 
with $B=\sqrt{-B_\m B^\m}$ such that $\eta_{\m\n} b^\m b^\n = -1$ and $u_\mu b^\mu=0$. 
%We adopt the matching conditions 
%\begin{eqnarray}
%\label{eq:frame-cond}
%\epsilon u^\nu =  u_\mu  T^{\mu\nu}
%\, , \quad
%B u^\nu = - b_\mu \tilde F^{\mu\nu} 
%\, . 
%\end{eqnarray}
%Those conditions kill the ambiguities of the hydrodynamic variables 
%$\epsilon, \, u^\mu  , \, B , \, b^\mu$ 
%in off-equilibrium states (see discussions in Appendix~\ref{sec:T1}). 
With this vector, we also introduce a projector $\Xi^{\m\n}=\Delta^{\m\n}+b^\m b^\n$ 
transverse to both $ u^\mu $ and $  b^\mu$ 
($\Xi^{\m\n} u_\m = 0 = \Xi^{\m\n} b_\m$). 
%and an antisymmetric cross projector $b_\star^{\m\n}=\ve^{\m\n\r\s}u_\r b_\s$. 
By the use of those tensors, we parameterize the constitutive relations as  
\begin{subequations}
\label{eq:TF_zeroth}
\begin{eqnarray}
\label{eq:T_zeroth}
T^{\mu\nu} &=& \epsilon u^\mu u^\nu - p_\perp \Xi^{\mu\nu} + p_\para b^\mu b^\nu + T^{\mu\nu}_{(1)}
\, ,
\\
\label{eq:Ft_zeroth}
\tilde F^{\mu\nu}  &=&   B^\mu u^\nu - B^\nu u^\mu 
+ \tilde F^{\mu\nu}_{(1)}
\, .
\end{eqnarray}
\end{subequations}
Here, we assume that $ p_\perp, \, p_\para$ 
are zeroth-order in derivatives. 
%are constrained by the thermodynamic laws. 
The first-order corrections are collectively denoted 
as $T^{\mu\nu}_{(1)}$ and $\tilde F_{(1)}^{\mu\nu}$. 
%Note that $ T^{\mu\nu} $ and $\tilde F^{\mu\nu} $ are symmetric and antisymmetric with respect to their Lorentz indices, respectively.
Considering a charge-neutral and parity-even plasma, we assume the charge-conjugation symmetry and require the energy-momentum tensor $T^{\mu\nu}$ not to have a charge-conjugation odd term proportional to $ u^{(\mu} b^{\nu)}$ at the zeroth-order in derivatives. 
Likewise, $\tilde F^{\mu\nu}$ could not have a term $\ve^{\mu\nu\a\b} u_\a b_\b$ at the zeroth-order because the resulting scalar coefficient for such a term is parity odd. Plugging those expressions into the equations of motion (\ref{eq:EoMs}) 
and contracting those with $\beta u_\n$ and $H_\n$, we obtain
\begin{subequations}
\label{eq:eqs}
\begin{align} 
 0 &=
 u_\nu \pd_\mu T^{\mu\nu} 
= D\epsilon +  (\epsilon+p_\perp ) \theta + (p_\perp - p_\para) b_\nu b^\mu \pd_\mu u^\nu 
+  u_\nu \pd_\mu T_{(1)}^{\mu\nu} 
\, ,
\\ 
 0 &= H_\nu \pd_\mu \tilde F^{\mu\nu} = 
- H_\mu D B^\mu - B^\mu H_\mu \theta + B^\mu H_\nu \pd_\mu u^\nu 
+ H_\nu \pd_\mu \tilde F_{(1)}^{\mu\nu} 
\, .
\end{align}
\end{subequations}
Here, we note that our definitions of 
$\e$, $u^\m$, and $B^\m$ lead to the matching conditions $u_\nu T_{(1)}^{\mu\nu} = 0 = u_\nu \tilde F_{(1)}^{\mu\nu} $ (see Appendix~\ref{sec:T1}).

Now, let us compute the divergence of the entropy current $ s^\mu = s u^\mu + s^\mu_{(1)} $. 
As in the previous section, we use the chain rule
%, $D s = \beta D \e - \beta H_\m D B^\m$, 
and the equations of motion~(\ref{eq:eqs}) 
to eliminate $D \e$ and $- H_\m D B^\m$. 
Then, we obtain 
\begin{eqnarray}
\pd_\mu s^\mu 
\= s \theta + Ds + \pd_\mu s_{(1)}^\mu 
\nn \\
\= s \theta + \beta D \e - \beta H_\m D B^\m + \pd_\mu s_{(1)}^\mu 
\nn \\
\= \beta (Ts - \epsilon - p_\perp + B^\mu H_\mu  )  \theta
- \beta b^\m [(p_\perp - p_\para) b^\nu + H^\nu ] \pd_\mu u_\nu 
\nnb
&&
+ T_{(1)}^{\mu\nu}  \pd_\mu ( \beta u_\nu )
+ \tilde F_{(1)}^{\mu\nu} \pd_\mu ( \beta H_\nu )
+ \pd_\mu ( s_{(1)}^\mu - \beta u_\nu T_{(1)}^{\mu\nu}  
- \beta H_\nu \tilde F_{(1)}^{\mu\nu}  )
\, .
\label{eq:entropy-current-MHD}
\end{eqnarray}
Applying the second law of local thermodynamics $\pt_\m s^\m \geq 0$ to Eq.~\eqref{eq:entropy-current-MHD}, 
we will derive the RMHD equation in the leading and next-to-leading orders in derivatives.

\subsubsection{Zeroth-order in derivatives: Nondissipative RMHD}
\begin{figure}
%\vspace{-1cm}
     \begin{center}
              \includegraphics[width=0.45\hsize]{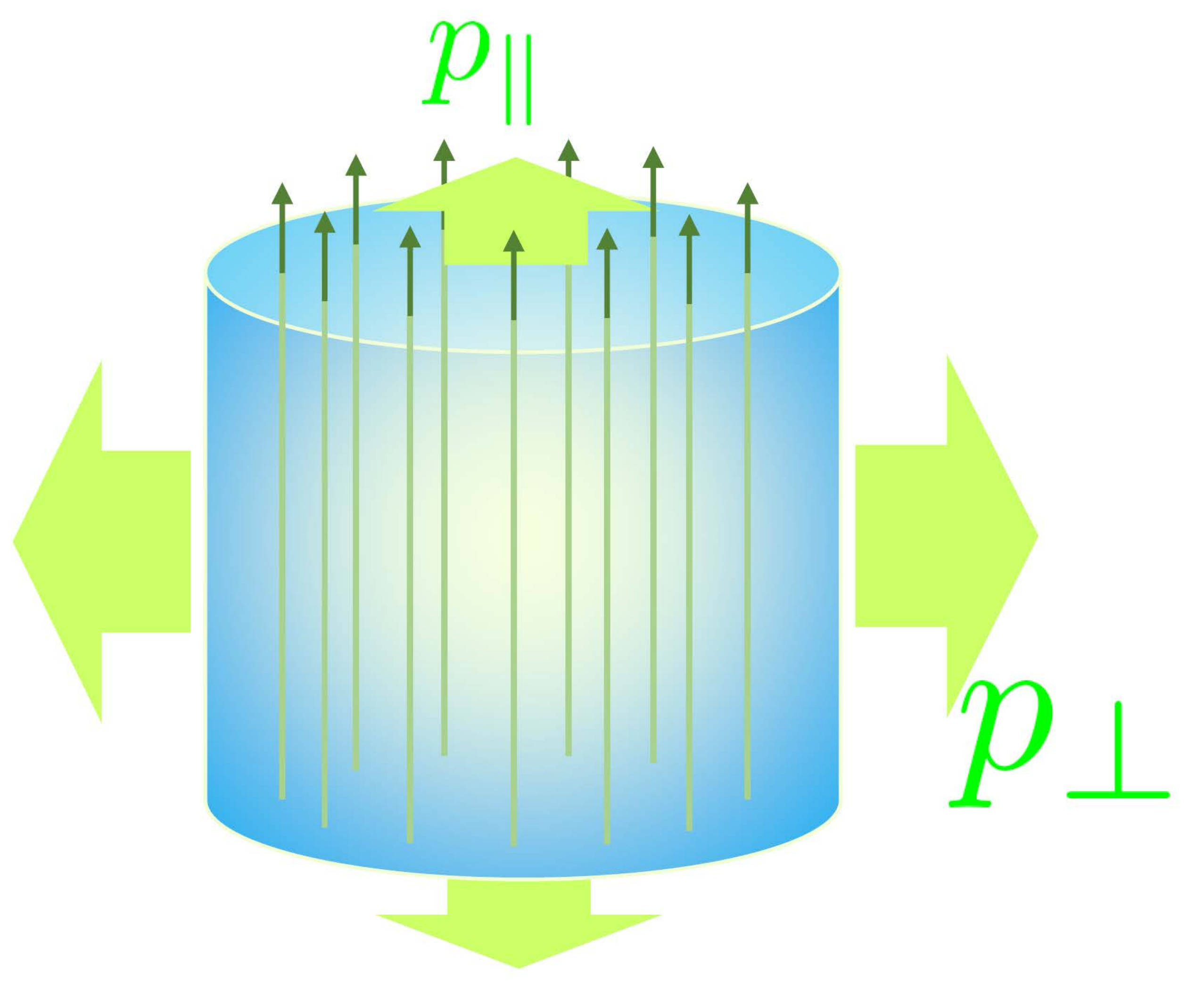}
     \end{center}
%\vspace{-1cm}
\caption{Anisotropic pressure induced by the magnetic field and the magnetization.}
\label{fig:pressure}
\end{figure}

At the leading order in derivatives, we find a set of constraints by 
requiring the absence of entropy production. 
For this requirement to be satisfied in any hydrodynamic configuration, 
all the leading-order terms should vanish independently 
because they are proportional to the irreducible tensor decomposition of $ \pd_\mu u_\nu $ 
and a non-derivative form of $ u_\nu $ 
in Eq.~\eqref{eq:entropy-current-MHD}. 
We thus find the following four constraints 
\begin{subequations}
\label{eq:constraints-LO}
\begin{eqnarray}
&& Ts + B^\mu H_\mu =  \epsilon + p_{\perp} \, ,
\\
&& (p_\perp - p_\para)  b^\nu + B H^\nu =0 \, .
\end{eqnarray}
\end{subequations}
%where we put $ p := p_\perp $. 
The first constraint serves as an extension of the thermodynamic relation with the magnetic flux.
The second constraint indicates that there is a pressure anisotropy $p_{\perp} - p_{\para} = B^\mu H_\mu (>0)$ induced by the finite magnetic flux (cf. Fig.~\ref{fig:pressure}). 
Besides, noting that the magnetic field $ H^\mu $ is parallel to $B^\mu $ 
and recalling that the (in-medium) magnetic field is parameterized by $ H^\mu = -\mu_m^{-1} B^\mu $ with the magnetic permeability $ \mu_m $, one identifies that the pressure difference gives the magnetic permeability, $\mu_m^{-1} = (p_{\perp} - p_{\para})/|B|^2$.
%$ p_\perp = p_\para + \mu_m^{-1} B^2$
% = p_\para - \mu_m^{-1} B_\mu B^\mu= p_\para + (\bB - {\bm M}) \cdot \bB $. 
%The third constraint requires that $ K^\mu $ be proportional to $ H^\mu $ in its arbitrary configurations, which, however, is not allowed by the fourth constraint unless $ K^\mu = 0 $.  

We conclude that the leading-order analysis leads to the following constitutive relations
\begin{subequations}
\label{eq:TF_zeroth-result}
\begin{eqnarray} 
T^{\mu\nu} _\zero \= 
\epsilon u^\mu u^\nu - p_{\perp} \Xi^{\mu\nu} 
+ (p_{\perp} - B^\a H_\a)
b^\mu b^\nu 
%+ (p+\mu_m^{-1} B^2) b^\mu b^\nu 
\, ,
\\ 
\tilde F^{\mu\nu}_\zero \=   B^\mu u^\nu - B^\nu u^\mu  
\, .
\end{eqnarray}
\end{subequations}
Note that the derived constitutive relation for $\tilde F^{\mu\nu}_\zero$ indicates that the electric field is absent at the leading order because it is given by 
$ E^\mu_\zero \equiv - \frac12 \epsilon^{\mu\nu\a\b} u_\nu \tilde F_{\zero \a\b} = 0 $. 
%the covariant form of an electric field is given by $ E^\mu = - \frac12 \epsilon^{\mu\nu\a\b} u_\nu \tilde F_{ \a\b}  $, 
%and is absent at the leading order, i.e., $ E^\mu_\zero = - \frac12 \epsilon^{\mu\nu\a\b} u_\nu \tilde F_{\zero \a\b} = 0 $ 
This result is consistent with our starting point treating the electric field as a non-hydrodynamic gapped variable.
We will find that an electric field (and corresponding electric current) appears as a first-order derivative correction. 
This will clearly show that they are slaved to the true hydrodynamic variables in the strict hydrodynamic limit as shown below in the form of the first-order constitutive relation.
%We will find that an electric field and current are only induced by the off-equilibrium dynamics beyond the ideal order, and thus are slaved to the hydrodynamic variables in the form of the constitutive relations. 

\subsubsection{First-order derivative corrections: Dissipative RMHD}
\label{sec:RMHD-1st}

We proceed to the derivation of the first-order corrections $T_{(1)}^{\mu\nu}, \, \tilde F_{(1)}^{\mu\nu}$.
Being equipped with the identified constraints \eqref{eq:constraints-LO}, 
we can simplify the entropy production rate~(\ref{eq:entropy-current-MHD}) as
\begin{eqnarray}
\pd_\mu s^\mu 
\= \beta T_{(1)}^{\mu\nu} \pd_\mu u_\nu 
+ \tilde F_{(1)}^{\mu\nu} \pd_\mu ( \beta H_\nu )
+  \pd_\mu \big( s_{(1)}^\mu - \beta u_\nu T_{(1)}^{\mu\nu}  
- \beta H_\nu \tilde F_{(1)}^{\mu\nu}  \big)
\, ,
\label{eq:entropy-current-MHD-simplified}
\end{eqnarray}
where we used $ T_{(1)}^{\mu\nu} u_\n = 0 $.
Noting that $u_\nu$ and $\beta H_\nu$ are independent variables, 
one can ensure the semi-positive entropy production, or the local second law $\pt_\m s^\m \geq 0$, in any hydrodynamic configuration by requiring that 
\begin{subequations} 
\label{eq:semi-positive}
\begin{align}
\label{eq:semi-positive-u}
\R_u &\equiv \beta T_{(1)}^{\mu\nu} \partial_\mu  u_\nu  \geq 0
\, ,
\\
\label{eq:semi-positive-E}
\R_E &\equiv  \tilde F_{(1)}^{\mu\nu} \partial_\mu(\beta H_\nu)   \geq 0
\, ,
\\
\label{eq:semi-positive-s}
s_{(1)}^\mu &= \beta u_\nu T_{(1)}^{\mu\nu} + \beta H_\nu \tilde F_{(1)}^{\mu\nu}
\, .
\end{align}
\end{subequations} 
As we will see, the first two conditions are satisfied by requiring each term to be a semi-positive bilinear form, which determines the first-order constitutive relations.
With the resulting constitutive relations, we can also find the derivative corrections to the entropy current from Eq.~\eqref{eq:semi-positive-s}.
Below, we will separately analyze the constitutive relations for
$T_{(1)}^{\mu\nu}$ and $\tilde F_{(1)}^{\mu\nu}$.
%This means that each term should be written in a semi-positive bilinear form. 

\paragraph{Electric field and resistivities:}

Let us first derive the constitutive relation for $\tilde F_{(1)}^{\mu\nu}$,
focusing on Eq.~\eqref{eq:semi-positive-E}.
To make $\R_E$ a semi-positive bilinear, we can express $\tilde F_{(1)}^{\mu\nu}$ as
\begin{equation}
 \label{eq:tilF-first}
 \tilde F_{(1)}^{\mu\nu} = T \rho^{\m\n\r\s} 
 \pt_{[\r} (\beta H_{\s]}).
\end{equation}
Here, we introduced a rank-four tensor $\rho^{\m\n\r\s}$, 
which will be identified as a resistivity tensor. 
As we will see, the semi-positivity of $\R_E$ and the anti-symmetric property in the Lorentz indices restrict possible structures of the resistivity tensor $\rho^{\m\n\r\s}$.

To identify the tensor structure of $\rho^{\m\n\r\s}$, we first recall 
that $\tilde F_{(1)}^{\mu\nu}$ is transverse to $u^\n$, $ \tilde F_{(1)}^{\mu\nu} u_\n = 0 $.
Thus, we can use $b^\m$ and $\Xi^{\m\n}$ to perform
the tensor decomposition of $\rho^{\m\n\r\s}$.
Moreover, we note that it is anti-symmetric with respect to the exchange of its Lorentz indices $\m \leftrightarrow \n$ and $\r \leftrightarrow \s$.
Taking into account these properties, we can write down the most general form for $\rho^{\m\n\r\s}$ in a neutral plasma as 
\begin{equation}
\label{eq:rho-tensor}
 \rho^{\m\n\r\s} = 
  2  \rho_\perp
  \left( b^\mu \Xi^{\nu[\rho} b^{\sigma]} -   b^\nu \Xi^{\mu[\rho} b^{\sigma]}
  \right) 
  + 2 \rho_\para  \Xi^{\mu[\rho} \Xi^{\sigma]\nu}
  . 
\end{equation}
The two coefficients $\rho_{\perp,\para}$ will be identified with two components of the electric resistivity (see Sec.~\ref{sec:currents-Hall}). Those tensor structures project out the gradient of the magnetic field in parallel and perpendicular to $b^\mu$ as opposed to what the subscripts of $\rho_{\perp,\para}$ denote. This is because the electric field is defined with  $ \tilF^{\mu\nu}_{(1)}$ and the antisymmetric tensor that swaps the directions (see Eq.~(\ref{eq:E-1st}) below). Note that $\rho^{\m\n\r\s}$ cannot have a tensor structure including $b_\star^{\m\n}:=\ve^{\m\n\a\b}u_\a b_\b$ because any such term would violate the  charge-conjugation symmetry. To respect the second law of local thermodynamics, we require semi-positivity of the two resistivities, $\rho_{\para} \geq 0 $ and $\rho_{\perp} \geq 0 $. 
In fact, substituting Eqs.~\eqref{eq:tilF-first} and \eqref{eq:rho-tensor} into Eq.~\eqref{eq:semi-positive-E}, we find 
\begin{equation}
 \R_E = 
 8T \rho_\perp
 \left(
  b^{[\nu} \Xi^{\mu]\rho} b^{\sigma}
  \pt_{[\r} \big( \beta H_{\s]} \big)
  \right)^2
  + 2 T \rho_\para
 \left(
  \Xi^{\mu[\rho} \Xi^{\sigma]\nu}
  \pt_{[\r} \big( \beta H_{\s]} \big)
 \right)^2,
\end{equation}
which is positive semi-definite. Equations~\eqref{eq:tilF-first} and \eqref{eq:rho-tensor} complete the constitutive relation for $\tilde F_{(1)}^{\mu\nu}$. The first-order corrections to $\tilde F_{(1)}^{\mu\nu}$ 
gives rise to an induced electric field 
\begin{eqnarray}
\label{eq:E-1st}
E_\one^\mu := - \frac12 \vep^{\mu\nu\a\b} u_\nu \tilde F_{\one \a\b}
=    (  \rho_\para b^\mu b^\nu  - \rho_\perp  \varXi^{\mu\nu} )
 T \vep_{\nu\lambda\alpha\beta} u^\lambda  \pt^{\a} (\beta H^{\b}) 
\,  ,
\end{eqnarray}
where we used an identity $ 2 b_\star^{\mu[\a} b^{\b]}
=  \varXi^\mu _\rho  \epsilon^{\rho\nu\alpha\beta}  u_\nu $ obtained from the Schouten identity.

\paragraph{Stress tensor and viscosities:}

One can derive the first-order corrections to the energy-momentum tensor $T^{\mu\nu}_{(1)}$ in the same manner. To ensure the semi-positivity of (\ref{eq:semi-positive-u}), the left-hand side should also be a positive semi-definite bilinear. 
Introducing the rank-four viscous tensor $\eta^{\m\n\r\s}$, 
we now express $ T_{(1)}^{\mu\nu}$ as 
\begin{equation}
\label{eq:T1-MHD}
 T_{(1)}^{\m\n} = \eta^{\m\n\r\s} \partial_\r u_\s,
\end{equation}
and identify the tensor structure of $\eta^{\m\n\r\s}$.

We first recall that we employ the Landau-Lifshitz frame (see Appendix~\ref{sec:T1} for a detailed discussion on the frame choice), in which $T_{(1)}^{\mu\nu}$ is transverse to $u^\n$ as $T_{(1)}^{\mu\nu} u^\nu =0$.
Thus, we can only use $b^\m$ and $\Xi^{\m\n}$ as a possible vector and tensor to decompose the viscous tensor $\eta^{\m\n\r\s}$. Note that $b_\star^{\m\n}$ cannot be used to decompose $\eta^{\m\n\r\s}$ because of charge-conjugation symmetry.
Moreover, recalling that $T^{\m\n}$ is the symmetric energy-momentum tensor, one finds that the viscous tensor should be symmetric with respect to the exchanges between its Lorentz indices $\m \leftrightarrow \n$ and $\r \leftrightarrow \s$.
These properties allow us to perform the tensor decomposition as
\begin{align}
\label{entropy:rank4tensor}
 \eta^{\m\n\r\s} 
 =& \zeta_\para b^\m b^\n b^\r b^\s
 + \zeta_\perp \Xi^{\m\n} \Xi^{\r\s}
 - \zeta_\times (b^\m b^\n  \varXi^{\r\s} + \Xi^{\m\n} b^\r b^\s)
 \nn \\
 &- 4\eta_\para b^{(\m} \varXi^{\n)(\r} b^{\s)}
 + \eta_\perp 
  \left( \Xi^{\mu\alpha} \Xi^{\nu \beta}
 + \Xi^{\nu\alpha} \Xi^{\mu \beta} 
 - \Xi^{\mu\nu} \Xi^{\alpha\beta} 
 \right),
\end{align}
where we introduced five viscosities --- 
three bulk viscosities $\zeta_{\para}, \zeta_{\perp}, \zeta_{\times}$
and two shear viscosities $\eta_{\para}, \eta_{\perp}$.
As we will specify, these viscosities must satisfy
a semi-positivity constraint to ensure the second law of local thermodynamics.

To get a physical intuition of the dissipative processes and find the semi-positivity constraint attached to each viscosity, it is useful to decompose the velocity gradient as 
\begin{eqnarray}
\label{eq:id-pu}
 \pt_{(\m} u_{\n)}
  = - \theta_\para b_\mu b_\nu
  + \frac12  \theta_\perp \Xi_{\mu\nu}
 - 2 b_\a b_{(\mu} \Xi_{\nu) \beta}  
 \pt^{(\a} u^{\b)} 
 + \pt_{\{\m} u_{\n\} },
 \label{eq:w_mn}
\end{eqnarray}
where we defined $\theta_\para \equiv (- b^\mu b^\nu) \pt_\m u_\n$, $\theta_\perp \equiv \Xi^{\mu\nu} \pt_\m u_\n $, and 
\begin{equation}
 \pt_{\{\m} u_{\n\}}
 \equiv 
 \frac{1}{2}
 \left( \Xi_{ \alpha \mu} \Xi_{\nu \beta}
 + \Xi_{ \alpha \nu} \Xi_{\mu \beta} 
 - \Xi_{\mu\nu} \Xi_{\alpha\beta} 
 \right)
 \pt^\a u^\b.
\end{equation}
Using this decomposition, we find the first-order derivative corrections to the constitutive relation \eqref{eq:T1-MHD} to be 
\begin{align}
\label{eq:T-1-decomposed}
 T_{(1)}^{\m\n}
 =& - ( \zeta_{\para} \theta_\para + \zeta_{\times} \theta_{\perp} ) 
 b^\mu b^\nu 
 + ( \zeta_{\perp} \theta_{\perp} + \zeta_{\times} \theta_{\para}) \Xi^{\m\n}
 - 4 \eta_{\para} b^{(\m} \Xi^{\nu) (\r} b^{\s)} \pt_\r u_\s
 + 2\eta_{\perp} \pt^{\{\m} u^{\n\}}.
\end{align}

\begin{figure}[t]
 \centering
 \includegraphics[width= 0.8 \linewidth ]{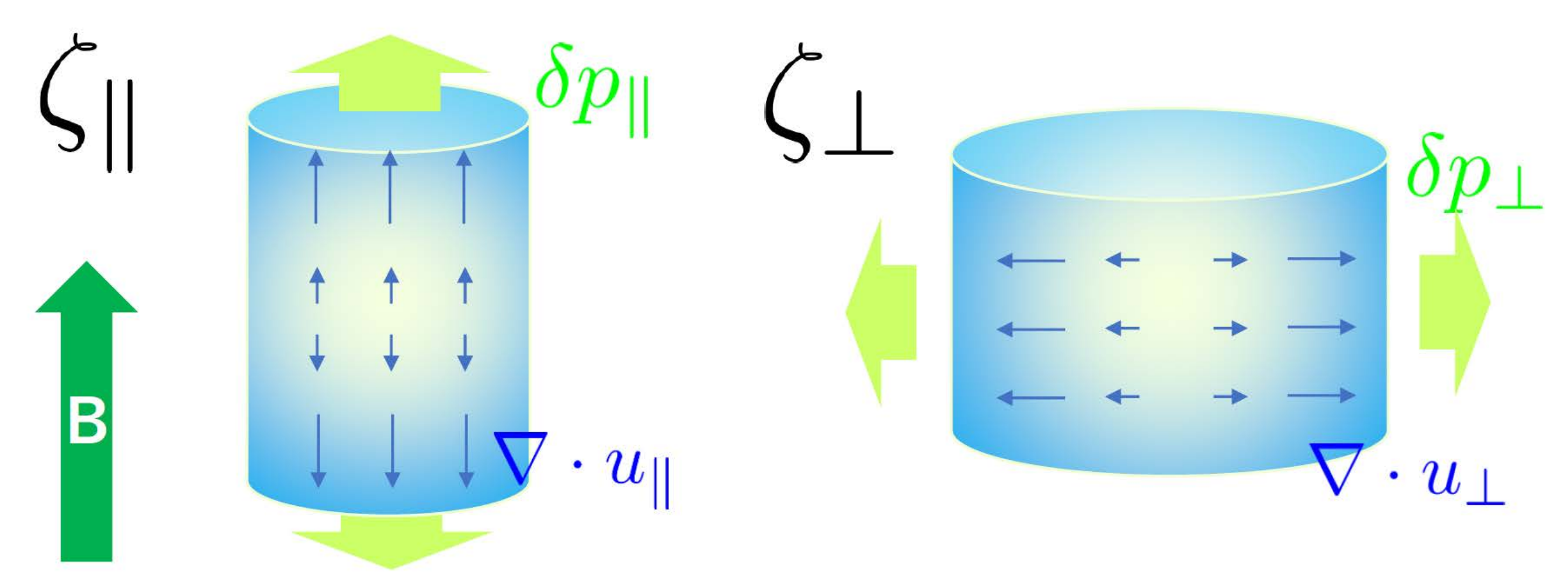}
 \caption{Longitudinal and transverse bulk viscosities that are the off-equilibrium responses of the pressure (green arrows)
in the direction of the expansion/compression (blue arrows) of the system.}
 \label{fig:bulk}
\end{figure}

\begin{figure}[t]
 \centering
 \includegraphics[width= 0.8 \linewidth]{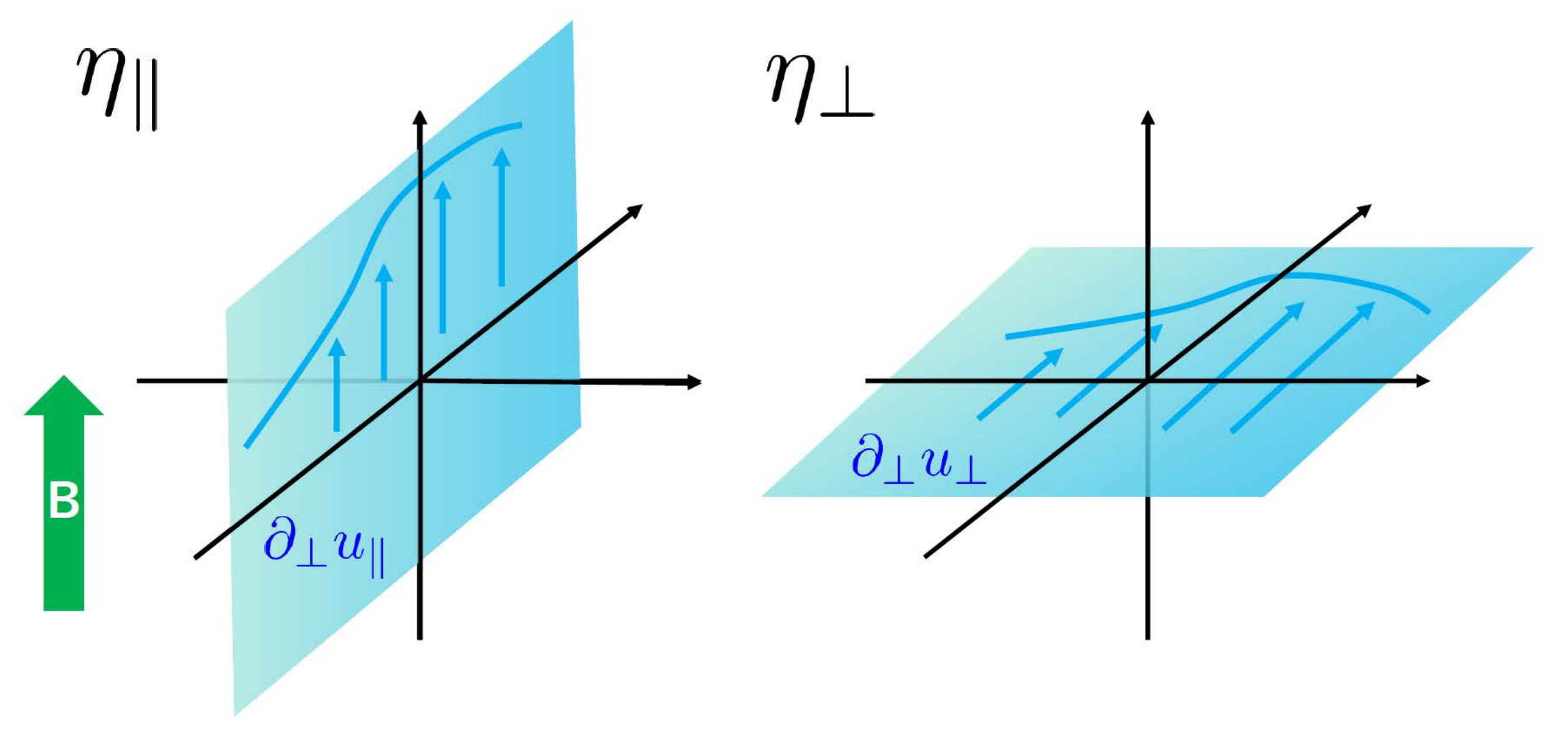}
 \caption{Shear deformations in and out of planes with respect to the magnetic field direction.}
 \label{fig:shear}
\end{figure}

\begin{figure}[h]
 \centering
 \includegraphics[width= 0.8 \linewidth ] {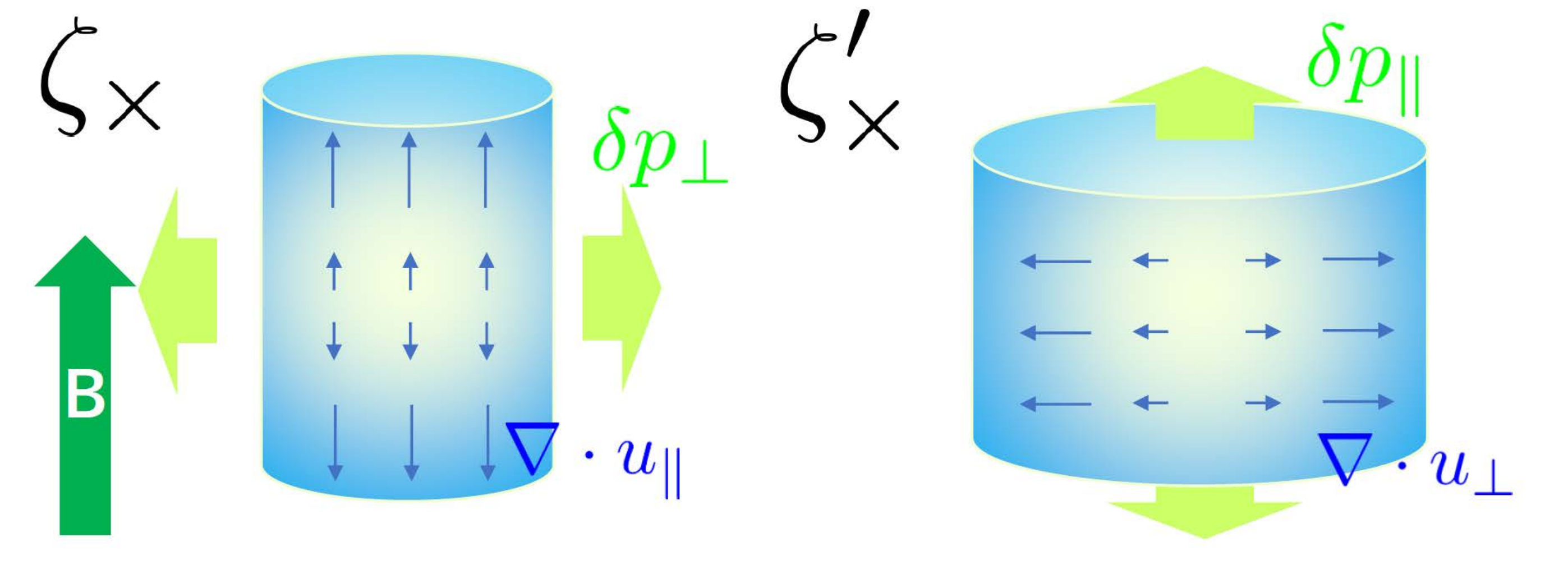}
 \caption{Cross bulk viscosities that are the response of the pressure (green arrows) in the orthogonal direction to the expansion/compression (blue arrows) of the system. 
 Those two cases are reciprocal processes to each other.
 Compare those two with Fig.~\ref{fig:bulk}.
}
\label{fig:cross}
\end{figure}

Equation \eqref{eq:T-1-decomposed} allows us to get a physical intuition on the dissipative process attached to each viscosity.
Without the magnetic flux, an expansion/compression rate is given as $\theta= \nabla_\mu u^\mu$.
In the presence of the magnetic flux,the expansion/compression in the parallel and perpendicular directions should be distinguished from each other, and thus, we have $\theta_\para$ and $\theta_\perp$.
Two viscosities $\zeta_{\para}$ and $\zeta_{\perp}$ describes a resistance to such two expansion/compression, respectively.
Similarly, the flow gradient for the shear deformation is projected
into the parallel and perpendicular directions, which leads to the two friction-like processes described by $\eta_{\para}$ and $\eta_{\perp}$ (cf. Figs.~\ref{fig:bulk} and \ref{fig:shear}). Based on these observations, we identify two coefficients $\zeta_\para$ and $\zeta_\perp$ with bulk viscosities and another two coefficients $\eta_\para$ and $\eta_\perp$ with shear viscosities.
Besides, there is an off-diagonal cross response proportional to $\zeta_\times$.
We also identify $\zeta_\times$ as one of the bulk viscosities since the associated term describes the cross response of the anisotropic pressures to the two  expansion/compression rates~(cf. Fig.~\ref{fig:cross}).
These processes are reciprocal to one another, and the associated transport coefficients should be the same according to the Onsager's reciprocal relation~\cite{Onsager,hooyman1954coefficients,deGroot1962,Grozdanov:2016tdf}. Putting $ \zeta_\para = \zeta_\perp =\zeta_\times $ 
and $ \eta_\para = \eta _\perp $, 
one can confirm that the anisotropic viscous corrections in Eq.~(\ref{eq:T-1-decomposed}) 
reduce to the isotropic form (\ref{eq:hydro:first}) 
by the use of $ \theta = \theta_\para + \theta_\perp $ 
and the identity (\ref{eq:id-pu}). 


We next investigate the positivity constraint required for the viscosities.
Substituting the decomposition \eqref{eq:w_mn} into Eq.~\eqref{eq:semi-positive-u}, we obtain 
\begin{align}
 \R_u 
 &= \beta \eta^{\m\n\r\s} \pt_\m u_\n \pt_\r u_\s
 \nn \\
 &= \beta
 \begin{pmatrix}
  \theta_{\para} & \theta_{\perp}
 \end{pmatrix}
 \begin{pmatrix}
  \zeta_{\para} & \zeta_{\times} \\
  \zeta_{\times} & \zeta_{\perp}
 \end{pmatrix}
 \begin{pmatrix}
  \theta_{\para} \\
  \theta_{\perp}
 \end{pmatrix}
 + 2\beta \eta_{\para} 
 \big( 2 b_{(\mu} \Xi_{\nu) \r} b_\s \pt^\r u^\s  \big)^2
 + 2\beta \eta_{\perp} 
 \big( \pt_{\{\m} u_{\n\}} \big)^2
\label{eq:Tw}
\, .
\end{align}
Firstly, the last two terms are positive semi-definite by requiring 
$\eta_{\para} \geq 0$ and $\eta_{\perp} \geq 0 $.  
On the other hand, for the first term to respect the local second law, 
the matrix composed of the bulk viscosities should be positive semi-definite. 
This means that the eigenvalue of that matrix needs to be non-negative:
$\zeta_\para+\zeta_\perp \geq  \pm \sqrt{ (\zeta_\para - \zeta_\perp)^2+4\zeta_\times^2 } $.
Notice also that the semi-positivity should be separately ensured 
in a parallel expansion/compression ($ \theta_\para \not =0, \ \theta_\perp=0 $)
and in a perpendicular one ($ \theta_\perp \not =0, \ \theta_\para=0 $) as necessary conditions, 
requiring that $ \zeta_\para \geq0 $ and $ \zeta_\perp \geq0 $.
Then, we find an inequality, $  \zeta_\para  \zeta_\perp -  \zeta_\times ^2 \geq 0 $, 
from the above eigenvalues. Summarizing, we found five inequalities 
\begin{eqnarray}
\label{eq:ineqalities}
\eta_\para \geq 0
\, , \ \ \
\eta_\perp \geq 0
\, , \ \ \
 \zeta_\para \geq 0
\, , \ \ \
 \zeta_\perp \geq 0
\, , \ \ \
 \zeta_\para  \zeta_\perp -  \zeta_\times ^2 \geq 0
\, .
\end{eqnarray}
Since the third and fifth (fourth and fifth) inequalities imply the fourth (third) inequality,
one can get rid of either third or fourth inequality.  
See Appendix~\ref{sec:MHD-viscosities} for more discussions 
and comparisons to the results in the literature.
We also note that the second law of local thermodynamics does {\it not}
require the sign of $ \zeta_\times  $ to be semi-positive,
but requires an inequality among $ \zeta_\times  $ and $ \zeta_{\para,\perp}$.

\section{Nonequilibrium statistical operator method for relativistic MHD}
\label{sec:mhd:qed}

From the underlying quantum field theory, one can also derive RMHD equations by generalizing the nonequilibrium statistical operator method which was initiated by some Japanese physicists in 1950s~\cite{Nakajima,Mori1}, further developed in 1960-1970s~\cite{McLennan,McLennan1,Kawasaki-Gunton,Zubarev:1979,Zubarev1,Zubarev2}, and sophisticated quite recently~\cite{Sasa2014,Becattini:2014yxa,Hayata:2015lga,Hongo:2016mqm,Hongo:2018nzb,Becattini:2019dxo,Hongo:2019rbd,Hongo:2020qpv}. We here review such an approach for deriving RMHD~(see Ref.~\cite{Hongo:2020qpv} for more details).

\subsection{Optimized perturbation with local Gibbs distribution}

The vital point in the nonequilibrium statistical operator method
is to correctly identify the appropriate form of the density operator.
As we already discussed, the dynamical variables in RMHD are the energy-momentum density $T^0_{~\mu}$ and the magnetic flux density $\tilF^{0\mu}$ in a given coordinate system. Let us first identify these operators by considering QED as an underlying quantum theory. The QED Lagrangian is given by
\begin{equation}
\label{eq:QED-Lagrangian}
 \Lcal_{\qed} = 
 \rmi \ppsi \gamma^{\mu} D_\mu \psi - m \ppsi \psi 
  - \frac{1}{4} \eta^{\mu\nu} \eta^{\alpha\beta} F_{\mu\alpha} F_{\nu\beta},
\end{equation}
where we introduced the Dirac field $\psi$ with electric charge $q$, its Dirac conjugate $\ppsi=\psi^\dag\gamma^0$, and the $\mathrm{U}(1)$ gauge field $A_\mu$. 
We also defined the covariant derivative
$D_\mu \psi \equiv \partial_\mu \psi - \rmi q A_\mu \psi$ and
the field strength tensor $F_{\mu\nu} \equiv \partial_\mu A_\nu - \partial_\nu A_\mu$. The Poincar\'e symmetry and the Bianchi identity enables us to find that the operators~\footnote{To keep the notations simple, we use the same symbol for operator $\tilde F^{\m\n}$ and its expectation value, $\lan\tilde F^{\m\n}\ran$, the meaning should be self-explained in the context.}, 
\begin{equation}
 \hT^\mu_{~\nu}
  = \rmi \ppsi \gamma^\mu D_\nu \psi + F^{\mu\rho} F_{\nu\rho} 
  - \delta^\mu_\nu \Lcal_{\qed}
  ,\quad 
 \tilF^{\mu\nu} = \frac{1}{2} \varepsilon^{\mu\nu\rho\sigma} F_{\rho\sigma},
 \label{eq:current-operators}
\end{equation}
satisfy the following Ward-Takahashi identities:
\begin{equation}
 \partial_\mu \hT^\mu_{~\nu} (t,\bx) = 0 , \quad 
  \partial_\mu \tilF^{\mu\nu} (t,\bx) = 0 .
  \label{eq:WT-identity}
\end{equation}
One may wonder how the Bianchi identity is related to the symmetry of QED.
In fact, it is not so obvious how the QED Lagrangian is equipped with the corresponding  symmetry --- the magnetic $\mathrm{U}(1)$ one-form symmetry ---  since it does not act on the local operator sitting in Eq.~\eqref{eq:QED-Lagrangian}. 
Rather, it acts on the line operator
$\displaystyle{\exp \left( \rmi \int_C \diff x^\m \tilA_\m \right)}$, called the t' Hooft line operator composed of a dual gauge field $\tilA_\m$ defined by 
$\tilF_{\m\n} = \pt_\m \tilA_\n - \pt_\n \tilA_\m $.
The insertion of the t'Hooft line is regarded as putting the test particle with the magnetic charge and is analogous to the insertion of the Wilson line~\cite{Gaiotto:2014kfa}.
Thus, considering Eq.~\eqref{eq:WT-identity} as the equations of motion is on the canonical line to set the starting point in constructing a symmetry-based effective theory.

We then identify $T^{\mu\nu}$ and $\tilF^{\mu\nu}$ in the previous section with expectation values of the above quantum operators $\hT^{\mu\nu}$ and $\tilF^{\mu\nu}$.
This identification motivates us to specify the appropriate density operator 
as the one describing fixed expectation values of our dynamical variables $\average{\hT^{0}_{~\nu} (t,\bx)}$ and $\average{\tilF^{0\mu}(t,\bx)}$.
The so-called local Gibbs (LG) distribution $\hrhoLG [\lambda_t; t]$ realizes such an density operator, which is parameterized by a set of the Lagrange multipliers $\lambda_t \equiv \{\beta^\mu (t,\bx) , \Hcal_\nu (t,\bx)\}$ as 
\begin{equation}
 \hrhoLG [\lambda_t; t] \equiv \exp ( - \hS [\lambda_t; t] ), 
\end{equation}
where we introduced the entropy functional operator 
\begin{equation}
 \hS [\lambda_t;t] \equiv 
  - \int \diff^3 x 
  \left[
   \hT^0_{~\nu} (t,\bx) \beta^\nu (t,\bx) 
   + \tilF^{0\nu} (t,\bx) \Hcal_\nu (t,\bx)
  \right]
  + \Psi [\lambda_t].
  \label{eq:entropy-operator}
\end{equation}
Here, the first argument $\lambda_t$ in $\hrhoLG [\lambda_t;t]$ and $\hS [\lambda_t;t]$ describes their functional dependence on the Lagrange multipliers, $\lambda_t \equiv \{\beta^\mu (t,\bx), \Hcal_\nu (t,\bx)\}$, while the second one $t$ represents the time argument for a set of the conserved charge-density operators $\hc_t \equiv \{\hT^{0}_{~\nu} (t,\bx), \tilF^{0\mu}(t,\bx) \}$. One finds that these Lagrange multipliers can be decomposed as 
$\beta^\mu (t,\bx) = \beta (t,\bx) u^\mu (t,\bx)$ and $\Hcal_\mu (t,\bx) = \beta (t,\bx) H_\mu (t,\bx)$ with the local inverse temperature 
$\beta (t,\bx)$, the fluid four-velocity $u^\mu (t,\bx)$, and the magnetic field $H_\mu (t,\bx)$.
In the following, we express the average of a quantum operator $\hOcal$ over the LG distribution as 
\begin{equation}
 \averageLG{\hOcal}_{\lambda_t} \equiv 
  \Tr ( \hrhoLG [\lambda_t; t] \hOcal ).
  \label{eq:average-LG-t}
\end{equation}

In Eq.~\eqref{eq:entropy-operator}, we also defined the local thermodynamic functional $\Psi [\lambda_t]$, called the Massieu-Planck functional, as a normalization factor of the LG distribution
\begin{equation}
 \Psi [\lambda_t] \equiv \log \Tr \exp 
  \left( 
   \int \diff^3 x 
  \left[
   \hT^{0}_{~\mu} (t,\bx) \beta^\mu (t,\bx) 
   + \tilF^{0\mu} (t,\bx) \Hcal_\mu (t,\bx)
  \right]
  \right).
  \label{eq:MP-functional-def}
\end{equation}
This functional is used to extract the average charge densities 
\begin{equation}
 \averageLG{\hT^0_{~\mu} (t,\bx)}
  = \frac{\delta \Psi [\lambda_t]}{\delta \beta^\mu (t,\bx)}, \quad 
 \averageLG{\tilF^{0\mu} (t,\bx)}
  = \frac{\delta \Psi [\lambda_t]}{\delta \Hcal_\mu (t,\bx)}.
\end{equation}
Besides, we define the entropy functional $S [c_t]$ by taking the average of Eq.~\eqref{eq:entropy-operator} over $\hrhoLG[\lambda_t;t]$ as
\begin{equation}
 S [c_t] 
  \equiv - \int \diff^3 x 
  \left[
   \averageLG{\hT^0_{~\nu} (t,\bx)}_{\lambda_t} \beta^\nu (t,\bx) 
   + \averageLG{\tilF^{0\nu} (t,\bx)}_{\lambda_t} \Hcal_\nu (t,\bx)
  \right]
  + \Psi [\lambda_t].
\end{equation}
In other words, the entropy functional $S [c_t]$ is defined by 
the Legendre transform of the Massieu-Planck functional $\Psi [\lambda_t]$, and thus, its argument is the averaged charge densities 
$c_t = \{ \averageLG{\hT^{0}_{~\nu} (t,\bx)}_{\lambda_t}, \averageLG{\tilF^{0\mu}(t,\bx)}_{\lambda_t} \}$.
One then finds that the Lagrange multipliers, or local thermodynamic variables conjugate to the conserved charge densities, as 
\begin{equation}
 \beta^\mu (t,\bx)
  = - \frac{\delta S[c_t]}{\delta T^0_{~\mu} (t,\bx)}, \quad
  \Hcal_\mu (t,\bx)
  = - \frac{\delta S[c_t]}{\delta \tilF^{0\mu} (t,\bx)},
  \label{eq:conjugate-variables}
\end{equation}
which are consistent with Eq.~\eqref{eq:1st-law-B} in the previous section.

To describe the dissipative transport phenomena with the nonequilibrium statistical operator method, we require a crucial assumption that the density operator $\hrho_0$ at the initial time $t_0$ be given by the LG distribution 
$\hrho_0 = \hrhoLG [\lambda_{t_0}; t_0]$~\cite{Zubarev:1979,Zubarev1,Zubarev2,Sasa2014,Hayata:2015lga,Hongo:2020qpv}. Using the Heisenberg picture, we define the expectation value of an arbitrary Heisenberg operator $\hOcal (t)$ at time $t (\geq t_0)$ as 
\begin{equation}
 \average{\hOcal (t)} \equiv 
  \Tr [ \hrho_0 \hOcal (t)]
  = \averageLG{ \hOcal (t) }_{\lambda_{t_0}},
  \label{eq:average-exact}
\end{equation}
where we expressed the rightmost side using the above assumption and Eq.~\eqref{eq:average-LG-t}.
Then, according to our identification, the averaged Ward-Takahashi identities,
\begin{equation}
 \partial_\mu \average{\hT^\mu_{~\nu} (t,\bx) } = 0 , \quad 
  \partial_\mu \average{\tilF^{\mu\nu} (t,\bx)} = 0 ,
\end{equation}
should provide the RMHD equations after an appropriate derivative expansion is employed.

From this microscopic point of view, we have already fixed the definition of expectation values $\average{\hT^\mu_{~\nu} (t,\bx) }$ and $\average{\tilF^{\mu\nu} (t,\bx)}$ by Eq.~\eqref{eq:average-exact}. 
Thus, the remaining problem is to derive the constitutive relations 
for Eq.~\eqref{eq:average-exact} based on our density operator $\hrho_0 = \hrhoLG [\lambda_{t_0}; t_0]$. 
However, recalling the result in the previous section, one realizes that this is a tough problem; We expect that the resulting constitutive relations should be expressed by the conjugate variables $\lambda_t$ at time $t$, whereas our density operator $\hrho_0 = \hrhoLG [\lambda_{t_0}; t_0]$ only contains those $\lambda_{t_0}$ at the initial time $t_0$.
In fact, the expectation value at time $t \, (>t_0)$ in the present setup is always defined by taking average over the initial density operator $\hrho_0 = \hrhoLG [\lambda_{t_0}; t_0]$, which only contains conjugate variables at the initial time.
Thus, it sounds impossible to express $\average{\hT^\mu_{~\nu} (t,\bx) }$ and $\average{\tilF^{\mu\nu} (t,\bx)}$ in terms of $\lambda_t$ at time $t$ as we did in the previous section.

The above observation implies that the initial density operator 
$\hrho_0 = \hrhoLG [\lambda_{t_0}; t_0]$ does not give a useful starting point to evaluate the expectation value at later time $t (>t_0)$.
Instead, one immediately finds that the better starting point is the local Gibbs distribution at time $t$.
The question is how we can shift into such a different distribution when we do not even have conjugate variables other than those at initial time $t_0$.

We can resolve this problem by invoking the optimized (or renormalized) perturbation theory~(see, e.g., Refs.~\cite{Stevenson:1981vj,Kleinert2009path,Jakovac2015resummation}).
Suppose that we know the configuration of the conserved charge densities at time $t$. 
With the help of the entropy functional $S[c_t]$, we first define the conjugate variable $\lambda_t$ by Eq.~\eqref{eq:conjugate-variables}.
One can show that this definition is equivalent to requiring the following matching conditions
\begin{equation}
 \average{T^0_{~\nu}} = \averageLG{T^0_{~\nu}}_{\lambda_t}, \quad
 \average{\tilF^{0\nu}} = \averageLG{\tilF^{0\nu}}_{\lambda_t}.
\end{equation}
Using the defined conjugate variables $\lambda_t$, we rearrange our density operator as 
\begin{equation}
 \hrho_0 = 
  \rme^{ - \hS [\lambda_t;t] + \hSigma [t,t_0;\lambda]}
  = \rme^{- \hS [\lambda_t;t]} \mathrm{T}_\tau \exp 
  \left( \int_0^1 \diff \tau \rme^{\tau\hS[\lambda_t;t]} 
   \hSigma [t,t_0;\lambda] \rme^{-\tau\hS[\lambda_t;t]} \right),
  \label{eq:density-operator-1}
\end{equation}
where we just added and subtracted the entropy operator $\hS [t;\lambda_t]$ 
at time $t$ and defined the entropy production operator $\hSigma [t,t_0;\lambda]$ as 
\begin{equation}
 \hSigma [t,t_0;\lambda]
  \equiv \hS [\lambda_t;t] - \hS [\lambda_{t_0};t_0].
\end{equation}
The rightmost side of Eq.~\eqref{eq:density-operator-1} gives a useful formula for the derivative expansion since the entropy production operator $\hSigma [t,t_0;\lambda]$ will be shown to be $O(\partial^1)$. 
Thus, we can regard $\hSigma [t,t_0;\lambda]$ as a derivative correction, and 
Eq.~\eqref{eq:density-operator-1} gives a familiar perturbative expansion formula in the interacting picture, which we learn in the elementary course of quantum mechanics.
As a result, we see that the nonequilibrium statistical operator method gives one useful expansion scheme relying on the new optimized (or renormalized) parameter $\lambda_t$.

Expanding the rightmost side of Eq.~\eqref{eq:density-operator-1} at the first-order in $\hSigma[t,t_0;\lambda]$, we obtain 
\begin{equation}
 \begin{split}
  \average{\hT^{\mu\nu} (t,\bx)} 
  &= \averageLG{\hT^{\mu\nu} (t,\bx)}_{\lambda_t}
  + \ip{\hT^{\mu\nu} (t,\bx), \hSigma[t,t_0;\lambda]}_t
  + O (\hSigma^2) ,
  \\
  \average{\tilF^{\mu\nu} (t,\bx)} 
  &= \averageLG{\tilF^{\mu\nu} (t,\bx)}_{\lambda_t}
  + \ip{\tilF^{\mu\nu} (t,\bx), \hSigma[t,t_0;\lambda]}_t
  + O (\hSigma^2) ,
 \end{split}
 \label{eq:expansion}
\end{equation}
where we introduced the Kubo-Mori-Bogoliubov inner product:
\begin{equation}
 \ip{\hA,\hB}_t \equiv 
  \int_0^1 \diff \tau 
  \averageLG{\rme^{\tau\hS[\lambda_t;t]} 
  \hA \rme^{-\tau\hS[\lambda_t;t]} \hB^{\dag} }_{\lambda_t}.
\end{equation}
Equation \eqref{eq:expansion} indicates that we have separated the problem into two parts: The first one is to evaluate the expectation value of currents 
with the LG distribution describing the local thermal equilibrium
\begin{equation}
 T^{\mu\nu}_{(0)} = \averageLG{\hT^{\mu\nu} (t,\bx)}_{\lambda_t} , \quad 
  \tilF^{\mu\nu}_{(0)} = \averageLG{\tilF^{\mu\nu} (t,\bx)}_{\lambda_t},
  \label{eq:LG-average}
\end{equation}
which will be shown to contain the leading-order terms in derivative. The second one is to find the dissipative corrections
\begin{equation}
 T^{\mu\nu}_{(1)} = \ip{\hT^{\mu\nu} (t,\bx), \hSigma[t,t_0;\lambda]}_t, 
  \quad 
  \tilF^{\mu\nu}_{(1)} = \ip{\tilF^{\mu\nu} (t,\bx), \hSigma[t,t_0;\lambda]}_t,
  \label{eq:dissipative-correction}
\end{equation}
by computing the entropy production.

\subsection{Evaluating the Local Gibbs averages}

Let us first investigate the LG-averaged currents given in Eq.~\eqref{eq:LG-average}.
In evaluating these expectation values, it is useful to put our system in the curved spacetime described by the vierbein $e_\mu^{~a}$ and introduce a background two-form gauge field $b_{\mu\nu}$ that couples to $\tilF^{\mu\nu}$.
In the presence of these background fields, one can show the following variational formulas~\cite{Hongo:2020qpv}
\begin{equation}
 \begin{split}
  \averageLG{\hT^{\mu\nu} (t,\bx)}_{\lambda_t}
  &= \frac{1}{\beta^{0} \sqrt{\gamma}} \frac{\delta \Psi[\lambda_t]}{\delta e_\mu^{~a} (t,\bx)} e^\nu_{~a},
  \\
  \averageLG{\tilF^{\mu\nu} (t,\bx)}_{\lambda_t}
  &= \frac{2}{\beta^{0} \sqrt{\gamma}}
  \left(
  \beta^{[\mu} \delta^{\nu]}_\rho
  \frac{\delta \Psi[\lambda_t]}{\delta \Hcal_\rho (t,\bx)}
  + \frac{\delta \Psi[\lambda_t]}{\delta b_{\mu\nu} (t,\bx)}
  \right),
 \end{split}
 \label{eq:variational-formula}
\end{equation}
where $e^\nu_{~a}$ is the inverse of vierbein and $\beta^0$ denotes the zeroth-component of the four vector $\beta^\m$. Here, we also used $\gamma \equiv \det g_{ij}$ with a spatial part of the metric $g_{ij} = e_i^{~a} e_j^{~b} \eta_{ab}$.
Thus, we can derive the LG expectation values $\averageLG{\hT^{\mu\nu} (t,\bx)}_{\lambda_t}$ and $\averageLG{\tilF^{\mu\nu} (t,\bx)}_{\lambda_t}$ once we get the form of the local thermodynamic functional $\Psi[\lambda_t]$ under the background fields.

To specify the form of the Massieu-Planck functional $\Psi[\lambda_t]$, 
we can rely on the path-integral formula~\cite{Hongo:2016mqm,Hongo:2020qpv}. 
Substituting Eq.~\eqref{eq:current-operators} into the definition 
\eqref{eq:MP-functional-def} of $\Psi[\lambda_t]$ and following the usual procedure of deriving the path-integral representation for the partition function, we obtain
\begin{equation}
 \Psi [\lambda_t]
  = \log \int \Dcal \psi \Dcal \ppsi \Dcal A_\mu \delta (F) 
  \det (\partial F /\partial \alpha) \exp 
  \left( \int_0^{\beta_{\mathrm{ref}}} \diff \tau \diff^3 x 
   \sqrt{-\tilg} \tilLcal_{\qed}
  \right),
  \label{eq:path-integral-formula}
\end{equation}
where $\beta_{\mathrm{ref}}$ is an arbitrary constant reference inverse temperature and $F$ denotes a gauge fixing condition $F = 0$ with a gauge parameter 
$\alpha$.
Due to the inhomogeneity of the local thermodynamic variables $\lambda_t$, 
we need to perform the path-integral not in the flat Euclidean spacetime but in the emergent curved spacetime with the two-form gauge field.
In fact, we find the Lagrangian density in Eq.~\eqref{eq:path-integral-formula} 
to be
\begin{equation}
 \tilLcal_{\qed}
  =  \rmi \ppsi \gamma^a \tile_a^{~\mu} \tilD_\mu \psi - m \ppsi \psi 
  - \frac{1}{4} \tilg^{\mu\nu} \tilg^{\alpha\beta} F_{\mu\alpha} F_{\nu\beta}
  + \frac{1}{2} \tilb_{\mu\nu} \tilF^{\mu\nu},
\end{equation}
and the background fields --- the thermal vierbein $\tile_\mu^{~a}$ 
and thermal two-form gauge field $\tilb_{\mu\nu}$ --- are given by 
\begin{equation}
 \tile_0^{~a} \equiv \rme^\sigma u^a, \quad 
  \tile_i^{~a} \equiv e_i^{~a}, \quad 
  \tilb_{0i} \equiv \rme^\sigma H_i, \quad 
  \tilb_{ij} \equiv b_{ij},
\end{equation}
where we defined $\rme^{\sigma (x)} = \beta(x)/\beta_{\mathrm{ref}}$. 
We also defined the thermal metric 
$\tilg_{\mu\nu} \equiv \tile_\mu^{~a} \tile_\nu^{~b} \eta_{ab}$ with
$\tilg \equiv \det \tilg_{\mu\nu}$, and the covariant derivative
\begin{equation}
 \tilD_\mu \psi \equiv \tilpartial_\mu \psi - \rmi q A_\mu \psi 
  - \frac{\rmi}{2} \tilomega_\mu^{~ab} \Sigma_{ab} \psi,
\end{equation}
with the representation matrix of the Dirac spinor under the Lorentz transformation $\Sigma_{ab} \equiv \rmi [\gamma_a,\gamma_b]/4$. 
The derivative in the thermal space is given as  $\tilpartial_\mu = (\rmi \partial_\tau , \partial_i)$ 
and the spin connection $\tilomega_\mu^{~ab}$ is determined by the thermal vierbein $\tile_\mu^{~a}$ as~\cite{Hongo:2016mqm}
\begin{equation}
 \tilomega_\m^{~ab}
 = \frac{1}{2} \tile^{a\n} \tile^{b\r} 
 (\tilC_{\n\r\m} - \tilC_{\r\n\m} - \tilC_{\m\n\r})
 \with 
 \tilC_{\m\n\r} \equiv \tile_\m^{~c} 
 \big( \tilpartial_\n \tile_{\r c} - \tilpartial_{\r} \tile_{\n c} \big).
\end{equation}

%We note the spatial components of background fields are given by the original background field $e_i^{~a}$ and $b_{ij}$, while the local thermodynamic variables determine temporal components.

In short, the Massieu-Planck functional is described by performing the path integral for QED in the presence of the curved and two-form backgrounds. 
This result is a generalization of the background field method to the locally thermalized system. 
We can read off the symmetry properties of the Massieu-Planck functional as follows.
To see this, note that the line element $\diff \tils^2$ and two-form gauge connection $\tilb$,
\begin{align}
 \diff \tils^2 
 &\equiv \tilg_{\mu\nu} \diff \tilx^\mu \otimes \diff \tilx^\nu
 = - \rme^{2\sigma} (\diff \tilt + a_i \diff x^i)^2 
 + g^{\prime}_{ij} \diff x^i \otimes \diff x^j, 
 \\
 \tilb
 &\equiv \frac{1}{2} \tilb_{\mu\nu} \diff \tilx^\mu \wedge \diff \tilx^\nu
 = \tilb_{0i} (\diff \tilt + a_j \diff x^j) \wedge \diff x^i
 + \frac{1}{2} \tilb_{ij}^{\prime} \diff x^i \wedge \diff x^j,
\end{align}
describe the backgrounds. 
Here, we defined $(\diff \tilt, \diff \tilbx) = (- \rmi \diff \tau, \diff \bx)$ and expressed $\diff \tils^2 $ and $\tilb$ using the Kaluza-Klein parameterization
\begin{equation}
 a_i \equiv - \rme^{-\sigma} u_i, \quad 
  g^\prime_{ij} \equiv g_{ij} + u_i u_j, \quad 
  \tilb_{ij}^{\prime} \equiv b_{ij} - \tilb_{0j} a_i - \tilb_{i0} a_j. 
\end{equation}
The apparently complicated Kaluza-Klein parameterization is, indeed, useful because the Massieu-Planck functional is invariant under the Kaluza-Klein gauge transformation 
\begin{equation}
 \begin{split}
  \tilt 
  &\to \tilt^{\prime} = \tilt + \chi (\bx) ,
  \\
  \bx 
  &\to \bx^{\prime} = \bx ,
  \\
  a_i (\bx) 
  &\to  a_i^{\prime} (\bx) = a_i (\bx) - \partial_i \chi (\bx) ,
 \end{split}
 \label{eq:KK}
\end{equation}
and $\rme^{\sigma}$, $\gamma_{ij}^{\prime}$ and $\tilb_{0i}$  and $\tilb_{ij}^{\prime}$ are all invariant under the Kaluza-Klein gauge transformation.
Besides, the Massieu-Planck functional is invariant under the spatial diffeomorphism and gauge transformation acting on $b_{ij}$,
\begin{align}
 x^i &\to x^{i\prime} = f^i (\bx) , 
 \label{eq:spatial-diffeo}
 \\
 \tilb_{ij} &\to \tilb_{ij} + \partial_i \theta_j - \partial_j \theta_i.
 \label{eq:one-form-gauge}
\end{align}

The crucial point here is that the Massieu-Planck functional has to respect the 
symmetries under the transformations \eqref{eq:KK}-\eqref{eq:one-form-gauge}. 
Using this symmetry property and relying on the derivative expansion, one can write down the most general form of the Massieu-Planck functional in an  order-by-order basis in derivative. 
In the leading-order expansion, we have two zeroth-order invariant scalars
$\rme^{\sigma(x)} = \beta (x) /\beta_{\mathrm{ref}}$ and 
$\sqrt{\big(\tilb_{0i}(x) \big)^2} = \rme^{\sigma (x)} H(x) = \Hcal (x)/\beta_{\mathrm{ref}} $, where we decomposed the magnetic field as
$H_\m = - H b_\m$ and $\Hcal_\m = -\Hcal b_\m$ with the normalized spatial vector $b^\m$.
Moreover, there is no invariant scalar at $O(\partial^1)$, and thus, the most general form of $\Psi [\lambda]$ in the leading-order expansion is given by
\begin{equation}
 \begin{split}
  \Psi [\lambda] 
  &= \int_0^{\beta_{\mathrm{ref}}} \diff \tau \diff^3 x 
  \sqrt{-\tilg} p (\beta,\Hcal) 
  + O (\partial^2)
  % \\
  % &
  = \int \diff^3 x \sqrt{\gamma} \beta^0 p (\beta,\Hcal) 
  + O (\partial^2).
 \end{split}
 \label{eq:leading-MP}
\end{equation}
Recalling the variational formula \eqref{eq:variational-formula}, we find that the functional derivative of Eq.~\eqref{eq:leading-MP} leads to
\begin{equation}
 \begin{split}
  \averageLG{\hT^{\mu\nu} (t,\bx)}_{\lambda_t}
  &= \epsilon u^\mu u^\nu - p_{\perp} (\eta^{\mu\nu} - u^\mu u^\nu + b^\mu b^\nu)
  + p_{\para} b^\mu b^\nu ,
  \\
  \averageLG{\tilF^{\mu\nu} (t,\bx)}_{\lambda_t}
  &= B (b^\mu u^\nu - b^\nu u^\mu) ,
 \end{split}
 \label{eq:leading-order}
\end{equation}
where we have taken the flat background limit and introduced a set of the scalar functions
\begin{equation}
 \epsilon = - \frac{\partial (\beta p)}{\partial \beta}, \quad 
  p_{\perp} = p , \quad 
  p_{\para} = p - \beta^{-1} \Hcal B, \quad 
  B = \beta \frac{\partial p}{\partial \Hcal}.
  \label{eq:scalars}
\end{equation}
All of them can be extracted from the single function $p (\beta,\Hcal)$.
Equations \eqref{eq:leading-order}-\eqref{eq:scalars} give the 
leading-order constitutive relations of RMHD, which agree with Eq.~\eqref{eq:TF_zeroth-result} in the previous section (recall $\beta^{-1} \Hcal B= H^\mu B_\mu$). In contrast to the entropy-current analysis, we now have the microscopic path-integral formula for $p (\beta,\Hcal)$. One can thus, in principle, compute all coefficient functions in Eq.~\eqref{eq:scalars}, or the equations of state, from the underlying microscopic theory, i.e., QED.

\subsection{Evaluating the dissipative corrections}

We next evaluate the dissipative corrections given in Eq.~\eqref{eq:dissipative-correction}.
For this purpose, we first rewrite the entropy production operator by using the Ward-Takahshi identity \eqref{eq:WT-identity} and performing integration by parts. 
The resultant expression reads 
\begin{align}
 \hSigma [t,t_0;\lambda] 
 &\equiv \hS [\lambda_t;t] - \hS [\lambda_{t_0};t_0]
 = \int_{t_0}^t \diff t' \partial_{t'} \hS[\lambda_{t'}; t']
 \nonumber \\
 &=  - \int_{t_0}^t \diff t' 
 \partial_{t'}
 \left[
 \int
 \diff^3 x^{\prime}
 \left(
 \beta^\nu (t',\bx^{\prime}) 
 \hT^{0}_{~\nu} (t',\bx^{\prime}) 
 + \Hcal_{\nu} (t',\bx^{\prime}) \tilF^{0\nu} (t',\bx^{\prime})
 \right)
 + \Psi [\lambda_{t'}]
 \right]
 \nonumber \\
 &= - \int_{t_0}^t \diff t' \diff^3 x^{\prime}
 \left[
 \delta \hT^{\mu\nu} (t^{\prime},\bx^{\prime}) 
 \partial_\mu \beta_\nu (t^{\prime},\bx^{\prime})
 + \delta \tilF^{\mu\nu} (t^{\prime},\bx^{\prime})
 \partial_\mu \Hcal_\nu (t^{\prime},\bx^{\prime})
 \right],
 \label{eq:Entropy-prod-1}
\end{align}
where we defined a deviation 
$\delta \hOcal (t) \equiv \hOcal (t) - \averageLG{\hOcal(t)}_{\lambda_t}$. 
To obtain the third line, we also used the following identity:
\begin{equation}
 \begin{split}
  \partial_t \Psi [\lambda_t] 
  &= - \baverageLG{
  \partial_t 
  \int \diff^3 x
  \left(
  \beta^\nu \hT^{0}_{~\nu} + \Hcal_{\nu} \tilF^{0\nu} 
  \right)}_{\lambda_t}
  \\
  &= \int \diff^3 x 
  \left(
  \averageLG{\hT^{\mu\nu}}_{\lambda_t} \partial_\mu \beta_\nu
  + \averageLG{\tilF^{\mu\nu}}_{\lambda_t} \partial_\mu \Hcal_\nu
  \right).
 \end{split}
\end{equation}
Note that the entropy production operator $\hSigma [t,t_0;\lambda]$ in Eq.~\eqref{eq:Entropy-prod-1} contains the time derivative of the parameters $\lambda_{t^{\prime}}$.
This can be explicitly seen by using the projection tensor, 
$\delta^\mu_\nu = u^\mu u_\nu + \Delta^\mu_\nu$, 
that decomposes a derivative as 
\begin{equation}
 \partial_\mu = u_\mu D + \nabla_{\mu} \with
  D \equiv u^\mu \partial_\mu , \quad
  \nabla_{\mu} \equiv \Delta^\nu_\mu \partial_\nu.
\end{equation}
Substituting this decomposition together with $\beta^\mu = \beta u^\mu$ and $\Hcal_\nu = - \Hcal b_\nu$, we obtain the following expression for the entropy production operator
\begin{equation}
\label{eq:Entropy-production-with-time-derivative}
 \begin{split}
  \hSigma [t,t_0;\lambda] 
  = - \int_{t_0}^t \diff t' \diff^3 x^{\prime}
  &\Big[
  u_\mu u_\nu \delta \hT^{\mu\nu} D \beta
  + u_\mu \delta \hT^{\mu\nu} \beta D u_\nu 
  - u_\mu b_\nu \delta \tilF^{\mu\nu} D \Hcal
  - \Hcal u_\mu \delta \tilF^{\mu\nu} D b_\nu
  \\
  &+ \beta \delta \hT^{\mu\nu} \nabla_{\mu} u_\nu 
  + u_\nu \delta \hT^{\mu\nu} \nabla_{\mu} \beta
  + \Hcal \delta \tilF^{\mu\nu} b^\rho u_\nu \nabla_{\mu} u_\rho
  + \delta \tilF^{\mu\nu} \Delta_\nu^\rho \nabla_{\mu} \Hcal_\rho
  \Big].
 \end{split}
\end{equation}
Here, it is important to eliminate time-derivative terms in the first line. 
Otherwise, the resulting Green-Kubo formulas pick up contributions from gapless linear hydrodynamic modes, which prevents us from obtaining 
convergent time integrals for the transport coefficients. 
In Ref.~\cite{Hongo:2020qpv}, we accomplish such a procedure by a formal manipulation. 
To complement that formal manipulation, we will here explicitly demonstrate this procedure.

To eliminate the time-derivative terms, we solve the leading-order RMHD equations following the leading-order constitutive relations~\eqref{eq:leading-order}.
Using Eq.~\eqref{eq:leading-order}, we find the leading-order equations of motion 
\begin{subequations}
 \begin{align}
  0&= 
  \pt_\m T^{\mu\nu}_{(0)}
  = \pt_\m
  \big[
  \epsilon u^\mu u^\nu 
  - p_{\perp} (\eta^{\mu\nu} - u^\mu u^\nu )
  + B^\mu H^\nu 
  \big]
  ,
  \\
  0 &= \pt_\m \tilF^{\mu\nu}_{(0)}
  = \pt_\m [B^\mu u^\nu - B^\nu u^\mu ],
 \end{align}
\end{subequations}
where we used $p_{\para} =%= p_{\perp} - H_\mu B^\mu = 
p_{\perp} - HB$ 
to express $p_{\para}$ in terms of $B^\mu = B b^\nu$ and $H^\nu = - H b^\nu$. 
After contracting these ideal RMHD equations with appropriate tensors such as 
$b_\m$, we obtain the following set of equations:
\begin{subequations}
 \label{eq:RMHD-ideal-eom}
 \begin{align}
  \label{eq:De-ideal}
  D \epsilon  
  &=  - (\epsilon + p_{\perp} ) \theta
  + HB \theta_{\para},
  \\
  \label{eq:Du-ideal}
  ( \epsilon + p_{\perp} ) D u_\nu 
  &= 
  \nabla_{\nu} p_{\perp}
  - H_\nu \partial_\mu B^\mu
  - \Delta_{\nu\rho} B^\mu \partial_\mu H^\rho,
  \\
  \label{eq:DB-ideal}
  D B &= 
  - B \theta_{\perp},
  \\
  \label{eq:Db-ideal}
  D b^\nu &= 
  \Xi^{\nu\rho} b^\mu \partial_\mu u_\rho
  + B^{-1} u^\nu \partial_\mu B^\mu,
 \end{align}
\end{subequations}
where we used $\theta = \theta_{\para} + \theta_{\perp}$ with 
$\theta_{\para} = - b^\mu b^\nu \partial_\mu u_\nu$ and 
$\theta_{\perp} = \Xi^{\mu\nu} \partial_\mu u_\nu$. Combining these equations with thermodynamic relations, we can further simplify $D u_\nu$ and also find the time derivatives of conjugate variables $D \beta$ and $D \Hcal$ as (see Appendix~\ref{sec:solving-ideal-RMHD} for a derivation)
\begin{subequations} 
\label{eq:eom-conjugate-parameters}
 \begin{align}
 \label{eq:Dbeta-ideal}
  D \beta 
  &= \beta 
  \left( 
  \frac{\partial p_{\perp}}{\partial \epsilon}
  - B \frac{\partial H}{\partial \epsilon}  
  \right) 
  \theta_{\para}
  + \beta \frac{\partial p_{\perp}}{\partial \epsilon}
  \theta_{\perp},
  \\
 \label{eq:DH-ideal}
  D \Hcal 
  &= - \beta \frac{\partial p_{\perp}}{\partial B} \theta_{\perp}
  - \beta 
  \left(
  \frac{\partial p_{\perp}}{\partial B}
  - B \frac{\partial H}{\partial B}
  \right)
  \theta_{\para},
  \\
 \label{eq:Du-ideal}
  D u_\nu 
  &= - \beta^{-1} \nabla_{\nu} \beta
  - \frac{1}{\epsilon + p_{\perp}} 
  \left[
  2 \beta^{-1} B b^\mu \nabla_{[\mu} \Hcal_{\nu]} 
  + \theta_{\para} H B u_\nu 
  \right],
  \\
  D b_\nu
  &= \Xi_{\nu}^{~\rho} b^\mu \nabla_\mu u_\rho
  + B^{-1} u_\nu \partial_\mu B^\mu .
 \end{align}
\end{subequations}

These leading-order equations of motion for thermodynamic parameters enable us to eliminate the time-derivative terms in the entropy production operator.
Substituting them into Eq.~\eqref{eq:Entropy-production-with-time-derivative} and rearranging terms, we eventually obtain the entropy production operator
\begin{equation}
 \label{eq:EntropyProd}
 \begin{split}
  \hSigma [t,t_0;\lambda] 
  = - \int_{t_0}^t \diff t' \diff^3 x 
  &\Big[
  \tildelta \hp_{\para} \beta \theta_{\para} 
  + \tildelta \hp_{\perp} \beta \theta_{\perp}
  + 2 \tildelta \hpi^{(\mu} b^{\nu)} \beta \nabla_{\mu} u_\nu
  + \tildelta \htau^{\mu\nu} \beta \nabla_{\mu} u_\nu
  \\
  &+ 2 \tildelta \hE^{[\mu} b^{\nu]} 
  \nabla_{\mu} \Hcal_\nu
  + \tildelta \hD^{\mu\nu} \nabla_{\mu} \Hcal_\nu
  + O (\nabla^2)
  \Big],
 \end{split}
\end{equation}
where we defined the projected components of 
the operators $\tildelta \hT^{\mu\nu}$ and $\tildelta \tilF^{\rho\sigma}$ as 
\begin{subequations}
\label{eq:projected-current}
 \begin{align}
  \tildelta \hp_\para 
  &= - b_\mu b_\nu \delta \hT^{\mu\nu}
  + u_\mu u_\nu \delta \hT^{\mu\nu} 
  \left( 
  \frac{\partial p_{\perp}}{\partial \epsilon}
  - B \frac{\partial H}{\partial \epsilon}  
  - \frac{HB}{\epsilon + p_{\perp}} 
  + u_\mu b_\nu \delta \tilF^{\mu\nu} 
  \frac{\partial p_{\para}}{\partial B}
  \right) ,
  \\
  \tildelta \hp_\perp 
  &= \frac{1}{2} \delta \hT^{\mu\nu} \Xi_{\mu\nu} 
  + u_\mu u_\nu \delta \hT^{\mu\nu} 
  \frac{\partial p_{\perp}}{\partial \epsilon}
  + u_\mu b_\nu \delta \tilF^{\mu\nu} \frac{\partial p_{\perp}}{\partial B} 
  , 
  \\
  \tildelta \hpi^\mu 
  &=  \left( 
  - b_{\sigma} \delta \hT^{\rho\sigma} 
  + H \delta \tilF^{\rho\sigma} u_\sigma
  \right)
  \Xi^{~\mu}_{\rho}    ,
  \\
  \tildelta \hat \tau^{\mu\nu} 
  &= \Big( \Xi^\mu_{~\rho} \Xi^\nu_{~\sigma} 
  - \frac{1}{2} \Xi^{\mu\nu} \Xi_{\rho\sigma} \Big) \delta \hT^{\rho\sigma}, \\
  \tildelta \hE^\mu 
  &= b_\rho \delta \tilF^{\rho\sigma} \Xi_{\sigma}^{~\mu}
  + \frac{2 B}{\epsilon + p_{\perp}} 
  u_\rho \delta \hT^{\rho\mu}  , 
  \\
  \tildelta \hD^{\mu\nu}
  &= \Xi^\mu_{~\rho} \Xi^\nu_{~\sigma} \delta \tilF^{\rho\sigma} .
 \end{align}
\end{subequations}

We then substitute the obtained entropy production operator into Eq.~\eqref{eq:dissipative-correction}.
Assuming that the correlation of projected operators $\tildelta \hT^{\mu\nu}$ and $\tildelta \tilF^{\mu\nu}$ decays with the microscopic scales, 
we perform the Markovian approximation for the integration kernel. As we emphasized, this approximation does not work if we do not solve the ideal RMHD equations to obtain the projected operator \eqref{eq:projected-current}. After this procedure, we eventually obtain the dissipative corrections 
to the constitutive relations 
\begin{subequations} 
 \label{eq:TJ-constitutive}
 \begin{align}
  T^{\mu\nu}_{(1)} 
  =&- ( \zeta_\parallel \theta_\parallel 
  + \zeta_{\times} \theta_\perp  ) b^\mu b^\nu
  + ( \zeta_\perp \theta_\perp
  + \zeta_{\times}' \theta_\parallel ) \Xi^{\mu\nu}  
  \nonumber
  \\
  &
  - 2 \eta_\parallel
  \left( b^\mu \Xi^{\nu(\rho} b^{\sigma)} 
  + b^\nu \Xi^{\mu(\rho} b^{\sigma)} \right) \nabla_{\rho} u_\sigma
  + 2 \eta_\perp \nabla^{\{ \mu} u^{\nu \}}
  ,
  \label{eq:T-constitutive}
  \\
  \tilF^{\mu\nu}_{(1)}
  =&  2 \rho_{\perp} T
  \left(b^\mu \Xi^{\nu[\rho} b^{\sigma]} -
  b^\nu \Xi^{\mu[\rho} b^{\sigma]}
  \right) \nabla_{\rho} \Hcal_\sigma
  + 2\rho_{\para} T \Xi^{\mu[\rho} \Xi^{\sigma]\nu} \nabla_{\rho} \Hcal_{\sigma}
  .
  \label{eq:J-constitutive}
 \end{align}
 \end{subequations}
 %\pp{[To be consistent with (\ref{eq:rho-tensor}) and  (\ref{eq:T-1-decomposed}), $\eta_\perp$ term should be plus and $ \Xi^{\mu[\rho} \Xi^{\sigma]\nu}$ term should be plus. (The two terms in the first term of F tilde was just interchanged as in (\ref{eq:rho-tensor}), which does not change the sign.)]}
Those tensor structures are the same as those in Eqs.~(\ref{eq:rho-tensor}) and  (\ref{eq:T-1-decomposed}). We defined a set of transport coefficients which are expressed in the form of the spacetime integral of the Kubo-Mori-Bogoliubov inner product: 
\begin{equation}
 \begin{split}
  \zeta_\parallel 
  &= \beta (t,\bx) \int_{-\infty}^{t} \diff t' \diff^3 x' 
  \ip{\tildelta \hp_\para(t,\bx), \tildelta \hp_\para (t',\bx')}_{t} 
  , \\
  \zeta_\perp
  &= \beta (t,\bx) \int_{-\infty}^{t} \diff t' \diff^3 x' 
  \ip{\tildelta \hp_\perp(t,\bx), \tildelta \hp_\perp (t',\bx')}_{t} ,
  \\
  \zeta_\times
  &= \beta (t,\bx) \int_{-\infty}^{t} \diff t' \diff^3 x' 
  \ip{\tildelta \hp_\para (t,\bx), \tildelta \hp_\perp (t',\bx')}_{t} , 
  \\
  \zeta_\times'
  &= \beta (t,\bx) \int_{-\infty}^{t} \diff t' \diff^3 x' 
  \ip{\tildelta \hp_\perp(t,\bx), \tildelta \hp_\para (t',\bx')}_{t} ,
  \\
  \eta_\parallel
  &= \frac{\beta (t,\bx)}{2} \int_{-\infty}^{t} \diff t' \diff^3 x' 
  \ip{\tildelta \hpi^{\mu}(t,\bx), \tildelta \hpi^{\nu} (t',\bx')}_{t} 
  \Xi_{\mu\nu} , 
  \\
  \eta_\perp
  &= \frac{\beta (t,\bx)}{4} \int_{-\infty}^{t} \diff t' \diff^3 x' 
  \ip{\tildelta \hat \tau^{\mu\nu}(t,\bx), \tildelta \hat \tau^{\rho\sigma} (t',\bx')}_{t} 
  \Xi_{\mu\rho} \Xi_{\nu\sigma}  ,  
  \\
  \rho_{\perp}
  &= \frac{\beta (t,\bx)}{2} \int_{-\infty}^{t} \diff t' \diff^3 x' 
  \ip{\tildelta \hE^{\mu}(t,\bx), \tildelta \hE^{\nu} (t',\bx')}_{t} 
  \Xi_{\mu\nu} , 
  \\
  \rho_{\para} 
  &= \frac{\beta (t,\bx)}{2} \int_{-\infty}^{t} \diff t' \diff^3 x' 
  \ip{\tildelta \hD^{\mu\nu}(t,\bx), \tildelta \hD^{\rho\sigma} (t',\bx')}_{t} 
  \Xi_{\mu\rho} \Xi_{\nu\sigma}.
 \end{split}
 \label{eq:GK}
\end{equation}
They are the Green-Kubo formulas~\cite{Green,Nakano,Kubo} for the seven transport coefficients --- three bulk viscosities ($\zeta_{\para}, \zeta_{\perp}, \zeta_{\times}$), 
two shear viscosities ($\eta_{\para}, \eta_{\perp})$, 
and two electric resistivities ($\rho_{\para},\rho_{\perp}$) in RMHD (see Sec.~\ref{sec:currents-Hall}).

Two remarks are in order.
In the previous section, we do not count $\zeta_{\times}^{\prime}$ as an independent transport coefficient. 
This is because the corresponding Green-Kubo formula in \eqref{eq:GK} respects
Onsager's reciprocal relation~\cite{Onsager}: $\zeta_{\times}^{\prime} = \zeta_{\times}$.
This can be shown by performing an expansion around the global equilibrium and using the charge-conjugation and time-reversal symmetries applied to the above Green-Kubo formula for $\zeta_{\times}^{\prime}$.
We also note that the Green-Kubo formulas \eqref{eq:GK} 
automatically provide a set of the simi-positivity constraints given by 
Eq.~\eqref{eq:ineqalities} specified in the previous section. This stems from the property of the Kubo-Mori-Bogoliubov inner product.
Here, it is worth emphasizing that the semi-positivity constraints (and Onsager's reciprocal relation) are not required but derived in the nonequilibrium statistical operator method.

While the expressions in Eq.~\eqref{eq:GK} are given in terms of the Kubo-Mori-Bogoliubov inner product, one can derive a set of more familiar Green-Kubo formulas in terms of the retarded Green's functions. 
For this purpose, we again expand Eq.~\eqref{eq:GK} on top of the global equilibrium. By inserting the convergence factor $\rme^{\epsilon (t'-t)}$ which will be eventually turned off by taking $\epsilon \to 0$ after the whole calculation, 
we generally obtain 
\begin{equation}
 \begin{split}
  % &\quad
  \int_{-\infty}^t \diff t' 
  \rme^{\epsilon (t'-t)} \ip{\hA(t), \hB(t')}_{\mathrm{eq}}
  =  - \frac{1}{\beta} 
  \int \frac{\diff \omega}{2\pi} 
  \mathcal{P} \left( \frac{1}{\omega} \right)
  \frac{\partial}{\partial \omega} G_R^{A,B} (\omega)
  - \frac{\rmi}{2\beta} 
  \lim_{\omega \to 0}
  \frac{\partial}{\partial \omega} G_R^{A,B} (\omega),
 \end{split}
\end{equation}
where $ \mathcal{P}$ stands for the principal value, $\ip{\hA(t), \hB(t')}_{\mathrm{eq}}$ is the global equilibrium limit of the inner product 
$\ip{\hA(t), \hB(t')}_{t}$, and $G_R^{A,B}$ is the retarded Green's function, $G_R^{A,B} (t-t') = -i\theta(t-t') \lan [ \hA(t) , \hB(t') ]\ran_\eq$. With the help of this identity, we can replace the inner products in Eq.~\eqref{eq:GK} with the retarded Green's functions $G_R^{A,B}$.

\section{Interlude: Connection to the conventional MHD}
\label{sec:comparison}

In previous sections, we have reviewed the recent formulation of the RMHD with only the energy-momentum conservation law and the Bianchi identity as relevant equations of motion from the very beginning. On the other hand, the RMHD can also be formulated in a conventional approach in which the Maxwell equation and electric charge conservation law enter as additional dynamical equations. In the following two sections~\ref{sec:kine} and \ref{sec:transport-coeff}, 
we will review hydrodynamics under the strong background magnetic field that is closer to the latter formulation though the magnetic field is non-dynamical.
Thus, in this intermediate section, we discuss the relation between the two formulations, focusing on the anisotropic pressure and the first-order constitutive relations. The second topic also serves as a basis for the following two sections.

\subsection{Anisotropic pressure} 

We start with the correspondence of the zeroth-order terms, 
especially the anisotropic pressure, between the conventional formulation and the formulation we discussed in Sec.~\ref{sec:entropy} and Sec.~\ref{sec:mhd:qed}. In the conventional formulation, one needs to separate the matter and EM components in the system, and connect the two components via the Maxwell equation $   \partial_\mu  F^{\mu\nu}  = J^\nu   $ and energy-momentum {\it non-conservation} equation   $\partial_\mu T^{\mu\nu}_{\rm matt} = F^{\nu\mu}J_\mu$, where $J^\m$ is the electric current and $T^{\mu\nu}_{\rm matt}$ is the matter energy-momentum tensor. To make a connection with our formulation in Sec.~\ref{sec:entropy} and Sec.~\ref{sec:mhd:qed}, let us decompose the total energy density and pressure presented in Sec.~\ref{sec:entropy} into the matter and magnetic components as $  \epsilon \sim \epsilon_{\rm matt} + |\bB|^2/2- {\bm M} \cdot \bB$
and $p \sim p_{\rm matt} + |\bB|^2/2 - {\bm M} \cdot \bB $, respectively~\footnote{
In the above expressions, the Lorentz scalars are given in the rest-frame expressions for clarity.
The energy density including the magnetization $ \epsilon_{\rm matt} - {\bm M} \cdot \bB $
corresponds to the definition of $ \epsilon $ in Ref.~\cite{Huang:2011dc} as stated there.
}. Then, inserting those expressions into Eq.~(\ref{eq:TF_zeroth-result}), we have 
\begin{eqnarray}
\label{eq:T_zeroth2}
T^{\mu\nu}_\zero 
&\simeq& ( \epsilon_{\rm matt} + \frac{1}{2} |{\bm B}|^2 - {\bm M} \cdot \bB) u^\mu u^\nu
\nnb
&&
- (p_{\rm matt} +  \frac{1}{2} |{\bm B}|^2 - {\bm M} \cdot \bB ) \varXi^{\mu\nu}
+ (p_{\rm matt} -  \frac{1}{2} |{\bm B}|^2 )  b^\mu b^\nu
\, .
\end{eqnarray}
The above expression agrees with those in Eqs.~(14) and (15) of Ref.~\cite{Huang:2011dc} (see also
Refs.~\cite{kluitenberg1954relativistic:IV, kluitenberg1954relativistic:V, lichnerowicz1967relativistic, Israel:1978up, gedalin1991relativistic} for classic works): the terms with $|\bm B|^2$ are combined into the Maxwell energy-momentum tensor (see below) and the rest terms give $T^{\m\n}_{{\rm matt}(0)}$.
In this way, the conventional ``ideal MHD limit'' is reproduced from the leading-order result in the new formulation that 
generalizes the conventional formulation in the following points.

\begin{itemize}

\item The right-hand side of the energy-momentum non-conservation law,
$  \partial_\mu T^{\mu\nu}_{\rm matt} = F^{\nu\mu} J_{\mu}$, describes the Joule heat and Lorentz force
which provide the source and/or dissipation of the energy and momentum, respectively.
Similarly, the electric current in the Maxwell equation $   \partial_\mu  F^{\mu\nu}  = J^\nu   $ provides a source of EM fields. This latter equation constrain the electric field as a gapped mode excited by the source term. The new formulation does not contain such a redundancy and can work in the strict hydrodynamic limit.

\item The pressures $p_\perp$ and $p_\parallel$ in Eq.~(\ref{eq:TF_zeroth-result}) satisfy $p_{\perp} - p_{\para} = B^\mu H_\mu (>0)$. Subtracting the magnetic pressures $p_{B\perp}=|\bm B|^2/2$ and $p_{B\parallel}=-|\bm B|^2/2$, we obtain the anisotropic matter pressures as $p_{{\rm matt}\parallel}=p_{\rm matt}$ and $p_{{\rm matt}\perp}=p_{\rm matt}-{\bm M}\cdot{\bm B}$, respectively. One thus finds that $p_{{\rm matt}\perp}-p_{{\rm matt}\parallel}=-\bm M\cdot\bm B$ in the rest frame of the fluid. This leads to $p_{{\rm matt}\perp}<p_{{\rm matt}\parallel}$ since the magnetic susceptibility is usually positive.

\item
These non-conservation equations can be combined together
into the form $  \partial_\mu  (T^{\mu\nu}_{\rm matt} + T^{\mu\nu}_{\rm Maxwell} ) = 0$,
where $  T^{\mu\nu}_{\rm Maxwell} $ is the Maxwell tensor~\footnote{
Explicitly, this means that
$
F^{\nu\mu}J_\mu = \partial_\mu ( F^{\mu \alpha} F^\nu_{\, \ \alpha} - g^{\mu\nu} F^{\alpha\beta} F_{\alpha\beta}/4)
=: - \partial_\mu T^{\mu\nu}_{\rm Maxwell}
$.
}. Therefore, this equation can be reduced from the first equation in Eq.~(\ref{eq:EoMs}) if one assumes a clear separation
between the matter and electromagnetic contributions to the energy-momentum tensor as in Eq.~(\ref{eq:T_zeroth2}). 
However, it would not be possible to separate those contributions in a strongly coupled system
where excitations are composed of mixture of matter and electromagnetic fields.
Moreover, hydrodynamic framework itself should not care such microscopic details of the system,
and the translational symmetry of the system only tells us the conservation of the {\it total} energy-momentum. 
Those facts should be respected in the formulation. 

\item Excluding an electric field from the set of hydrodynamic variables,
one does not need to assume an ``infinite electric conductivity'' as in the conventional formulation
(see, e.g., Refs.~\cite{lichnerowicz1967relativistic,davidson2002introduction}).
Note that the electric conductivity is a dimensionful quantity
and is, moreover, not defined {\it a priori} in the formulation of hydrodynamics.
If it implies an infinitesimally short relaxation time, there would be also a conceptual conflict when
one tries to include (finite) dissipative effects such as a viscosity in the derivative corrections.
The formulation in Sec.~\ref{sec:entropy} and Sec.~\ref{sec:mhd:qed} is free of such a dilemma.

\end{itemize}

\subsection{First-order constitutive relations including Hall transports}
\label{sec:currents-Hall}

We next discuss the correspondence between the two formulations at the first-order in derivatives
and show how the electric field and current arises in the new formulation.
We also include the charge-conjugation odd terms in $ T_\one^{\mu\nu} $
and $ \tilde F_\one^{\mu\nu} $ to identify how the Hall components could appear
when a finite charge density is allowed in a finite time scale. When charge-conjugation odd terms are allowed in the tensor decomposition (\ref{eq:rho-tensor}),
we have an additional term as
\begin{equation}
\label{eq:rho-tensor-Hall}
 \rho^{\m\n\r\s} =
  2 \rho_\perp
  \left( b^\mu \Xi^{\nu[\rho} b^{\sigma]} -   b^\nu \Xi^{\mu[\rho} b^{\sigma]}
  \right)
  + 2 \rho_\para  \Xi^{\mu[\rho} \Xi^{\sigma]\nu}
  + 2 \rho_\Hall    \left(  b^\mu b_\star^{\nu[\rho} b^{\sigma]}  -  b^\nu b_\star^{\mu[\rho} b^{\sigma]} \right)
 ,
\end{equation}
where $ b^{\mu\nu}_\star = \vep^{\mu\nu\alpha\beta} u_\alpha  b_\beta $. The charge-conjugation odd term does not create entropy in Eq.~(\ref{eq:semi-positive-E}).
Thus, the additional coefficient $\rho_\Hall$ can be both positive and negative values, while $\rho_{\para,\perp}$ should be positive semi-definite as we have seen.
The first-order correction $ \tilde F_\one^{\mu\nu} $ provides
the constitutive relation of the electric field
\begin{eqnarray}
\label{eq:E-1st-Hall}
E_\one^\mu \= -  [  \rho_\para  (-b^\mu b^\rho) +  \rho_\perp   \varXi^{\mu\rho} + \rho_\Hall b_\star^{\mu\rho} ]
T \epsilon_{\rho\nu\alpha\beta} u^\nu  \pt^{\a} (\beta H^{\b})
 \, ,
\end{eqnarray}
where we used identities $b_\star^{\m\l}b^\n_{\star\l}=\Xi^{\m\n}$,
$ 2b_\star^{\mu[\a} b^{\b]}
=   \varXi^{\mu\rho}  \epsilon_{\rho\nu\alpha\beta} u^\nu $,
and $2 \varXi^{\mu[\a} b^{\b]}
= - b_\star^{\mu\rho} \epsilon_{\rho\nu\a\b}  u^\nu $.
This is an extension of Eq.~(\ref{eq:E-1st}) with the Hall term.
On the other hand, the in-medium Maxwell equation provides the relation between
the zeroth-order field strength tensor and the free electric current (also called conduction current) as~\footnote{Using $H^\m=-\m_m^{-1} B^\m$ with $ \mu_m $ the magnetic permeability , one can find that the free current $J_{f\one}^\m$ is related to the total current $J_\one^\m =-(1/2)\vep^{\nu\mu\alpha\beta}
\partial_\nu  (B_\a u_\b - B_\b u_\a )$ by $J_{f\one}^\m=\m_m^{-1} J^\m_\one+\vep^{\mu\nu\alpha\beta}  H_\a u_\b \nabla_\nu \ln\mu_m$.
}
\begin{eqnarray}
\label{eq:Maxwell-eq}
J^\mu_{f\one} \=  \frac{1}{2} \vep^{\nu\mu\alpha\beta}
\partial_\nu  (H_\a u_\b - H_\b u_\a )
\nnb
\= - T \vep^{\mu\nu\alpha\beta}  u_\nu  \nabla_\a (\beta H_\b)
+ \vep^{\mu\nu\alpha\beta} H_\n \nabla_\a u_\b
-  \vep^{\mu\nu\alpha\beta}  u_\n  H_\a (  D u_\b -  \nabla_\b \ln T)
\, .
\end{eqnarray}
The first term is generated by a nonzero curl of the magnetic field. The second term is along $u^\m$ direction and thus generates a non-zero charge density in the rest frame of the fluid, $u_\m J_{f\one}^\m=\vep^{\mu\nu\alpha\beta} u_\m H_\n \nabla_\a u_\b =-2\omega_\m H^\m$ with the vorticity vector  $\omega^\mu:=(1/2)\vep^{\mu\nu\alpha\beta} u_\n \nabla_\a u_\b$.
The rest of terms are Hall-like currents
that are driven by the acceleration and the temperature gradient, which, upon using Eq.~(\ref{eq:Du-ideal}), can be re-written as
\begin{eqnarray}
\label{eq:Maxwell-eq:3rd}
-  \vep^{\mu\nu\alpha\beta}  u_\n  H_\a (  D u_\b -  \nabla_\b \ln T)&=&\frac{2T}{\e+p_\perp}\vep^{\mu\nu\alpha\beta}  u_\n  H_\a B^\l\nabla_{[\l}(\b H_{\b]})  \nnb
%&&=-\frac{T}{\e+p_\perp}(\vep^{\nu\alpha\beta\l} B^\m+\vep^{\alpha\beta\l\m} B^\n+\vep^{\beta\l\m\n} B^\a)  u_\n  H_\a \nabla_{\l}(\b H_{\b}) \nnb
%&&=\frac{T}{\e+p_\perp}(-\vep^{\l\n\a\b} B^\m+\vep^{\m\n\a\b} B^\l)  u_\n  H_\l \nabla_{\a}(\b H_{\b}) \nnb
%&&=\frac{TBH}{\e+p_\perp}(\vep^{\m\n\a\b} +\vep^{\l\n\a\b} b^\m  b_\l)  u_\n  \nabla_{\a}(\b H_{\b}) \nnb
&=&\frac{TBH}{\e+p_\perp}\Xi^\m_\l\vep^{\l\n\a\b}  u_\n  \nabla_{\a}(\b H_{\b}).
\end{eqnarray}
Therefore, we have
\begin{eqnarray}
\label{eq:Maxwell-eq:Jf}
&&- T \vep^{\mu\nu\alpha\beta}  u_\nu  \nabla_\a (\beta H_\b)=\frac{\e+p_\perp}{\e+p_\perp-BH}\Xi^\m_\l J^\l_{f\one}-b^\m b_\l J^\l_{f\one}.
\end{eqnarray}
Plugging those expressions,
one can rewrite the entropy production rate (\ref{eq:semi-positive-E}) as
\begin{eqnarray}
\R_E = -  \beta E_{(1)}^{\mu} \left[ \frac{\e+p_\perp}{\e+p_\perp-BH}\Xi_{\m\l} J^\l_{f\one}-b_\m b_\l J^\l_{f\one} \right] + \order(\pd^3)
\label{eq:E-bilinear2}
\, .
\end{eqnarray}
The origin of the entropy production $\R_E$ is identified with the Joule heat due to the induced electric field and current. 

To get a direct relation between the electric field and current,
we eliminate the magnetic fields in Eq.~(\ref{eq:E-1st-Hall}) using Eq.~(\ref{eq:Maxwell-eq}) to find~\footnote{The structure of Eq.~(\ref{eq:Ohmic:Efield}) becomes more transparent if we write it in three-vector form in frame $u^\m=(1, \bm 0)$. The result is $$\bm E_\one=\tilde{\rho}_\perp\bm J_{f\one}+\frac{\tilde{\rho}_\para-\tilde{\rho}_\perp}{|\bm B|^2}(\bm B\cdot\bm J_{f\one})\bm B+\frac{\tilde{\rho}_H}{|\bm B|}\bm J_{f\one}\times\bm B,$$ where $\bm J_{f\one}=\m_m^{-1}\left(\bm J_\one-\bm\nabla\ln\m_m\times\bm B\right)$ in terms of the total current $\bm J_\one$. When order-zero effective background charges are present, $\bm E_\one$ may contain also terms proportional to the gradient of the effective background charge potential. Similar results have recently been obtained by using the method of effective field theory~\cite{Vardhan:2022wxz}.}
\begin{eqnarray}
\label{eq:Ohmic:Efield}
E^\mu_\one  \= [ - \tilde{\rho}_\para b^\mu b^\nu  + \tilde{\rho}_\perp \varXi^{\mu\nu}
  + \tilde{\rho}_\Hall  b_\star^{\mu\nu}] J_{f\one\nu}\, ,
\end{eqnarray}
where $\tilde{\rho}_\para:=\rho_\para, \tilde{\rho}_\perp:=\rho_\perp(\e+p_\perp)/(\e+p_\perp-BH)$, and $\tilde{\rho}_H:=\rho_H(\e+p_\perp)/(\e+p_\perp-BH)$. Note that, same as $\rho_\para$ and $\rho_\perp$, $\tilde{\rho}_\para$ and $\tilde{\rho}_\perp$ are positive semi-definite as well. It is more convenient to express the free electric current in terms of the electric fields and we obtain
\begin{eqnarray}
\label{eq:Ohmic}
J^\mu_{f\one}  \=-2\omega\cdot H u^\m+ [ - \sigma_\para b^\mu b^\nu  + \sigma_\perp \varXi^{\mu\nu}
  + \sigma_\Hall  b_\star^{\mu\nu}] E_{\one\nu}
\, .
\end{eqnarray}
The three coefficients are given by the coefficients $\tilde{\rho}_{\perp,\para,\Hall}$ as
\begin{eqnarray}
\sigma_\para = \frac{1}{\tilde{\rho}_\para} , \quad
\sigma_\perp = \frac{\tilde{\rho}_\perp}{\tilde{\rho}_\perp^2 + \tilde{\rho}_\Hall^2} , \quad
\sigma_\Hall =  \frac{\tilde{\rho}_\Hall}{\tilde{\rho}_\perp^2 + \tilde{\rho}_\Hall^2}
\, .
\end{eqnarray}
The Hall term $\propto \sigma_\Hall  $ identically vanishes
in Eq.~(\ref{eq:E-bilinear2}) and does not create entropy.
The other coefficients $  \sigma_\para $ and $ \sigma_\perp $
are the parallel and perpendicular components of the Ohmic conductivity with respect to the direction of the magnetic flux.
The Ohmic conductivities should be positive semi-definite $ \sigma_{\para,\perp} \geq 0 $,
while the Hall conductivity $ \sigma_\Hall $ can be both positive and negative values.

Next, we discuss the Hall components in the viscous tensor.
One can divide the energy-momentum tensor into the dissipative and nondissipative components as
\begin{eqnarray}
 T_{(1)}^{\mu\nu} = T_{(1) {\rm dis}}^{\mu\nu}  +  T_{(1) \star}^{\mu\nu}.
\end{eqnarray}
In the preceding sections,
we have already discussed the dissipative component $ T_{(1) {\rm dis}}^{\mu\nu} $ that creates an entropy.
Here, we focus on the nondissipative component $  T_{(1) \star}^{\mu\nu} $ that can be further decomposed as
\begin{eqnarray}
\label{eq:Hall-viscous-tensor}
T_{(1) \star}^{\mu\nu} =
\big[ \, 2\eta_{\Hall\para}  (- b^{\alpha}b^{(\mu} ) b_\star^{\nu)\beta}
+ 2 \eta_{\Hall\perp} \varXi^{\alpha (\mu} b_\star^{\nu)\beta} \, \big] w_{\a\b}
\, ,
\end{eqnarray}
where $w_{\a\b} = \nabla_{(a} u_{\b)} $. This term does not create entropy~\footnote{
We immediately notice that $  b^{\alpha}b^{(\mu}  b_\star^{\nu)\beta} w_{\alpha\beta} w_{\mu\nu} =
2 b_\star^{\nu\beta} (b^{\alpha} w_{\alpha\beta})( b^{\mu}  w_{\mu\nu})  = 0$,
and then that $ \varXi^{\alpha (\mu} b_\star^{\nu)\beta} w_{\alpha\beta} w_{\mu\nu}
= g^{\alpha (\mu} b_\star^{\nu)\beta} w_{\alpha\beta} w_{\mu\nu}  = 0 $.},
and the Hall viscous coefficients $ \eta_{\Hall_\para, \perp} $ can take both positive and negative values.
%as projected by $ b^\mu $ and $ \varXi^{\mu\alpha} $
The presence of the antisymmetric tensor $b_\star^{\mu\nu} $ implies that
the direction of the stress is orthogonal to both the flow velocity and the magnetic field.

\begin{figure}[t]
	\begin{center} %\hspace*{-5mm}
		\includegraphics[width=\hsize]{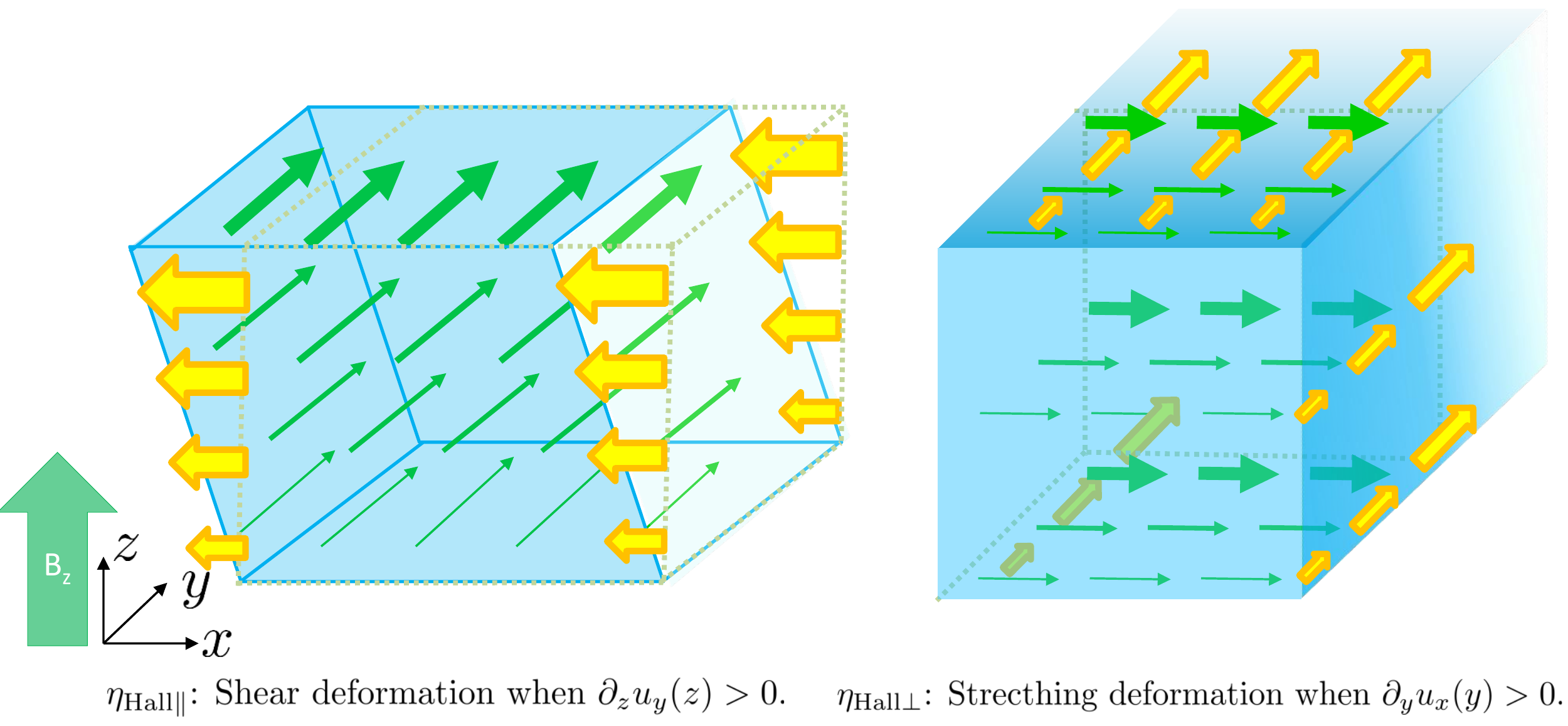}
	\end{center}
%\vspace{-1cm}
\caption{Hall viscosities induced by the Lorentz force exerting on charged fluid cells.
The flow velocity and the Lorentz force are shown with green and orange arrows, respectively.
The length and thickness of the arrows denote the magnitudes of the vectors.
The direction of a magnetic field is taken along the $  z$ direction.
}
\label{fig:Hall_viscosities}
\end{figure}

We briefly demonstrate the mechanisms
that generate the Hall viscosities
in magnetic fields (assuming that there is a possible electric-charge density).
%The left and right panels in Fig.~\ref{fig:Hall_viscosities} show
%the stresses generated by the Hall viscous terms proportional to
%the $  \xi_\Hall^\para $ and $  \xi_\Hall^\para $, respectively.
In Fig.~\ref{fig:Hall_viscosities}, the direction of a magnetic field is taken along the $  z$ direction,
and the flow velocity and the Lorentz force are shown with green and orange arrows, respectively.
When there is a gradient of the transverse flow along the magnetic field $ \partial_z u_x(z)> 0 $,
the Lorentz force exerting on the fluid volume is oriented in the $ y $ direction
and has a gradient along the magnetic field (see the left panel).
This configuration corresponds to the term proportional to $  \eta_{\Hall\para} $.
The gradient of the Lorentz force gives rise to a shear stress in the plane
orthogonal to the flow-velocity and magnetic fields.
On the other hand, when there is a flow gradient in the transverse plane $\partial_y u_x(y) >0 $,
the Lorentz force is oriented in the $  y$ direction and has a gradient in the same direction (see the right panel).
Therefore, the fluid volume is stretched in this direction
according to the term proportional to $  \eta_{\Hall\perp} $.
This term also induces a shear deformation in the $  x$-$y$ plane in response to
an expansion/compression in the $  x$ direction, $ \partial_x u_x(x) \not = 0$.
This effect can be understood in a similar manner.

In the nonrelativistic theory,
it has been known for some time that there are seven viscous coefficients~\cite{hooyman1954coefficients, Landau10:kinetics} (see also Ref.~\cite{deGroot1962}).
%This counting is based on the number of independent tensor structures under the symmetries of the system.
%The magnetic field explicitly breaks the spatial rotational symmetry
%down to the residual one around the direction of the magnetic field.
%Then, there are eight viscous coefficients which, however, reduces to the seven coefficients owing to
%the Onsager relation as a consequence of the microscopic time-reversal property~\cite{de1952thermodynamics, hooyman1954coefficients}.
In this paper, we have obtained the breakdown of the seven viscous coefficients in the relativistic extension.
Including the five dissipative viscosities discussed in the previous sections,
we identified three bulk, two shear, and two Hall composnents.
The relativistic extension was also carried out earlier in Refs.~\cite{Huang:2011dc, Hernandez:2017mch, Grozdanov:2016tdf, Hongo:2020qpv}.
However, those authors use different tensor bases.
For readers' convenience, we provide an explicit comparison among those tensor bases in Appendix~\ref{sec:MHD-viscosities}.

\section{Towards relativistic MHD from kinetic equations}\label{sec:kine}

In addition to hydrodynamics, kinetic theory is also often used for the studies of many-body systems (See Refs.~\cite{Landau10:kinetics,deGroot:book} for the standard textbooks of kinetic theory). It is valid in the regime where the system permits well-defined particles (or quasi-particles) and when the densities of these particles are low enough. More precisely, let the typical microscopic scales be set by the interaction range $r_c\sim \sqrt{\s}$ with $\s$ the scattering cross section, the inter-particle distance $r_d\sim n^{-1/3}$ with $n$ the number density, and the mean-free path $\l_{\rm mfp}\sim 1/(n\s)$. The applicability of kinetic theory requires $\l_{\rm mfp}\gg r_d\gg r_c$ (dilute condition).  The scatterings of the particles drive the system to evolve towards global thermal equilibrium, a process called kinetic thermalization. This process is usually associated with the arising of a new, macroscopic scale, $L$, over which the macroscopic properties of the system vary. In the late stage of the thermalization, the scale $L$ will be clearly separated from the microscopic scales. This gives a characteristic parameter ${\rm Kn}=\l_{\rm mfp}/L$ (the Knudsen number). When ${\rm Kn}\ll 1$ or equivalently when the derivative $\pt$ is much smaller than $1/\l_{\rm mfp}$, the macroscopic properties become insensitive to the microscopic details of the system and the hydrodynamic description is expected to arise. Therefore, when the kinetic theory is applied to the regime where ${\rm Kn}\ll 1$, its solution is expected to be expressed by the local hydrodynamic variables and their derivatives. This gives a systematical way to derive the hydrodynamic constitutive relations (including expressing the transport coefficients in terms of kinetic-theory parameters) and EOMs by expanding the kinetic equation and the distribution functions in ${\rm Kn}$ or $\pt$. This is the essential idea of the Chapman-Enskog method~\cite{Chapman:book} which we will now discuss. Another frequently utilized approach to hydrodynamics from kinetic theory is the Grad's method of moments~\cite{Grad:19491} which we will explore in Sec.~\ref{sec:mom}.

\subsection{Chapman-Enskog method}\label{sec:chap}

Before studying RMHD using kinetic theory, in order to demonstrate the methodology, let us consider a simpler case in which the EM fields are absent. The starting point is the relativistic Boltzmann equation for on-shell particles:
\begin{eqnarray}
\label{eq:boltz}
k^\m\pt_\m f_{\vk}=C[f], 
\end{eqnarray}
where $k^\m=(k^0=\sqrt{\vk^2+m^2},\vk)$ with $m$ being the mass of the particles, $f_\vk(x)$ is the distribution function, and all external forces are omitted. For the sake of simplicity, we consider only one species of particles. We assume that the collisional process conserves the energy-momentum and also the particle numbers, so that the collision kernel satisfies
\begin{eqnarray}
\label{eq:boltz:NTint}
\int dK C[f]=\int dK k^\m C[f]=0, 
\end{eqnarray}
where $dK=g d^3\vk/[(2\pi)^3k^0]$ is the invariant momentum-space measure with $g=2S+1$ the spin degenerate factor with $S$ spin of the particles. By using Eq.~(\ref{eq:boltz}), Eq.~(\ref{eq:boltz:NTint}) leads to the conservation of particle-number current $N^\m$ and energy-momentum tensor $T^{\m\n}$, 
\begin{subequations}
\begin{eqnarray}
\label{eq:kin:Ncons}
\pt_\m N^\m &=&0, \\
\label{eq:kin:Tcons}
\pt_\m T^{\m\n} &=&0,
\end{eqnarray}
\end{subequations}
where $N^\m$ and $T^{\m\n}$ are expressed by
\begin{subequations}
\begin{eqnarray}
\label{eq:kin:N}
N^\m &=&\lan k^\m\ran=n u^\m+\n^\m, \\
\label{eq:kin:T}
T^{\m\n} &=&\lan k^\m k^\n\ran=\e u^\m u^\n-\Delta^{\m\n}(p+\Pi)+h^\m u^\n+h^\n u^\m +\pi^{\m\n},
\end{eqnarray}
\end{subequations}
with $\lan\cdots\ran=\int dK(\cdots)f_\vk$. Here, we introduced a timelike unit vector field $u^\m$ which will be identified as the fluid velocity. Using $u^\m$, the momentum can be decomposed into two parts (following the notations of Ref.~\cite{Denicol:2012cn}): $k^\m=E_\vk u^\m  + k^{\lan\m\ran}$, with $E_\vk=u\cdot k$ and $k^{\lan\m\ran}=\Delta^{\m\n}k_\n$. We have also introduced the number density $n$, the number diffusion current $\n^\m$, the energy density $\e$, the thermodynamic pressure $p$, the viscous pressure $\Pi$, the shear viscous tensor $\pi^{\m\n}$, and the heat flux $h^\m$,
\begin{eqnarray}
\label{eq:kin:variousquantities}
&n = \lan E_\bk \ran,\; \n^\m=\lan k^{\lan\m\ran}\ran,\; \e=\lan E_\vk^2\ran,\; p+\Pi=-\frac{1}{3}\Delta^{\m\n}\lan k_\m k_\n\ran,\; \pi^{\m\n}= \lan k^{\lan\m}k^{\n\ran}\ran,\;  h^\m =\lan E_\vk k^{\lan\m\ran}\ran,\nonumber\\
\end{eqnarray}
where $A^{\lan\m\n\ran}=(1/2)\Delta^\m_\a\Delta^\n_\b(A^{\a\b}+A^{\b\a})-(1/3)\Delta^{\m\n}\Delta_{\a\b}A^{\a\b}$ is the spatial traceless symmetrization of $A^{\m\n}$. The expression for $p$ itself will be determined later. To specify the rest frame of the fluid, we use the Landau-Lifshitz choice so that $T^{\m\n}u_\n=\e u^\m$. This imposes the constraint
\begin{eqnarray}
\label{eq:kin:LLconstraint}
h^\m =\lan E_\vk k^{\lan\m\ran}\ran=0.
\end{eqnarray}

To proceed, one takes the Chapman-Enskog expansion of the distribution function, $f_\vk=f^{(0)}_{\vk}+f_{\vk}^{(1)}+\cdots$, and of the Boltzmann equation (\ref{eq:boltz}) order by order in ${\rm Kn}$ or equivalently in derivatives. At zeroth order in the derivative expansion, Eq.~(\ref{eq:boltz}) reads
\begin{eqnarray}
\label{eq:kin:ce0}
0=C[f^{(0)}].
\end{eqnarray}
Its solution is called the local-equilibrium distribution $f^{(0)}_\vk=f_{0\vk}$ because it saturates the local detailed balance. We assume that $f_{0\vk}=1/[\exp(\b E_\vk-\a)+a]$ with $a=1, 0$ and $-1$ for fermions, classical Boltzmann particles, and bosons. Here, $\b$ and $\a=\b\m$ are the local inverse temperature and the ratio of the chemical potential to temperature. Their values are fixed by matching conditions~\cite{Stewart:1971,deGroot:book}:  
\begin{eqnarray}
\label{eq:kin:match}
\e=\lan E_\vk^2\ran=\lan E_\vk^2\ran_0,\;\; n=\lan E_\vk \ran=\lan E_\vk\ran_0,
\end{eqnarray}
where $\lan\cdots\ran_0=\int dK(\cdots)f_{0\vk}$. Substituting $f_{0\vk}$ into Eqs.~(\ref{eq:kin:N}) and (\ref{eq:kin:T}), one obtains the zeroth-order $N^\m$ and $T^{\m\n}$:
\begin{subequations}
\begin{eqnarray}
\label{eq:kin:N:0}
N_{(0)}^\m &=& n u^\m, \\
\label{eq:kin:T:0}
T_{(0)}^{\m\n} &=&\e u^\m u^\n-p\Delta^{\m\n},
\end{eqnarray}
\end{subequations}
where $p=-\frac{1}{3}\Delta^{\m\n}\lan k_\m k_\n\ran_0$, and Eqs.~(\ref{eq:kin:Ncons}) and (\ref{eq:kin:Tcons}) become the ideal hydrodynamic equations.

At first order in derivative expansion, the collision kernel becomes a linear integral operator acting on $f_\vk^{(1)}$, $C^{(1)}[f]=\hat{L}f_\vk^{(1)}$. The Boltzmann equation (\ref{eq:boltz}) at $O({\rm Kn})$ thus reads
\begin{eqnarray}
\label{eq:kin:ce1}
k^\m\pt_\m f_{0\vk}=\hat{L}f_\vk^{(1)}.
\end{eqnarray}
Once $f_\vk^{(1)}$ is solved out from this equation, the first-order constitutive relations are then obtained: 
\begin{eqnarray}
\label{eq:kin:firstconstitutive}
\n^\m=\lan k^{\lan\m\ran}\ran_1,\; \Pi=-\frac{1}{3}\Delta^{\m\n}\lan k_\m k_\n\ran_1,\; \pi^{\m\n}= \lan k^{\lan\m}k^{\n\ran}\ran_1,
\end{eqnarray}
with $\lan\cdots\ran_1=\int dK(\cdots)f^{(1)}_{\vk}$. To see how this procedure works in practice, let us consider a collision kernel in relaxation-time approximation (RTA)~\footnote{Note that this collision kernel does not automatically conserve particle number and energy-momentum, Eq.~(\ref{eq:boltz:NTint}). However, these conservation laws are recovered once the matching conditions (\ref{eq:kin:match}) and Landau-Lifshitz frame constraint (\ref{eq:kin:LLconstraint}) are imposed.},
\begin{eqnarray}
\label{eq:kin:rta}
C[f]=-E_\vk\frac{f_\vk-f_{0\vk}}{\tau_R}.
\end{eqnarray}
Writing $f^{(1)}_\vk$ as $f^{(1)}_\vk=f_{0\vk}\tilde{f}_{0\vk}\phi_\vk$ with $\phi_\vk$ at $O({\rm Kn})$ and $\tilde{f}_\vk=1-af_\vk$, one obtains ${\hat L}f^{(1)}_\vk=-\tau_R^{-1}E_\vk f_{0\vk}\tilde{f}_{0\vk}\phi_\vk$. From Eq.~(\ref{eq:kin:ce1}), one finds
\begin{eqnarray}
\label{eq:kin:phik}
\phi_\vk=\frac{\b\tau_R}{E_\vk}\bigg[\left(\frac{1}{3}\Delta_{\m\n}k^\m k^\n+E^2_\vk\frac{\pt p}{\pt\e}\Big|_n+E_\vk\frac{\pt p}{\pt n}\Big|_\e\right)\theta+T\left(\frac{nE_\vk}{\e+p}-1\right)k^\m\nabla_\m\a+k^{\lan\m}k^{\n\ran}\xi_{\m\n}\bigg],\nonumber\\
\end{eqnarray}
where we defined $\xi_{\m\n}:=w_{\lan\m\n\ran}$ which is the shear tensor~\footnote{The symbol $\xi_{\m\n}$ for shear tensor is used only in this section in order to simplify the equations. In other sections, we simply use $w_{\lan\m\n\ran}$ or $\nabla_{\lan\m}u_{\n\ran}$ to denote the shear tensor.}, and the terms are organized in such a way that they are mutually orthogonal under integral $\int dK (\cdots)F(E_\vk)$ with $F(E_\vk)$ an arbitrary converging function of $E_\vk$. Substituting $\phi_\vk$ into Eq.~(\ref{eq:kin:firstconstitutive}), one finds
\begin{subequations}
\begin{eqnarray}
\label{eq:kin:conductivity}
\n^\m&=&\kappa\, \nabla^\m \a,\quad{\rm with}\quad \kappa=\tau_R T\left(\frac{2}{3}\frac{\e-3p}{m^2}-\frac{n^2}{\e+p}+\frac{m^2}{3}\left\lan \frac{1}{E_\vk^{2}}\right\ran_0\right),\,\\
\label{eq:kin:bulkvis}
\Pi &=&-\zeta\, \theta,\quad {\rm with}\quad \zeta = \tau_R\left[\left(\frac{1}{3}-c_s^2\right)(\e+p)-\frac{2}{9}(\e-3p)-\frac{m^4}{9}\left\lan \frac{1}{E_\vk^{2}}\right\ran_0\right],\,\\
\label{eq:kin:shearvis}
\pi^{\m\n} &=& 2\eta\, \xi^{\m\n},\quad {\rm with}\quad \eta= \tau_R\left[\frac{4}{5}p+\frac{1}{15}(\e-3p)-\frac{m^4}{15}\left\lan \frac{1}{E_\vk^{2}}\right\ran_0\right],\,
\end{eqnarray}
\end{subequations}
where $c_s^2=\pt p/\pt\e|_{s/n}=\pt p/\pt\e|_{n}+(\e+p)^{-1}n\,\pt p/\pt n|_{\e}$ is the sound velocity squared, $\kappa, \zeta, \eta$ are conductivity of number diffusion current, bulk viscosity, and shear viscosity. Note that when $m=0$, the bulk viscosity vanishes because $\e=3p$ and $c_s^2=1/3$ in massless limit. Thus, at $O({\rm Kn})$, we recover the relativistic Navier-Stokes hydrodynamics. 

With the above preparation, let us now consider RMHD in the Chapman-Enskog method. In this case, the Boltzmann equation contains the EM force term: 
\begin{eqnarray}
\label{mhd:boltz}
k^\m\pt_\m f_{\vk}+qF^{\m\n}k_\n\pt_{k^\m}f_\vk=C[f], 
\end{eqnarray}
where $q$ is the charge of the particles (we consider only one species of particles and assume $q>0$). We assume that the particles are under binary elastic collisions and the colliding processes are not interfered by the external EM field $F^{\m\n}$. Hence the collision kernel preserves the particle number (and the electric charges as well) and energy-momentum, \ie, Eq.~(\ref{eq:boltz:NTint}) still holds which implies  
\begin{eqnarray}
\label{mhd:kin:numbercons}
\pt_\m N^\m&=&\pt_\m \int dK k^\m f_\vk=0,\\ 
\label{mhd:kin:emcons}
\pt_\m T^{\m\n}&=&\pt_\m \int dK k^\m k^\l f_\vk= q F^{\n\l}N_\l=F^{\n\l}J_\l, 
\end{eqnarray}
where $J^\m=q N^\m$ is the charge current. We emphasize that the EM fields appear in our kinetic approach as external fields which is very different from Sec.~\ref{sec:entropy} and Sec.~\ref{sec:mhd:qed} in which the dynamics of the EM fields plays important role. This leads to different leading-order constitutive relations, but they share the same structures of the dissipative constitutive relations for the fluid (namely, the viscous and conductivity tensors). 

Let us clarify a significant distinction between how the EM fields are treated in this section and how they are treated in Sec.~\ref{sec:entropy} and Sec.~\ref{sec:mhd:qed}. In Sec.~\ref{sec:entropy} and Sec.~\ref{sec:mhd:qed}, the EM fields are dynamical and the energy-momentum tensor includes contributions from the EM fields as well, resulting in a conserved energy-momentum tensor. The RMHD derived in Sec.~\ref{sec:entropy} and Sec.~\ref{sec:mhd:qed} is therefore a strict hydrodynamic theory.
This is not the case with this section, where the EM fields are external fields (as clearly seen in Eq.~(\ref{mhd:boltz})), hence $N^\m$ and $T^{\m\n}$ obtained from the kinetic theory only include contributions from the matter. As shown in the right-hand side of Eq.~(\ref{mhd:kin:emcons}), matter exchanges energy and momentum with EM fields, causing the $T^{\m\n}$ to be non-strictly conserved. For this reason, the pressure and energy density include only the matter contributions.

The presence of the EM fields introduces new scales into the kinetic equation (and also the hydrodynamic equations). So let us clarify the range of scales that we will focus in this case. The typical microscopic scales are still set by the interaction range $r_c$, inter-particle distance $r_d$, and the mean-free path $\l_{\rm mfp}$. In terms of ${\rm Kn}$ or the gradient $\pt$ of the conserved densities, the electric field in the rest frame of the fluid, $E^\m=F^{\m\n}u_\n$, is considered as $O(\pt)$, while the magnetic field in the rest frame of the fluid, $B^\m=\varepsilon^{\m\n\a\b}u_\n F_{\a\b}/2$, can be an $O(1)$ quantity, as we already emphasized in Sec.~\ref{sec:entropy}. This amounts to the fact that the electric field is screened by the gradient of charge distribution in a plasma, while magnetic field is not. The magnetic field makes the motion of charged particles curvilinear with radius $R_L=k_\perp/(qB)$ (the Larmor radius) with $k_\perp$ being the momentum of the particles transverse to the magnetic field. For hot relativistic plasma, $k_\perp\sim T$; so $R_T=T/(qB)$ sets the magnetic cyclotron scale. We will always assume that the thermal wavelength $\b=1/T$ of the particles and the interaction range $r_c$ are much smaller than $R_T$, $\b\ll R_T$ and $r_c\ll R_T$. The first inequality means  that the Landau quantization effect is not significant and we can safely use a classical treatment. The second inequality means that in the collision process the magnetic field can be neglected so we can use a magnetic-field independent collision operator C (but the distribution function $f_\vk$ can certainly depend on $B^\m$). The situation with an even stronger magnetic field, $R_T\ll\b$, that can make the Landau quantization effect significant, will be discussed in Sec.~\ref{sec:transport-coeff}. %We do not constrain the ratio $\xi=\l_{\rm mfp}/R_T$. But in some situations, we will focus on the case with $\xi\gg1$ so that the magnetic field gives dramatic influence to the dissipative phenomena (the MHD in this case is sometimes called Braginskii MHD~\cite{Braginskii:1965}). %(For example, at high temperature limit, the ratio $MB/P$ of free charged fermions goes to zero.) So we will essentially treat a classical Boltzmann equation in which the spin appears merely as a degeneracy factor and an indicator of the type of the distribution function (Boltzmann, Bose-Einstein, or Fermi-Dirac).
%\pp{[Do those conditions constrain the derivative order of the magnetic field?]}\huang{It seems not. I think the derivative order is controlled by screening scale: the magnetic field is not screened, so we assign $O(1)$ for it. But the scales given by $1/T$ and $r_c$ are much shorter than the screening scale. So we further require $qB$ to be much weaker than $T^2$ and $T/r_c$.}

We will always use the Landau-Lifshitz frame for the fluid and use the matching conditions for $\e$ and $n$. The zeroth-order distribution function is still chosen as $f^{(0)}_\vk=f_{0\vk}=1/[\exp(\b E_\vk-\a)+a]$ with $a=1, 0$ and $-1$ for fermions, classical Boltzmann particles, and bosons. Choosing such an equilibrium distribution amounts to assume that the magnetization pressure $-MB$ is much smaller than thermodynamic pressure $p$ and thus is omitted. This also implies that the magnetization current is vanishing and we thus do not distinguish the free electric current and the total electric current in this section. Therefore, the ideal-fluid constitutive relations are still given by Eqs.~(\ref{eq:kin:N:0}) and (\ref{eq:kin:T:0}). To obtain $f^{(1)}_\vk$, let us again take the RTA for the collision kernel,
\begin{eqnarray}
\label{mhd:rta}
C[f]=-E_\vk\frac{f_\vk-f_{0\vk}}{\tau_R}.
\end{eqnarray}
First, we consider a situation in which both $E^\m$ and $B^\m$ in Eq.~(\ref{mhd:boltz}) are of order $O(\pt)$, namely, when both the electric and magnetic fields are very weak. This is not in the MHD region but let us make a case study first. Writing $f^{(1)}_\vk$ as $f^{(1)}_\vk=f_{0\vk}\tilde{f}_{0\vk}\phi_\vk$ with $\phi_\vk$ at $O({\pt})$ and $\tilde{f}_\vk=1-af_\vk$, one obtains from  Eq.~(\ref{mhd:boltz}),
\begin{eqnarray}
\label{mhd:phik}
\phi_\vk &=&-\frac{\tau_R}{E_\vk}(f_{0\vk}\tilde{f}_{0\vk})^{-1}\left(k^\m\pt_\m f_{0\vk}+qF^{\m\n}k_\n\pt_{k^\m}f_{0\vk}\right)
\nonumber\\
&=&
\frac{\b\tau_R}{E_\vk}\bigg[
T\left(\frac{nE_\vk}{\e+p}-1\right)k^\m\left(\nabla_\m\a+\b qE_\m\right)
\nonumber\\
&&
+\left(\frac{1}{3}\Delta_{\m\n}k^\m k^\n+E^2_\vk\frac{\pt p}{\pt\e}\Big|_n+E_\vk\frac{\pt p}{\pt n}\Big|_\e\right)\theta
+k_\perp^{\lan\m}k_\perp^{\n\ran}\xi_{\m\n}\bigg].
\end{eqnarray}
Compared with Eq.~(\ref{eq:kin:phik}), the only new term is the electric-field term which is always accompanied with $\nabla_\m\a$ and the magnetic field drops out. To obtain $\phi_\vk$, we have used the EOMs at $O(1)$ order (ideal hydrodynamic equations):
\begin{eqnarray}
\label{mhd:ideal:u}
Du^\m &=&\frac{1}{\e+p}\left(\nabla^\m p+n q E^\m\right),\\
\label{mhd:ideal:a}
D\a &=& -\b \frac{\pt p}{\pt n}\Big|_\e\h,\\
\label{mhd:ideal:b}
D\b &=& \b \frac{\pt p}{\pt \e}\Big|_n\h.
\end{eqnarray}
Substituting $f^{(1)}_\vk$ into Eqs.~(\ref{eq:kin:firstconstitutive}), we find that the $\Pi$ and $\pi^{\m\n}$ are still given by Eqs.~(\ref{eq:kin:bulkvis}) and (\ref{eq:kin:shearvis}) but $\n^\m$ is given by $\n^\m=\kappa\left(\nabla^\m\a+\b qE^\m\right)$ with the number diffusion constant $\kappa$ given in Eq.~(\ref{eq:kin:conductivity}). The charge diffusion current $j^\m=q\lan k^{\lan\m\ran}\ran_1$ is
\begin{eqnarray}
\label{mhd:chapman:jmu}
j^\m=q\n^\m=q\kappa\left(\nabla^\m\a+\b qE^\m\right).
\end{eqnarray}
This relation gives that the electric conductivity (the coefficient in front of $E^\m$) is determined by number diffusion conductivity, $\sigma=\b q^2\kappa$,  a relation representing the Wiedemann–Franz law.

The above situation with $B^\m\sim O(\pt)$ is not in the MHD region. Let us now consider the MHD region in which $E^\m$ is at $O(\pt)$ but $B^\m$ is at $O(1)$. The Boltzmann equation at $O(\pt)$ reads
\begin{eqnarray}
\label{mhd:boltz2}
k^\m\pt_\m f_{0\vk}-qE^\n k_\n u^\m\pt_{k^\m}f_{0\vk}=-qB b_\star^{\m\n}k_\n\pt_{k^\m}f^{(1)}_\vk-\frac{E_\vk}{\tau_R}f^{(1)}_\vk, 
\end{eqnarray}
where $B=\sqrt{-B^\m B_\m}$ is the strength of the magnetic field and $b_\star^{\m\n}=\varepsilon^{\m\n\a\b}u_\a b_\b$ is the cross projector. 
%\pp{[Should we suppress the repeated definitions or just leave them as the paper is long?]}\huang{Can we keep them? That may be convenient for the readers.}
The solution for $f^{(1)}_\vk$ is $f^{(1)}_\vk=f_{0\vk}\tilde{f}_{0\vk}\phi_\vk$ with 
\begin{eqnarray}
\label{mhd:chapman:f1}
\phi_\vk &=& \phi_\parallel +\frac{\tau_R}{E_\vk}\frac{\b}{1+\xi_\vk^2}\left(V^\m +\xi_\vk b_\star^{\m\n}V_\n\right)k^\perp_\m + \frac{\tau_R}{E_\vk}\frac{\b}{1+4\xi_\vk^2}\left(w_\perp^{\m\n} +2\xi_\vk b_\star^{\m\l}w^\n_{\perp\l}\right)k^\perp_{\{\m}k^\perp_{\n\}}, \\
\phi_\parallel &=&\frac{\b\tau_R}{E_\vk}\bigg[\left(\frac{1}{2}\Xi_{\a\b}k^\a k^\b+E^2_\vk\frac{\pt p}{\pt\e}\Big|_n+E_\vk\frac{\pt p}{\pt n}\Big|_\e\right)\theta_\perp+\left(-k_\parallel^2+E^2_\vk\frac{\pt p}{\pt\e}\Big|_n+E_\vk\frac{\pt p}{\pt n}\Big|_\e\right)\theta_\parallel\nonumber\\ 
&&-Tk_\parallel\left(\frac{nE_\vk}{\e+p}-1\right) b^\m\left(\nabla_\m\a+\b qE_\m\right)\bigg],\\
V^\m &=& \left[ T\left(\frac{nE_\vk}{\e+p}-1\right)\Xi^{\m\l}\left(\nabla_\l\a+\b qE_\l\right)-2 k_\parallel \Xi^{\m\l} w_{\l\rho}b^\rho+\frac{ qB}{\e+p}E_\bk b_\star^{\m\l}\n_\l\right],
\end{eqnarray}
where $\xi_\vk\equiv qB\tau_R/E_\vk$ and we have introduced
\begin{eqnarray}
\begin{split}
\label{mhd:chapman:notation}
&k_\parallel = b\cdot k,\;\; \Xi^{\m\n}=\Delta^{\m\n}+b^\m b^\n,\;\; k_\perp^\m =\Xi^{\m\n}k_\n,  \;\; k_\perp^{\{\m}k_\perp^{\n\}}=k_\perp^\m k_\perp^\n-\frac{1}{2}\Xi^{\m\n}\Xi_{\a\b}k^\a k^\b,&\; \\
&w_{\m\n}=\frac{1}{2}(\nabla_\m u_\n+\nabla_\n u_\m),\;\; \h_\perp=\Xi_{\a\b}w^{\a\b},\;\; \h_\parallel=-b_\m b_\n w^{\m\n},\;\;  w_\perp^{\m\n}=\left(\Xi^{\m\a}\Xi^{\n\b}-\frac{1}{2}\Xi^{\m\n}\Xi^{\a\b}\right) w_{\a\b}.&
\end{split}
\end{eqnarray}
Note that to derive Eq.~(\ref{mhd:chapman:f1}) we use again Eqs.~(\ref{mhd:ideal:u})-(\ref{mhd:ideal:b}) but Eq.~(\ref{mhd:ideal:u}) is replaced by 
\begin{eqnarray}
\label{mhd:ideal:u2}
Du^\m &=&\frac{1}{\e+p}\left(\nabla^\m p+n q E^\m + qB b_\star^{\m\l}\n_\l\right),
\end{eqnarray}
because now $B^\m\sim O(1)$ and is kept in the ideal EOMs. 

We can check explicitly that the matching conditions $\lan E_\vk\ran_1=0, \lan E_\vk^2\ran_1=0$ are satisfied with $f^{(1)}_\vk$. The Landau-Lifshitz frame-fixing condition $\lan E_\vk k^{\lan\m\ran}\ran_1=0$ is, however, not satisfied. In fact, because $f^{(1)}_\vk$ depends on $\n^\mu$ explicitly, the Landau-Lifshitz condition must be coupled with the defining condition for $\n^\m$ to determine $u^\m$ and $\n^\m$. A simpler way to solve out $\n^\m$ from these complicated coupled equations is by considering the frame-independent vector (See Appendix \ref{sec:T1} for the discussion of the transformation among different choices of the fluid velocity)
\begin{eqnarray}
\label{eq:kin:lmu}
l^\m=\lan k^{\lan\m\ran}\ran_1-\frac{n}{\e+p}\lan E_\vk k^{\lan\m\ran}\ran_1,
\end{eqnarray}
which depends on $\n^\m$ and should equal to $\n^\m$ once the Landau-Lifshitz condition is fulfilled. Therefore, $\n^\m$ in the Landau-Lifshitz frame should be determined by 
\begin{eqnarray}
\label{eq:kin:lmunu}
\n^\m=\lan k^{\lan\m\ran}\ran_1-\frac{n}{\e+p}\lan E_\vk k^{\lan\m\ran}\ran_1.
\end{eqnarray}
This is a linear equation for $\n^\m$ and can be directly solved. To demonstrate this and for simplicity of the discussion, we assume that the magnetic cyclotron frequency $\omega_B=1/R_T=qB/T$ is much larger than the collision rate $\omega_{\rm col}=1/\tau_R$ (or equivalently, $\xi=\l_{\rm mfp}/R_T\gg 1$). In this case, $\xi_\vk\sim \xi\gg 1$ and we can expand $f^{(1)}_\vk$ in $1/\xi_\vk$. Solving Eq.~(\ref{eq:kin:lmunu}) order by order in $1/\xi$, we obtain
\begin{eqnarray}
\label{eq:kin:chapnumu}
\n^\m &=& \kappa^{\m\n}\left(\pt_\n\a+\b qE_\n\right),\\
\label{eq:kin:chapkmn}
\kappa^{\m\n} &=& \kappa_\perp\Xi^{\m\n}-\kappa_\parallel b^\m b^\n +\kappa_H b_\star^{\m\n},
\end{eqnarray}
where the conductivities read
\begin{subequations}
\label{eq:kin:chapcond}
\begin{eqnarray}
\label{eq:kin:chapcond1}
\kappa_\perp &=&\frac{1}{\tau_R(qB)^2}\frac{\e+p}{\b}\left(\frac{J_{31}J_{51}}{J_{41}^2}-1\right) +O(\xi^{-4})=\frac{1}{\tau_R(qB)^2}\frac{J_{31}D_{41}}{J^2_{41}}+O(\xi^{-4}),\\
\label{eq:kin:chapcond2}
\kappa_\parallel &=& \tau_R\left( J_{11}-\frac{n}{\e+p}J_{21}\right)=\tau_R\frac{D_{21}}{J_{31}},\\
\label{eq:kin:chapcond3}
\kappa_H &=& \frac{1}{qB}\left( \frac{n}{\b}-\frac{J_{31}^2}{J_{41}}\right)+O(\xi^{-3})=\frac{1}{qB}\frac{D_{31}}{J_{41}}+O(\xi^{-3}).
\end{eqnarray}
\end{subequations}
We have introduced the thermodynamic functions $J_{nq}$ and $D_{nq}$ to simplify the expressions. For the later use, we also introduce another two thermodynamic functions $I_{nq}$ and $G_{nm}$. They are defined by~\cite{Denicol:2012cn}
\begin{subequations}
\begin{eqnarray}
\label{eq:notation:I}
I_{nq}&=&\frac{(-1)^q}{(2q+1)!!}\int dK E_\vk^{n-2q}(\Delta_{\a\b}k^\a k^\b)^q f_{0\vk},\\
\label{eq:notation:J}
J_{nq}&=&\frac{(-1)^q}{(2q+1)!!}\int dK E_\vk^{n-2q}(\Delta_{\a\b}k^\a k^\b)^q f_{0\vk}\tilde{f}_{0\vk},\\
\label{eq:notation:D}
D_{nq}&=& J_{n+1,q}J_{n-1,q}-J_{nq}^2,\\
\label{eq:notation:G}
G_{nm}&=& J_{n0}J_{m0}-J_{n-1,0}J_{m+ 1,0},
\end{eqnarray}
\end{subequations}
where $J_{nq}$ and $I_{nq}$ are related by $\beta J_{nq}=I_{n-1,q-1}+(n-2q)I_{n-1,q}$. Note that $\kappa_\parallel =\kappa$ in Eq.~(\ref{eq:kin:conductivity}) (which can be directly shown using Eq.~(\ref{eq:notation:J})), meaning that the charge diffusion along the magnetic field is unaffected by the magnetic field. Note that the leading-order Hall conductivity $\kappa_H$ is independent of the relaxation time $\tau_R$; it is purely due to the Lorentz force. The first term in $\kappa_H$ gives the classical Hall conductivity for a stationary material while the second term is due to the fluid flow and thus depends on how we approximate the dynamic kinetic and hydrodynamic equations.

Substituting $f^{(1)}_\vk$ into $\tau^{\m\n}:=\lan k^{\lan\m\ran} k^{\lan\n\ran}\ran_1=\pi^{\m\n}-\Delta^{\m\n}\Pi$, we obtain the viscous stress tensor $\tau^{\m\n}$ as
\begin{eqnarray}
\label{eq:kin:taumn}
\tau^{\m\n}&=&\sum_{i=1}^7\eta_i\eta^{\m\n\rho\sigma}_i w_{\rho\sigma},\end{eqnarray}
where the tensor forms of $\eta^{\m\n\r\s}_i$ are already defined in Eq.~(\ref{entropy:rank4tensor}) and Eq.~(\ref{eq:Hall-viscous-tensor}) but listed here for convenience 
\begin{subequations}
\label{eq:kin:etatensor}
\begin{eqnarray}
\eta^{\m\n\rho\sigma}_1 &=& b^\m b^\n b^\rho b^\sigma,\\
\eta^{\m\n\rho\sigma}_2 &=& \Xi^{\m\n}\Xi^{\rho\sigma},\\
\eta^{\m\n\rho\sigma}_3 &=& -\Xi^{\m\n}b^\rho b^\sigma-\Xi^{\rho\sigma}b^\m b^\n,\\
\eta^{\m\n\rho\sigma}_4 &=& -2\left[b^{(\m}\Xi^{\n)\rho}b^\sigma +b^{(\m}\Xi^{\n)\sigma}b^\rho\right],\\
\eta^{\m\n\rho\sigma}_5 &=& 2\,\Xi^{\rho(\m}\Xi^{\n)\sigma}-\Xi^{\m\n}\Xi^{\rho\sigma},\\
\eta^{\m\n\rho\sigma}_6 &=& -b^{(\m}b_\star^{\n)\rho}b^\sigma -b^{(\m}b_\star^{\n)\sigma}b^\rho,\\
\eta^{\m\n\rho\sigma}_7 &=& \Xi^{\rho(\m}b_\star^{\n)\sigma}+\Xi^{\sigma(\m}b_\star^{\n)\rho},
\end{eqnarray}
\end{subequations}
and the viscosities are given by
\begin{subequations}
\label{chap:b:vis}
\begin{eqnarray}
\label{chap:b:vis1}
\eta_1 &\equiv&\zeta_\parallel=\b\tau_R\left(3J_{32}-J_{31}\frac{\pt p}{\pt\e}\Big|_n-J_{21}\frac{\pt p}{\pt n}\Big|_\e\right),\\
\eta_2 &\equiv&\zeta_\perp=\b\tau_R\left(2J_{32}-J_{31}\frac{\pt p}{\pt\e}\Big|_n-J_{21}\frac{\pt p}{\pt n}\Big|_\e\right),\\
\eta_3 &\equiv&\zeta_\times=\b\tau_R\left(J_{32}-J_{31}\frac{\pt p}{\pt\e}\Big|_n-J_{21}\frac{\pt p}{\pt n}\Big|_\e\right),\\
\eta_4 &\equiv&\eta_\parallel=\frac{\b}{\tau_R(qB)^2}J_{52} +O(\xi^{-4}),\\
\eta_5 &\equiv&\eta_\perp=\frac{\b}{4\tau_R(qB)^2}J_{52} +O(\xi^{-4}),\\
\eta_6 &\equiv&\eta_{H\parallel}=\frac{2\b}{qB}J_{42} +O(\xi^{-3}),\\
\eta_7 &\equiv&\eta_{H\perp}=\frac{\b}{2qB}J_{42} +O(\xi^{-3}),
\end{eqnarray}
\end{subequations}
where $\zeta$'s and $\eta$'s are bulk and shear viscosities whose physical meaning has been discussed in Sec.~\ref{sec:MHD}. 
The above result automatically satisfies the Onsager relation for $\zeta_\times$ that was assumed in Sec.~\ref{sec:MHD}.
Note that $\eta_{H\parallel}$ and $\eta_{H\perp}$ do not depend on $\tau_R$ at the leading order in $1/\xi$; they arise due to the Lorentz force and are thus called Hall viscosities; see more discussions in Sec.~\ref{sec:currents-Hall}.
 
For massless Boltzmann gas, using $J_{nq}=(n+1)!\b^{2-n}p/[2(2q+1)!!]$~\cite{Denicol:2018rbw} and $\tau_R=9\l_{\rm mfp}/4$, the transport coefficients become
\begin{eqnarray}
\begin{split}
\label{chap:massless}
&\kappa_\parallel=\frac{3}{16}\l_{\rm mfp}n,\;\; \kappa_\perp=\frac{16}{45}\frac{\l_{\rm mfp}n}{\xi^2},\;\; \kappa_H=\frac{1}{5}\frac{n}{\b qB},\;\; \zeta_\parallel=\frac{12}{5}\l_{\rm mfp}p,\;\; \zeta_\perp=\frac{3}{5}\l_{\rm mfp}p,&\\
& \zeta_\times=-\frac{6}{5}\l_{\rm mfp}p,\;\; \eta_\parallel=\frac{32}{3}\frac{\l_{\rm mfp}p}{\xi^2}, \;\; \eta_\perp=\frac{8}{3}\frac{\l_{\rm mfp}p}{\xi^2},\;\; \eta_{H\parallel}=\frac{8p}{\b qB}, \;\; \eta_{H\perp}=\frac{2p}{\b qB}.&
\end{split}
\end{eqnarray}
Note that $\zeta_\parallel\zeta_\perp-\zeta_\times^2=0$ due to the conformal invariance in 3-dimensional space in the massless limit. %We also note that these results are obtained with the single particle energy and the equilibrium state without any influence of the magnetic field. This is, however, not the true equilibrium state under strong magnetic field. In Sec.~\ref{sec:transport-coeff}, we will discuss a more realistic calculation of some of these transport coefficients and show that $\zeta_\parallel$ in strong magnetic field vanishes at massless limit due to the conformal invariance in the dimensionally reduced 1+1 spacetime.

Recently, there have been intensive use of the Chapman-Enskog method to study the constitutive relations and transport coefficients in RMHD; see Refs.~\cite{Mohanty:2018eja,Das:2019ppb,Dey:2019axu,Das:2019pqd,Chen:2019usj,Dash:2020vxk,Rath:2020beo,Panda:2020zhr,Panda:2021pvq,Rath:2021ryd,Rath:2022oum} (also, e.g. Refs.~\cite{Satapathy:2021cjp,Ghosh:2020wqx} for related Kubo-formula calculations). Our discussions here serve as a simple overview and summary of how the Chapman-Enskog method is used to derive the constitutive relations of the first-order RMHD. For more information, the reader is referred to the list of references for recent works.

The above iterative procedure of solving Boltzmann equation can continue to the higher orders in gradient, but the calculations become much more involved. For example, at the second order, one obtains the Burnett-type equations~\cite{Burnett:1935,Burnett:1936}. However, the relativistic EOMs from the Chapman-Enskog expansion are in general unstable~\cite{Hiscock:1983zz,Hiscock:1985zz,Hiscock:1987zz,Denicol:2008ha,Pu:2009fj,Biswas:2020rps} (though there could exist special rest frames where the first-order relativistic dissipative hydrodynamics can be stable; see Refs.~\cite{Bemfica:2017wps,Bemfica:2019knx,Kovtun:2019hdm,Bemfica:2020zjp,Hoult:2020eho,Armas:2022wvb}). For the Burnett equations, even at the non-relativistic limit, they suffer from the so-called Bobylev instability~\cite{Bobylev:1982}. For this reason, instead, the method of moments are widely used in recent years because it can avoid this problem~\cite{Betz:2009zz,Denicol:2010xn,Denicol:2010br,Denicol:2012cn,Denicol:2012es}.

\subsection{Grad's method of moments}\label{sec:mom}

This approach was first established by Grad for non-relativistic systems~\cite{Grad:19491} that can be viewed as an expansion of the distribution function in terms of momenta $k^\n$ and was then generalized to relativistic systems by Israel and Stewart~\cite{Stewart:1971,Israel:1976213,Israel:1979wp,Israel:19792} among others. Recently, a variant of this method was developed and widely adopted to derive the causal second-order hydrodynamics~\cite{Betz:2009zz,Denicol:2010xn,Denicol:2010br,Denicol:2012cn,Denicol:2012es,Jaiswal:2013npa,Bazow:2013ifa,Denicol:2014vaa,Molnar:2016vvu,Molnar:2016gwq,Denicol:2018rbw,Denicol:2019iyh,Tinti:2018qfb,Most:2021uck,Wagner:2022ayd}. The following discussions closely follow Refs.~\cite{Denicol:2012cn,Denicol:2012es,Denicol:2018rbw,Denicol:2019iyh}. We first discuss the simpler situation without EM fields and then consider the effect of EM fields.

The ideal hydrodynamics is still generated by the local equilibrium distribution $f_{0\vk}$. Write the dissipative part of the distribution function as $\d f_\vk=f_\vk-f_{0\vk}=f_{0\vk}\tilde{f}_{0\vk}\phi_\vk$. In the method of moments, $\phi_\vk$ is expanded as a series of the irreducible tensors $1, k^{\lan\m\ran}, k^{\lan\m}k^{\n\ran}, k^{\lan\m}k^\n k^{\l\ran}, \cdots$, $\phi_\vk=\sum_{l=0}^M \l^{\lan\m_1\cdots\m_l\ran}(E_\vk)k_{\lan\m_1}\cdots k_{\m_l\ran}$. These irreducible tensors are defined by $k^{\lan\m_1}\cdots k^{\m_l\ran}=\Delta^{\m_1\cdots\m_l\n_1\cdots\n_l}k_{\n_1}\cdots k_{\n_l}$ and form a complete and orthogonal set in the spatial momentum space, satisfying~\cite{Denicol:2012cn} 
\begin{eqnarray}
\label{eq:moment:ortho}
\int dK F(E_\vk)k^{\lan\m_1}\cdots k^{\m_m\ran}k^{\lan\n_1}\cdots k^{\n_n\ran}=\frac{m!\d_{nm}}{(2m+1)!!}\Delta^{\m_1\cdots\m_m\n_1\cdots\n_m}\int dK F(E_\vk)(\Delta_{\a\b}k^\a k^\b)^m,\quad\quad\quad
\end{eqnarray}
where $(2m+1)!!$ is the double factorial, $F$ is an arbitrary converging scalar function, and $\Delta^{\m_1\cdots\m_m\n_1\cdots\n_m}$ is the projection $\Delta^{\m\n}$ when $m=1$ and the  symmetrized spatial traceless projection when $m\geq2$, \eg, $\Delta^{\m\n\a\b}=\frac{1}{2}(\Delta^{\m\a}\Delta^{\n\b}+\Delta^{\m\b}\Delta^{\n\a})-\frac{1}{3}\Delta^{\m\n}\Delta^{\a\b}$. The coefficients $\l^{\lan\m_1\cdots\m_l\ran}(E_\vk)$ may be further expanded in $E_\vk$, conveniently with an orthogonal basis of polynomials $P_n^{(l)}(E_\vk)$, $\l^{\lan\m_1\cdots\m_l\ran}(E_\vk)=\sum_{n=0}^{N_l}c_n^{\lan\m_1\cdots\m_l\ran}P_n^{(l)}(E_\vk)$~\footnote{Here, we have assumed that the coefficients $\l^{\lan\m_1\cdots\m_l\ran}(E_\vk)$ have no singularity at $E_\vk\rightarrow 0$ so that they are Taylor expandable (A singularity may appear when the system evolves towards, \eg, a condensate near $E_\vk\rightarrow 0$). }. The functions $P_n^{(l)}(E_\vk)$ are $n$-th order polynomials in $E_\vk$, $P_n^{(l)}(E_\vk)=\sum_{r=0}^n a_{nr}^{(l)}E_\vk^r$, where the coefficients $a_{nr}^{(l)}$ can be determined order by order using the orthonormality condition
\begin{eqnarray}
\nn
\label{eq:ortho:cond}
\frac{{\mathcal N}^{(l)}}{(2l+1)!!}\int dK (\Delta_{\a\b}k^\a k^\b)^l f_{0\vk}\tilde{f}_{0\vk}P_n^{(l)}(E_\vk)P_m^{(l)}(E_\vk)=\d_{nm},
\end{eqnarray} 
where ${\mathcal N}^{(l)}=(2l+1)!!\left[\int dK(\Delta_{\a\b}k^\a k^\b)^l f_{0\vk}\tilde{f}_{0\vk}\right]^{-1}$ are the normalization factors~\footnote{Here we list a few of the coefficients $a_{nr}^{(l)}$ which are useful for the following calculations: $a_{00}^{(l)}=1$ is by setting, $a_{11}^{(l)}=\pm J_{2l,l}/\sqrt{D_{2l+1,l}}$, $a_{10}^{(l)}=\mp J_{2l+1,l}/\sqrt{D_{2l+1,l}}$, $a_{22}^{(0)}=\pm\sqrt{J_{00}D_{10}}/\sqrt{J_{20}D_{20}+J_{30}G_{12}+J_{40}D_{10}}$, $a_{21}^{(0)}=G_{12}a_{22}^{(0)}/D_{10}, a_{20}^{(0)}=D_{20} a_{22}^{(0)}/D_{10}$. For given $n$, the sign of $a_{nr}^{(l)}$ can take either the upper or the lower convention without changing $\delta f_\vk$; we will use the upper-sign convention. More details about $P_n^{(l)}(E_\vk)$ can be found in Refs.~\cite{Denicol:2012cn,Denicol:2012es}. }. The remaining coefficients $c_n^{\lan\m_1\cdots\m_l\ran}$ can be more conveniently re-expressed using the irreducible moments,
\begin{eqnarray}
\label{eq:moment:rho}
\r_r^{\m_1\cdots\m_l}\equiv\lan E_\vk^r k^{\lan\m_1}\cdots k^{\m_l\ran}\ran_\d=\int dK E_\vk^r k^{\lan\m_1}\cdots k^{\m_l\ran} \d f_\vk,
\end{eqnarray} 
as $c_n^{\lan\m_1\cdots\m_l\ran}=\frac{{\mathcal N}^{(l)}}{l!}\sum_{r=0}^n\r_r^{\m_1\cdots\m_l}a_{nr}^{(l)}$. Finally, the distribution function is expressed as
\begin{eqnarray}
\label{eq:dist:mom}
f_\vk=f_{0\vk}\left(1+\tilde{f}_{0\vk}\sum_{l=0}^M\sum_{n=0}^{N_l}\sum_{r=0}^{n}\frac{{\mathcal N}^{(l)}}{l!}a_{nr}^{(l)}P_n^{(l)}(E_\vk)\r_r^{\m_1\cdots\m_l}k_{\lan\m_1}\cdots k_{\m_l\ran}\right).
\end{eqnarray} 
In principle, $M$ and $N_l$ should be infinite; but in practice, they are chosen to be finite in order to truncate the expansion. Thus, the distribution function is solely determined by the moments $\r_r^{\m_1\cdots\m_l}$ and, after substituting it into the Boltzmann equation, the Boltzmann equation is turned into a set of coupled EOMs for the moments. For our purpose, we focus on the three leading-order moments, $\r_0$, $\r_0^\m$, and $\r_0^{\m\n}$, because they are related to the dissipative currents,
\begin{eqnarray}
\label{eq:dissip:mom}
\r_0=-\frac{3}{m^2}\Pi,\;\; \r_0^\m=\n^\m,\;\; \r_0^{\m\n}=\pi^{\m\n}.
\end{eqnarray}  
To get the first relation to $\Pi$, we used the matching condition for $\e$. 
Together with the matching condition for $n$ and the Landau-Lifshitz frame condition, we find that 
\begin{eqnarray}
\label{eq:mom:constraints}
\r_1=\r_2=0,\;\; \r_1^\m=0.
\end{eqnarray}  

To derive the EOMs for $\r_0$, $\r_0^\m$, and $\r_0^{\m\n}$ or, equivalently, for $\Pi$, $\n^\m$, and $\pi^{\m\n}$, we re-write the Boltzmann equation (\ref{eq:boltz}) as
\begin{eqnarray}
\label{eq:boltz:delta}
\d \dot{f}_\vk=-\dot{f}_{0\vk}-\frac{1}{E_\vk}k^\m\nabla_\m f_{\vk}+\frac{1}{E_\vk}C[f],
\end{eqnarray}
where $\dot{A}\equiv dA/d\tau\equiv D A$. Substituting this expression into Eq.~(\ref{eq:kin:variousquantities}), one obtains a set of exact dynamical equations for $\Pi, \n^\m$, and $\pi^{\m\n}$: 
\begin{subequations}
\begin{eqnarray}
\label{eq:dyna:Pi}
\dot{\Pi}&\equiv&-\frac{1}{3}\frac{d}{d\tau}\int dK \Delta_{\m\n}k^\m k^\n \d f_\vk=-\frac{m^2}{3}\int dK \d \dot{f}_{\vk}\nonumber\\
&=&-\frac{m^2}{3}C-\b_\Pi\h-\zeta_{\Pi\Pi}\Pi\h+\zeta_{\Pi\pi}\pi^{\m\n}\xi_{\m\n}-\zeta_{\Pi\n}\pt_\m\n^\m+\frac{m^4}{9}\r_{-2}\h+\frac{m^2}{3}\nabla_\m\r_{-1}^\m+\frac{m^2}{3}\r_{-2}^{\m\n}\xi_{\m\n},\qquad\qquad\\
\label{eq:dyna:nu}\dot{\n}^{\lan\m\ran}&\equiv&\Delta^{\m}_\r\dot{\n}^\r = \int dK k^{\lan\m\ran}\d \dot{f}_\vk\nonumber\\
&=&C^\m+\b_\n\nabla^\m\a-\n^\m\h-\n_\l\left(\omega^{\l\m}+\frac{3}{5}\xi^{\l\m}\right)+\frac{1}{h}\left(\Pi\dot{u}^\m-\nabla^\m\Pi+\Delta^{\m}_\r\pt_\l\pi^{\l\r}\right)
\nonumber\\
&&-\frac{m^2}{3}\left(\nabla^\m\r_{-1}+\r_{-2}^\m\h+\frac{6}{5}\r_{-2}^\l \xi^{\m}_{~\l}\right)-\Delta^\m_\r\nabla_\l\r_{-1}^{\r\l}-\r_{-2}^{\m\r\l}\xi_{\r\l}, \\
\label{eq:dyna:pi}
\dot{\pi}^{\lan\m\n\ran}&\equiv&\Delta^{\m\n}_{~~\;\a\b}\frac{d}{d\tau}\int dK k^{\lan\a}k^{\b\ran}\d f_\vk=\int dK k^{\lan\m}k^{\n\ran}\d\dot{f}_\vk\nonumber\\
&=&C^{\m\n}+2\b_\pi \xi^{\m\n}-\frac{4}{3}\pi^{\m\n}\h-\frac{10}{7}\pi^{\r\lan\m} \xi^{\n\ran}_{~~\r}+2\pi^{\lan\m}_\r\omega^{\n\ran\r}+\frac{6}{5}\Pi \xi^{\m\n}
-\frac{2m^4}{15}\r_{-2}\xi^{\m\n}-\frac{2m^2}{5}\nabla^{\lan\m}\r_{-1}^{\n\ran}\nonumber\\
&&
-\frac{m^2}{3}\r_{-2}^{\m\n}\h-\frac{4m^2}{7}\r_{-2}^{\l\lan\m}\xi^{\n\ran}_{~~\l}-\Delta^{\m\n}_{~~\;\a\b}\nabla_\l\r_{-1}^{\a\b\l}-\r_{-2}^{\m\n\a\b}\xi_{\a\b},
\end{eqnarray}
\end{subequations}
where $\omega^{\m\n}=(\nabla^\m u^\n-\nabla^\n u^\m)/2$ is the vorticity. We have adapted the Landau-Lifshitz frame (\ref{eq:kin:LLconstraint}) and the matching conditions (\ref{eq:kin:match}) and defined the collision terms 
\begin{eqnarray}
\label{eq:mom:collision}
C^{\m_1\cdots\m_l}=\int dK \frac{1}{E_\vk}k^{\lan\m_1}\cdots k^{\m_l\ran}C[f].
\end{eqnarray}
To derive Eqs.~(\ref{eq:dyna:Pi})-(\ref{eq:dyna:pi}), the exact EOMs (\ref{eq:kin:Ncons})-(\ref{eq:kin:Tcons}) are used to eliminate the time derivative of $\a,\b$, and $u^\m$: 
\begin{subequations}
\begin{eqnarray}
\label{eq:mom:alpha}
\dot{\a}&=&\frac{1}{D_{20}}\left[ -J_{30}(n\h+\pt_\m\n^\m) + J_{20}(\e+p+\Pi)\h-J_{20}\pi^{\m\n}\xi_{\m\n}\right],\\
\label{eq:mom:beta}
\dot{\b}&=&\frac{1}{D_{20}}\left[ -J_{20}(n\h+\pt_\m\n^\m) + J_{10}(\e+p+\Pi)\h-J_{10}\pi^{\m\n}\xi_{\m\n}\right],\\
\label{eq:mom:umu}
\dot{u}^\m&=&\frac{1}{\b}\left(\frac{1}{h}\nabla^\m\a-\nabla^\m\b\right)-\frac{1}{\e+p}\left(\Pi \dot{u}^\m-\nabla^\m \Pi+\Delta^\m_\n\pt_\l\pi^{\n\l}\right),
\end{eqnarray}
\end{subequations}
where $h=(\e+p)/n$ is the enthapy per particle.
The coefficients in Eqs.~(\ref{eq:dyna:Pi})-(\ref{eq:dyna:pi}) are all thermodynamic quantities given by~\cite{Denicol:2010xn,Denicol:2012cn,Denicol:2012es}
\begin{subequations}
\begin{eqnarray}
\label{eq:mom:betaPi}
\b_\Pi &=& \frac{m^2}{3}\left[I_{01}-I_{00}-\frac{n}{D_{20}}(h G_{20}-G_{30})\right]=\left[\left(\frac{1}{3}-c_s^2\right)(\e+p)-\frac{2}{9}(\e-3p)-\frac{m^4}{9}\left\lan \frac{1}{E_\vk^{2}}\right\ran_0\right],\nonumber\\ \\
\zeta_{\Pi\Pi}&=&\frac{2}{3}-\frac{m^2}{3}\frac{G_{20}}{D_{20}}=1-\left(\frac{\pt p}{\pt\e}\right)_n,\\
\zeta_{\Pi\pi}&=&-\frac{m^2}{3}\frac{G_{20}}{D_{20}}=\frac{1}{3}-\left(\frac{\pt p}{\pt\e}\right)_n,\\
\zeta_{\Pi\n}&=&\frac{m^2}{3}\frac{G_{30}}{D_{20}}=-\left(\frac{\pt p}{\pt n}\right)_\e,\\
\beta_\n&=& J_{11}-\frac{1}{h}J_{21}=T\left(\frac{2}{3}\frac{\e-3p}{m^2}-\frac{n^2}{\e+p}+\frac{m^2}{3}\left\lan \frac{1}{E_\vk^{2}}\right\ran_0\right),\\
\b_\pi&=& I_{21}-I_{22}=\frac{4}{5}p+\frac{1}{15}(\e-3p)-\frac{m^4}{15}\left\lan \frac{1}{E_\vk^{2}}\right\ran_0.
\end{eqnarray}
\end{subequations}
Note that $\b_\Pi, \b_\n, \b_\pi$ agree with the bulk viscosity, conductivity, and shear viscosity given in Eq.~(\ref{eq:kin:conductivity})-(\ref{eq:kin:shearvis}) obtained from the Chapman-Enskog method up to the overall factor of $\tau_R$. This is not surprising because they have the same origin: they both are from the gradient of $f_{0\vk}$ in the Boltzmann equation.

The EOMs (\ref{eq:dyna:Pi})-({\ref{eq:dyna:pi}}) are not closed because they contain the moments other than $\Pi, \n^\m, \pi^{\m\n}$. To close them, we have to truncate the expansion (\ref{eq:dist:mom}) for $\d f_{0\vk}$ so that it is expressed by the dissipative currents $\Pi, \n^\m, \pi^{\m\n}$. The minimal truncation scheme adjusts $M=2$, $N_0=2, N_1=1$, and $N_2=0$, so that the distribution function is expanded in terms of 14 moments which are $\r_0=-3\Pi/m^2, \r_0^\m=\n^\m, \r_0^{\m\n}=\pi^{\m\n}, \r_1=0, \r_2=0, \r_1^\m=0$ (The last two scalar and one vector moments vanish due to the matching condition and Landau-Lifshitz frame choice). This scheme is called the 14-moment approximation, introduced first by Israel and Stewart. Generalization to include other moments are possible, but it does not guarantee a better performance of the method. Now, using Eq.~(\ref{eq:moment:rho}) and Eq.~(\ref{eq:mom:collision}), all the moments and the collision terms are expressed by $\Pi, \n^\m, \pi^{\m\n}$ (In particular, $\r_r^{\m_1\cdots\m_l}$ with $l\geq 3$ are vanishing), and thus Eq.~(\ref{eq:dyna:Pi})-({\ref{eq:dyna:pi}}) are closed.

To see how the 14-moment approximation works in practice. Let us consider a massless Boltzmann gas (\ie, $a=0$ for $f_{0\vk}$) and consider the RTA for the collision kernel, \ie, Eq.~(\ref{eq:kin:rta}). In this case, $I_{nq}=J_{nq}=\frac{(n+1)!}{(2q+1)!!}\b^{2-n}p/2$ and analytical expressions can be obtained for all the transport coefficients. In the massless limit, the bulk viscous pressure vanishes and each term in Eq.~(\ref{eq:dyna:Pi}) vanishes as well. The dissipative part of the distribution function reads~\footnote{Though $\Pi$ vanishes in the massless limit, $\r_0=-3\Pi/m^2$ may not vanish and $\d f_\vk$ still contain terms in the form of $f_{0\vk}\tilde{f}_{0\vk}\left[c_0+c_1 P_1^{(0)}(E_\vk)+c_2 P_2^{(0)}(E_\vk)\right]$ with $c_{r}\propto \Pi/m^2$. But the contributions of these terms to Eqs.~(\ref{eq:dyna:Pi})-(\ref{eq:dyna:pi}) vanish at $m^2\rightarrow 0$ limit. So we omit them.},
\begin{eqnarray}
\label{eq:mom:deltaf}
\d f_\vk &=& f_{0\vk}\tilde{f}_{0\vk}\left[-\frac{1}{p}\left(1-2 P_1^{(1)}(E_\vk)\right)\n^\m k_\m+\frac{\b^2}{8p}\pi^{\m\n}k_{\lan\m}k_{\n\ran}\right],
\end{eqnarray}
where $P_1^{(1)}(E_\vk)=\b E_\vk/2-2$. The collision terms are thus given by
\begin{eqnarray}
\label{eq:mom:collisionterms}
C^\m = -\frac{\n^\m}{\tau_R},\;\;\; C^{\m\n} = -\frac{\pi^{\m\n}}{\tau_R}.
\end{eqnarray}
Similarly, all the moments appearing in Eqs.~(\ref{eq:dyna:nu})-(\ref{eq:dyna:pi}) are easily obtained. The final results are~\cite{Denicol:2010xn,Denicol:2012cn,Denicol:2012es}
\begin{subequations}
\begin{eqnarray}
\label{eq:mom:nu}
\tau_R\dot{\n}^{\lan\m\ran}+\n^\m
&=&\kappa\nabla^\m\a-\tau_R\n_\l\omega^{\l\m}-\d_{\n\n}\n^\m\h-\l_{\n\n}\n_\l \xi^{\l\m}+l_{\n\pi}\Delta^{\m}_\r\nabla_\l\pi^{\l\r}\nonumber\\&&+\tau_{\n\pi}\pi^{\m\l}\dot{u}_\l-\l_{\n\pi}\pi^{\m\n}\nabla_\n\a, \qquad\;\;\\
\label{eq:mom:pi}
\tau_R\dot{\pi}^{\lan\m\n\ran}+\pi^{\m\n}
&=&2\eta \xi^{\m\n}+2\tau_R\pi^{\lan\m}_\r\omega^{\n\ran\r}-\d_{\pi\pi}\pi^{\m\n}\h-\tau_{\pi\pi}\pi^{\lan\m}_\r \xi^{\n\ran\r},
\end{eqnarray}
\end{subequations}
where the transport coefficients are
\begin{eqnarray}
\begin{split}
\label{eq:mom:transcoeff}
&\kappa=\tau_R\b_\n=\frac{\tau_R}{12}n,\;\; \d_{\n\n}=\tau_R,\;\; \l_{\n\n}=\frac{3}{5}\tau_R,\;\; l_{\n\pi} = \l_{\n\pi} = \frac{\tau_R}{20}\b,\;\; \tau_{\n\pi}=\frac{\tau_R}{5}\b,&\; \\
&\eta =\tau_R\b_\pi=\frac{4\tau_R}{5}p,\;\; \d_{\pi\pi}=\frac{4\tau_R}{3},\;\; \tau_{\pi\pi}=\frac{10\tau_R}{7}.&
\end{split}
\end{eqnarray}
Here, $\kappa$ and $\eta$ are the conductivity and the shear viscosity, respectively; they are the first-order transport coefficients. All the other ones on the right-hand sides are the second-order transport coefficients. The Navier-Stokes limit is reached when $t\gg \tau_R$, \ie, $\n^\m$ and $\pi^{\m\n}$ relax to their Navier-Stokes values at the late time. Such relaxation dynamics is very important for practical applications of relativistic dissipative hydrodynamics, because it cures the numerical instability stemming from the instantaneous responses represented in the usual first-order Navier-Stokes constitutive relations. 

With the above preparation, we now turn to the Grad's method of moments for RMHD. The basic logic is the same as above, so that we will skip some of the intermediate steps. We will always use the Landau-Lifshitz choice for the rest frame of the fluid. First, the EM fields induce the Joule heat/Lorentz force term in Eq.~(\ref{mhd:kin:emcons}), 
and the exact EOMs of $\a, \b$, and $u^\m$, Eqs.~(\ref{eq:mom:alpha})-(\ref{eq:mom:umu}), 
are replaced as 
\begin{subequations}
\begin{eqnarray}
\label{B:mom:alpha}
\dot{\a}&=&\frac{1}{D_{20}}\left[ -J_{30}(n\h+\pt_\m\n^\m) + J_{20}(\e+p+\Pi)\h-J_{20}\pi^{\m\n}\xi_{\m\n}+J_{20}qE^\m\n_\m\right],\\
\label{B:mom:beta}
\dot{\b}&=&\frac{1}{D_{20}}\left[ -J_{20}(n\h+\pt_\m\n^\m) + J_{10}(\e+p+\Pi)\h-J_{10}\pi^{\m\n}\xi_{\m\n}+J_{10}q E^\m \n_\m\right],\\
\label{B:mom:umu}
\dot{u}^\m&=&\frac{1}{\b}\left(\frac{1}{h}\nabla^\m\a-\nabla^\m\b\right)-\frac{1}{\e+p}\left(\Pi \dot{u}^\m-\nabla^\m \Pi+\Delta^\m_\n\pt_\l\pi^{\n\l}-n qE^\m-qB b_\star^{\m\n}\n_\n\right).\;\;
\end{eqnarray}
\end{subequations}
These equations contain unknown dynamical variables $\Pi, \n^\m, \pi^{\m\n}$ and are thus not closed. The EOMs of $\Pi, \n^\m, \pi^{\m\n}$ can be derived in the same manner as in the case without EM fields. They can be conveniently expressed as the irreducible moments of rank 0, 1, and 2, which are $\rho_0, \rho_0^\m, \rho_0^{\m\n}$.  These exact EOMs read~\cite{Denicol:2018rbw,Denicol:2019iyh}
\begin{subequations}
\begin{eqnarray}
\label{B:dyna:Pi}
\dot{\Pi}
&=& \frac{m^2}{3}\frac{G_{20}}{D_{20}}qE_\m\n^\m-\frac{m^2}{3}qE_\m\r_{-2}^\m
\\
&&
+ \Big[ -\frac{m^2}{3}C-\b_\Pi\h-\zeta_{\Pi\Pi}\Pi\h+\zeta_{\Pi\pi}\pi^{\m\n}\xi_{\m\n}-\zeta_{\Pi\n}\pt_\m\n^\m+\frac{m^4}{9}\r_{-2}\h+\frac{m^2}{3}\nabla_\m\r_{-1}^\m+\frac{m^2}{3}\r_{-2}^{\m\n}\xi_{\m\n} \Big] , \nonumber\\
\label{B:dyna:nu}\dot{\n}^{\lan\m\ran}
&=& \b_\n \b qE^\m
+qB b_{\star\; \n }^{\m}\left(-\frac{1}{h} \n^\n+\r_{-1}^\n\right)+\left(\frac{2}{3}\r_0+\frac{m^2}{3}\r_{-2}\right)qE^\m+\r^{\m\n}_{-2}qE_\n
\\
&& + \Big[ C^\m
%+\b_\n(\nabla^\m\a+\b qE^\m)
+ \b_\n \nabla^\m\a
-\n^\m\h-\n_\l\left(\omega^{\l\m}+\frac{3}{5}\xi^{\l\m}\right)+\frac{1}{h}\left(\Pi\dot{u}^\m-\nabla^\m\Pi+\Delta^{\m}_\r\pt_\l\pi^{\l\r}\right)\nonumber\\
&&-\frac{m^2}{3}\left(\nabla^\m\r_{-1}+\r_{-2}^\m\h+\frac{6}{5}\r_{-2}^\l \xi^{\m}_{~\l}\right)-\Delta^\m_\r\nabla_\l\r_{-1}^{\r\l}-\r_{-2}^{\m\r\l}\xi_{\r\l} \Big] , \nonumber\\
\label{B:dyna:pi}
\dot{\pi}^{\lan\m\n\ran}&=&
 2 qB \rho_{-1}^{\lambda(\mu}  b_{\star\  \lambda}^{\nu)}
+ qE_\a  \rho_{-2}^{\a\mu\nu} 
+ \frac25 m^2  qE^{\langle\mu}\rho_{-2}^{\nu\rangle}
+ \frac85 q E^{\langle\mu} \nu^{\nu\rangle} 
\\
&&+ \Big[ C^{\m\n}+2\b_\pi \xi^{\m\n}-\frac{4}{3}\pi^{\m\n}\h-\frac{10}{7}\pi^{\lan\m}_\r \xi^{\n\ran\r}+2\pi^{\lan\m}_\r\omega^{\n\ran\r}+\frac{6}{5}\Pi \xi^{\m\n}-\frac{2m^4}{15}\r_{-2}w^{\lan\m\n\ran}-\frac{2m^2}{5}\pt^{\lan\m}_\perp\r_{-1}^{\n\ran}\nonumber\\
&&-\frac{m^2}{3}\r_{-2}^{\m\n}\h-\frac{4m^2}{7}\r_{-2}^{\l\lan\m}\xi^{\n\ran}_{\l}-\Delta^{\m\n}_{~~\;\a\b}\nabla_\l\r_{-1}^{\a\b\l}-\r_{-2}^{\m\n\a\b}\xi_{\a\b}
\Big] . \nonumber
\end{eqnarray}
\end{subequations}
Note that only the EM related terms differ from Eqs.~(\ref{eq:dyna:Pi})-(\ref{eq:dyna:pi}), 
and the other terms between the square brackets are the same. Within the RTA, the collision terms are given in very simple forms: 
\begin{eqnarray}
C = \frac{3}{\tau_R m^2}\Pi, \quad C^\m = -\frac{\n^\m}{\tau_R}, \quad C^{\m\n} = -\frac{\pi^{\m\n}}{\tau_R}.
\end{eqnarray}
The above equations, together with Eqs.~(\ref{B:mom:alpha})-(\ref{B:mom:umu}), are still not closed because they involve the moments other than $\r_0, \r^\m_0, \r^{\m\n}_0$. To close them, we again take the 14-moment approximation in which the moment expansion of $\d f_\vk$ is truncated by choosing the first 14 moments (i.e., by taking $M=2$, $N_0=2, N_1=1$, and $N_2=0$). In this scheme, the independent moments are $\r_0=-3\Pi/m^2, \r_0^\m=\n^\m, \r_0^{\m\n}=\pi^{\m\n}, \r_1=0, \r_2=0, \r_1^\m=0$ (The last two scalar and one vector moments vanish due to the matching conditions for $n$ and $\e$ and Landau-Lifshitz frame choice). The other moments can thus be expressed in terms of the first 14 moments:
\begin{subequations}
\begin{eqnarray}
\label{B:other:rank0}
\r_{-r}&=&-\frac{3}{m^2}\gamma_r^{(0)}\Pi, \\
\label{B:other:rank1}
\r^\m_{-r} &=& \gamma_r^{(1)}\n^\m, \\
\label{B:other:rank2}
\r^{\m\n}_{-r} &=& \gamma_r^{(2)}\pi^{\m\n}, 
%\label{B:other:rank3}
%\r^{\m\n\l}_{-r} &=& 0, 
\end{eqnarray}
\end{subequations}
and the moments of rank higher than 2 are  all vanishing. 
The coefficients $\gamma_r^{(l)}$ are thermodynamic functions given by~\cite{Denicol:2012cn}
\begin{eqnarray}
\label{B:other:gammar}
\gamma_r^{(l)}=\frac{(-1)^{l} }{(2l+1)!!}\frac{1}{J_{2l,l}}\sum_{n=0}^{N_l}a_{n0}^{(l)}\int dK f_{0\vk}\tilde{f}_{0\vk}E_\vk^{-r}P_n^{(l)}(E_\vk)(\Delta_{\a\b} k^\a k^\b)^l.
\end{eqnarray}
With Eqs.~(\ref{B:other:rank0}) -(\ref{B:other:rank2}), Eqs.~(\ref{B:dyna:Pi})-(\ref{B:dyna:pi}) and Eqs.~(\ref{B:mom:alpha})-(\ref{B:mom:umu}) are closed with given EM fields. 
%\cgd{The residual terms in Eqs.~(\ref{B:other:rank0}) -(\ref{B:other:rank2}) stem from the truncation of the moment expansion.}

The peculiar feature of Eqs.~(\ref{B:dyna:Pi})-(\ref{B:dyna:pi}) is that they show the relaxation behaviors of $\Pi, \n^\m, \pi^{\m\n}$ towards the hydrodynamic constitutive relations at time scales much larger that $\tau_R$.  
In order to make a comparison with the results from the Chapman-Enskog method, we take the Navier-Stokes limit for Eqs.~(\ref{B:dyna:Pi})-(\ref{B:dyna:pi}). In this limit, we regard the dissipative fluxes $\Pi\sim \n^\m\sim \pi^{\m\n}\sim O(\pd)$ as well as $E^\m\sim O(\pd), B^\m\sim O(1)$ and discard all second-order terms in Eqs.~(\ref{B:dyna:Pi})-(\ref{B:dyna:pi}). This leads to 
\begin{subequations}
\begin{eqnarray}
\label{B:dyna:Pi2}
\Pi &=& -\tau_R\b_\Pi\h, \\
\label{B:dyna:nu2}
\n^\m &=& \tau_R\b_\n(\nabla^\m\a+\b qE^\m)+qB\tau_R b_\star^{\m\n}\left(\gamma_1^{(1)}-\frac{1}{h}\right)\n_\n, \\
\label{B:dyna:pi2}
\pi^{\m\n}&=&2\tau_R\b_\pi \xi^{\m\n}+qB\tau_R\gamma_1^{(2)}(b_\star^{\m\l}\pi^\n_{~\l}+b_\star^{\n\l}\pi^\m_{~\l}).
\end{eqnarray}
\end{subequations}
After solving Eqs.~(\ref{B:dyna:nu2})-(\ref{B:dyna:pi2}) for $\n^\m$ and $\pi^{\m\n}$ and combining $\pi^{\m\n}$ and $\Pi$ into $\tau^{\m\n}=\pi^{\m\n}-\Pi\Delta^{\m\n}$, we obtain the following constitutive relations at $O(\pd)$:
\begin{subequations}
\begin{eqnarray}
\label{B:mom:con:nu}
\n^\m &=& \kappa^{\m\n}\left(\pt_\n\a+\b qE_\n\right),\\
\label{B:mom:con:tau}
\tau^{\m\n}&=&\sum_{i=1}^7\eta_i\eta^{\m\n\rho\sigma}_i w_{\rho\sigma},
\end{eqnarray}
\end{subequations}
with $\kappa^{\m\n}$ and $\eta^{\m\n\rho\sigma}_i$ given in Eq.~(\ref{eq:kin:chapkmn}) and Eq.~(\ref{eq:kin:etatensor}) and the transport coefficients given by
\begin{subequations}
\label{B:mom:con:transp}
\begin{eqnarray}
\kappa_\perp &=&\frac{\tau_R\b_\n}{1+(qB\tau_R)^2(\gamma_1^{(1)}-1/h)^2}=\frac{1}{(qB)^2\tau_R}\frac{J_{31}D^2_{31}}{J_{41}^2D_{21}}+O(\xi^{-4}), \\
\kappa_\parallel &=&\tau_R\b_\n=\tau_R\frac{D_{21}}{J_{31}}, \\
\kappa_\times &=&\frac{qB\tau^2_R\b_\n(\gamma_1^{(1)}-1/h)}{1+(qB\tau_R)^2(\gamma_1^{(1)}-1/h)^2}=\frac{1}{qB}\frac{D_{31}}{J_{41}} +O(\xi^{-3}), \\
\eta_1 &\equiv&\zeta_\parallel=\tau_R\left(\b_\Pi+\frac{4}{3}\b_\pi\right),\\
\eta_2 &\equiv&\zeta_\perp=\tau_R\left(\b_\Pi+\frac{1}{3}\b_\pi\right),\\
\eta_3 &\equiv&\zeta_\times=\tau_R\left(\b_\Pi-\frac{2}{3}\b_\pi\right),\\
\eta_4 &\equiv&\eta_\parallel=\frac{\tau_R\b_\pi}{1+(\gamma_1^{(2)}qB\tau_R)^2}= \frac{\b}{(qB)^2\tau_R}\frac{J_{42}^2}{J_{32}}+O(\xi^{-4}),\\
\eta_5 &\equiv&\eta_\perp=\frac{\tau_R\b_\pi}{1+4(\gamma_1^{(2)}qB\tau_R)^2}= \frac{\b}{4(qB)^2\tau_R}\frac{J_{42}^2}{J_{32}}+O(\xi^{-4}),\\
\eta_6 &\equiv&\eta_{H\parallel}=\frac{2qB\tau^2_R\b_\pi \gamma_1^{(2)}}{1+(\gamma_1^{(2)}qB\tau_R)^2}= \frac{2\b}{qB}J_{42} +O(\xi^{-3}),\\
\eta_7 &\equiv&\eta_{H\perp}=\frac{2qB\tau^2_R\b_\pi \gamma_1^{(2)}}{1+4(\gamma_1^{(2)}qB\tau_R)^2}= \frac{\b}{2qB}J_{42} +O(\xi^{-3}),
\end{eqnarray}
\end{subequations}
where we have used the relations $\gamma_1^{(1)}=(J_{11}J_{41}-J_{21}J_{31})/D_{31}, \gamma_1^{(2)}=J_{32}/J_{42}$ which can be directly calculated from Eq.~(\ref{B:other:gammar}) and $\b_\pi=I_{21}-I_{22}=\b J_{32}$. Compared with Eqs.~(\ref{eq:kin:chapcond}) and (\ref{chap:b:vis}), we find that the results from the method of moments in the Navier-Stokes limit coincide with the results from Chapman-Enskog method at the leading-order in $1/\xi$  except for $\kappa_\perp, \eta_\parallel$ and $\eta_\perp$. For Boltzmann gas in the massless limit, they are given by
\begin{eqnarray}
\begin{split}
\label{chap:massless2}
\kappa_\perp=\frac{16}{75}\frac{\l_{\rm mfp}n}{\xi^2},\;\; \eta_\parallel=\frac{80}{9}\frac{\l_{\rm mfp}p}{\xi^2}, \;\; \eta_\perp=\frac{20}{9}\frac{\l_{\rm mfp}p}{\xi^2},
\end{split}
\end{eqnarray}
and all the other transport coefficients are the same as those given in Eq.~(\ref{chap:massless}). We note that the results of method of moments here are in complete consistence with Refs.~\cite{Denicol:2018rbw,Denicol:2019iyh}, but one should be careful when making the detailed comparison because Refs.~\cite{Denicol:2018rbw,Denicol:2019iyh} did not use the RTA which makes the relaxation time $\tau_\pi$ for the shear viscous tensor different from ours. %$20/27$ smaller than ours.

In this section, we construct the RMHD constitutive relations at $O(\pt)$ and gives explicit expressions for the anisotropic transport coefficients 
using the RTA. In the next section, we discuss a more rigorous calculation of these transport coefficients in QED or QCD going beyond the RTA. To end this section, we note that for practical calculations using kinetic equations in, e.g., astrophysics, one may use the particle-in-cell method to perform the full kinetic calculations going beyond the RTA; see, e.g., Refs.~\cite{Marcowith:2020vho,Pohl:2019nvw,Baumgarte-Shapiro2010,Baumgarte-Shapiro2021}.

\section{Perturbative evaluation of transport coefficients}
 \label{sec:transport-coeff}

In the previous section, we discussed the kinetic theory with the relaxation-time approximation (RTA). This simple model provides us with an insight into how 
the balance between the driving forces and the collision term is realized. 
However, this model has a few drawbacks that should be improved 
so that one can push forward the kinetic theory. For example, the RTA does not automatically satisfy the conservation laws of energy, momentum, particle number, etc, since symmetries of the system are not implemented in the collision term. In addition, the RTA collision kernel completely ignores the potential effects of the magnetic fields on the scattering processes. 

In case of weak-coupling theories, one can achieve more realistic evaluation of the transport coefficients with the perturbation theories that respect the symmetries 
and can take into account effects of the magnetic fields on the scattering processes~\cite{Hattori:2016lqx, Hattori:2016cnt, Hattori:2017qih, Li:2017tgi, Fukushima:2017lvb, Fukushima:2019ugr} 
(see also Refs.~\cite{Critelli:2014kra, Li:2018ufq} for the strong-coupling methods). 
In this section, we demonstrate the perturbative evaluation of the transport coefficients in the strong magnetic fields where the energy eigenstates of the scattering particles are subject to the Landau quantization. 
One finds that the chiral symmetry constrains the scattering amplitudes 
among the particles in the ground states called the lowest Landau level (LLL). 
This serves as a prominent example of the aspects that are not captured in the RTA treatment in last section. 

\subsection{Transport coefficients in strong magnetic fields}
\label{sec:transport-coeff:lll}

\begin{figure}[t]
%\vspace{-1cm}
     \begin{center}
              \includegraphics[width=0.4\hsize]{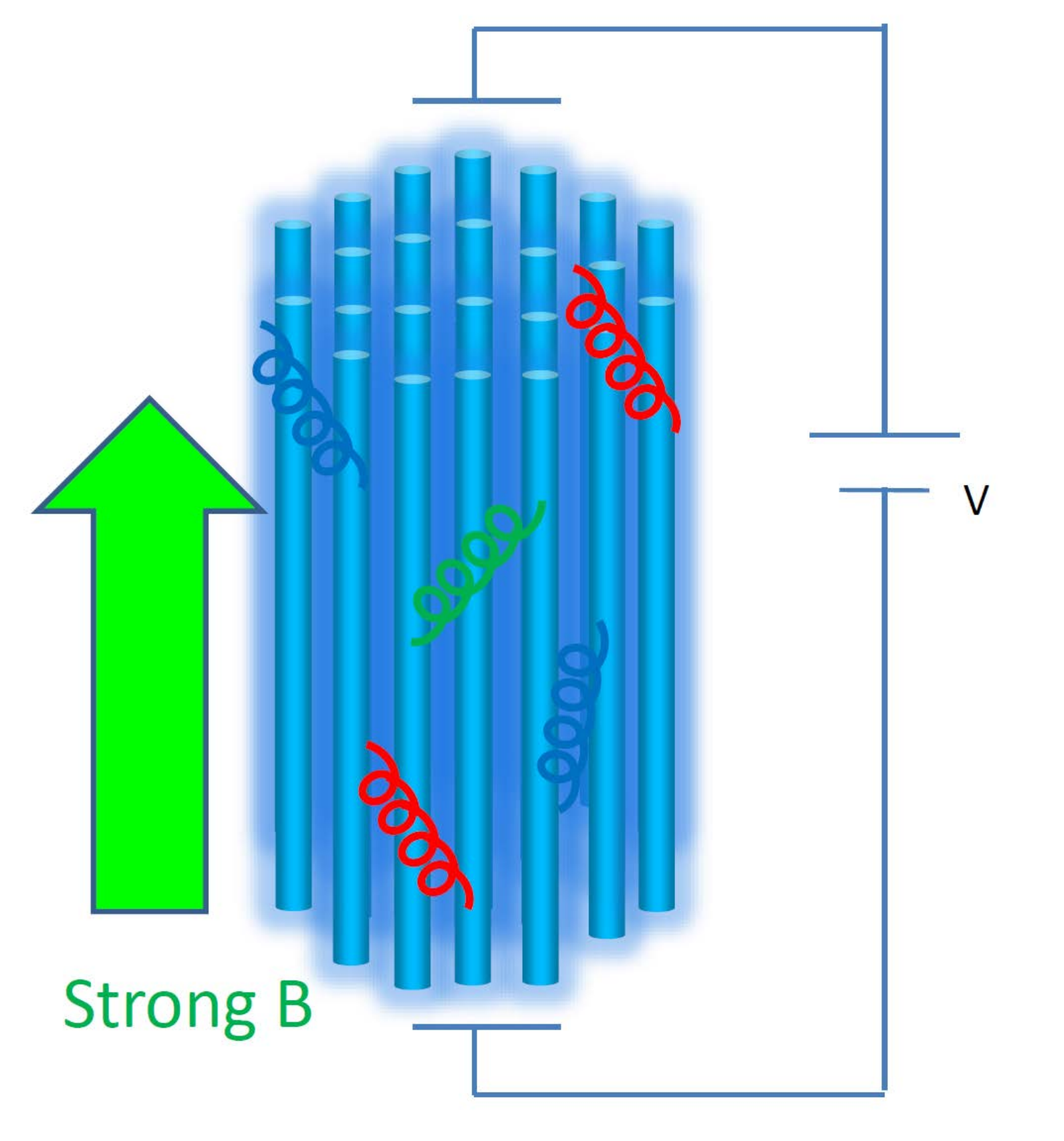}
     \end{center}
\vspace{-1cm}
\caption{Induced currents in a strong magnetic field.
Fermion carriers live in the (1+1) dimensions (blue tubes), while photon scatterers in the (3+1) dimensions.}
\label{fig:conductivity}
\end{figure}

We provide a concise description of perturbative evaluation of 
the transport coefficients in the strong magnetic field limit. 
We focus on the contributions from a single fermion with an electric charge $ e $ 
and mass $ m $ for simplicity.
Notice that the cyclotron radius is inversely proportional to the magnetic-field strength. 
Therefore, it is expected that the charged fermions can only serve as transport carriers 
for the currents along the magnetic field in the strong field limit (cf. Fig.~\ref{fig:conductivity}). 
Indeed, one can show that the fermions in the LLL can only contribute to 
the longitudinal electric current $ J_z $ and the longitudinal pressure $ \delta p_\para $ 
\cite{Hattori:2016cnt, Hattori:2016lqx, Hattori:2017qih}. 
Therefore, only the longitudinal conductivity $  \sigma_\para$ and bulk viscosity $ \zeta_\para $ take
nonzero values in the strong-field limit, while the other transport coefficients vanish in the strong-field limit.
 
To see the induced currents, we apply an electric field $ \bE =(0,0,E_z) $ and a flow perturbation
with an expansion/compression along the magnetic field, $ {\bm u} \sim ( 0, 0, u_z(z) ) $.
In the linear-response regime where the induced currents are linear in those driving forces, 
the longitudinal transport coefficients $  \sigma_{\para}$ and $ \zeta_\para $ %in a charge-neutral plasma
are given by~\footnote{
A factor of $ 1/3 $ was attached to $ \zeta_\para $ in Ref.~\cite{Hattori:2017qih}
according to the convention in Ref.~\cite{Huang:2011dc}.
Here, we follow the present convention without this numerical factor (see Eq.~(\ref{eq:correspondences}) for the correspondences). 
Note also that the color factors are included in Ref.~\cite{Hattori:2017qih} for the QCD plasma, 
while we focus on the QED plasma for simplicity. }
\begin{subequations}
\begin{eqnarray}
&&
\sigma_{\para} = \frac{ J_z}{E_z} = e  \frac{e B }{ 2\pi } \cdot
2 \int_{-\infty}^{\infty}  \frac{dk_z}{2\pi} \frac{k_z}{\epsilon_{k} } \left[ \frac{\delta f(k_z)}{E_z} \right]
\label{eq:conductivity}
\, ,
\\
&&
\zeta_\para = -  \frac{ \tilde p_{\para} }{\partial _z u_z}
= -  \frac{ e B }{ 2\pi } \cdot
2 \int_{-\infty}^{\infty}  \frac{dk_z}{2\pi}  
\frac{p_k ^2 -  \Theta_\beta  \epsilon_{k} ^2}{ \epsilon_{k} }
\left[  \frac{ \delta f(k_z) }{\partial _z u_z} \right]
\label{eq:bulk-viscosity}
\, ,
\end{eqnarray}
\end{subequations}
where $\Theta_\beta =(\partial p_\para/\partial \epsilon)_B$~\footnote{Here, the dominant contribution to the pressures comes from the matter part, \eg, $p_\para=p_{{\rm matt}\para}$. To simplify the notations, we omit the subscript `$\rm matt$' in this section.} and $\e_k$ is the one-particle energy in the LLL 
(see Eq.~(\ref{eq:Landau-levels}) for an explicit form). 
We have assumed without loosing generality 
that a homogeneous magnetic field is oriented in the $  z$ direction with $ eB >0 $. The two-dimensional transverse phase space is degenerate 
with the Landau degeneracy factor $ eB / (2\pi)  $ since the energy eigenvalue does not depend 
on the position of the cyclotron motion in a homogeneous magnetic field.
The momentum integrals come from standard definitions of the currents 
with the one-particle distribution functions $ f(k_z) $
where $ \delta f(k_z) \equiv f(k_z)-f_{\rm eq}(k_z)$ denotes the displacement from
the equilibrium distribution function caused by the driving force. 
In neutral plasmas, particles and antiparticles provide the same contributions,
which results in the factor of $ 2 $ in front of the integrals.
There is no spin degeneracy factor 
because the LLL is the unique ground state with respect to the spin direction.

To extract  the off-equilibrium component of the pressure $   \tilde p_{\para}$, one needs to subtract the equilibrium component 
because the pressure varies even in an adiabatic expansion/compression~\cite{Arnold:2006fz}.
In the operator form, the off-equilibrium component of the pressure is
$ \tilde p_\para = T^{\mu\nu} b_\mu b_\nu - \Theta_\beta T^{\mu\nu} u_\mu u_\nu=  T^{33}  - \Theta_\beta T^{00} $. In the massless limit, the adiabatic expansion rate is given by $  \Theta_\beta = 1$ that agrees with an inverse of the number of spatial dimensions $ (=1) $
in the presence of the effective dimensional reduction~\cite{Hattori:2017qih}.
The bulk viscosity vanishes in the massless limit due to the absence of a characteristic scale of the dimensionally reduced system,
where the scale invariance preserves the equilibrium state at any step of the expansion/compression.
Therefore, the bulk viscosity is sensitive to the fermion mass that breaks the scale invariance.
In the massive case, one finds a deviation as $  \Theta_\beta = 1 - 3m^2/(\pi^2 T^2)$~\cite{Hattori:2017qih}.

The crucial part of the computation is evaluation of the off-equilibrium components of the distribution functions $ \delta f(k_z) $ 
that will be obtained as solutions of the kinetic equations. In the presence of the electric field $ E_z $ and the expansion/compression flow $u_z (z) $,
the explicit forms of the kinetic equations are, respectively, given as 
\begin{subequations}
\begin{eqnarray}
\label{eq:Boltzmann1}
&&
e E_z \frac{ \partial f (k_z)}{\partial_{k_z}}  = C[f]
\, ,
 \\
&&
\label{eq:Boltzmann2}
(\partial_t +v_z \partial_z ) f(k_z; t,z) = C[f]
\, ,
\end{eqnarray}
\end{subequations}
where $v_z \equiv k_z/\epsilon_k$ is the carrier velocity in the direction of magnetic field.
We have taken the steady and homogeneous limits in Eq.~(\ref{eq:Boltzmann1}).
The equilibrium distribution functions are 
given by $f_{\text{eq}}(k_z) = [ \exp (\beta \epsilon_k) +1 ]^{-1}$ in the absence of the flow 
and by $f_{\text{eq}}(k_z,t,z) = [ \exp\{\beta(t)\gamma_u (\epsilon_k-k_z u_z) \}+1]^{-1}$
in the presence of the flow $u_z$ with the gamma factor $\gamma_u = (1-u_z^2)^{-1/2}$.
Note that one should take into account the time dependence of temperature in Eq.~(\ref{eq:Boltzmann2})
since the energy density decreases during the adiabatic expansion/compression of the system~\cite{Arnold:2006fz}.

\subsection{Roles of the chiral symmetry}\label{sec:transport-coeff:chira}

The electric and momentum currents reach steady states
when the external driving force and the collisional effects are balanced with each other.
The transport coefficients in the zero frequency and momentum limits
characterize such steady-state currents in response to the external forces.
Therefore, the collision term $C[f]$ plays a crucial role in determining
what steady state the system reaches. 
The collision term provides a bridge between microscopic interaction properties
and  macroscopic hydrodynamic frameworks.

We consider the scattering processes among fermions and photons according to the QED Lagrangian. 
To compute the collision term $C[f]$, we use perturbation theory 
with respect to the coupling constant $  e$. 
As one can imagine from the familiar cyclotron radiation,
the leading-order contributions stem from 1-to-2 (2-to-1) scattering processes.
This contrasts to the ordinary perturbative expansion in the absence of an external magnetic field
that starts from the 2-to-2 processes (see, e.g., Ref.~\cite{Arnold:2000dr}).
A simple reason for the absence of the 1-to-2 processes is the kinematics;
The invariant masses in the initial and final states do not match each other.

\begin{figure}
%		\vspace{-0.3cm}
		\begin{center}
			\includegraphics[width=0.9\hsize]{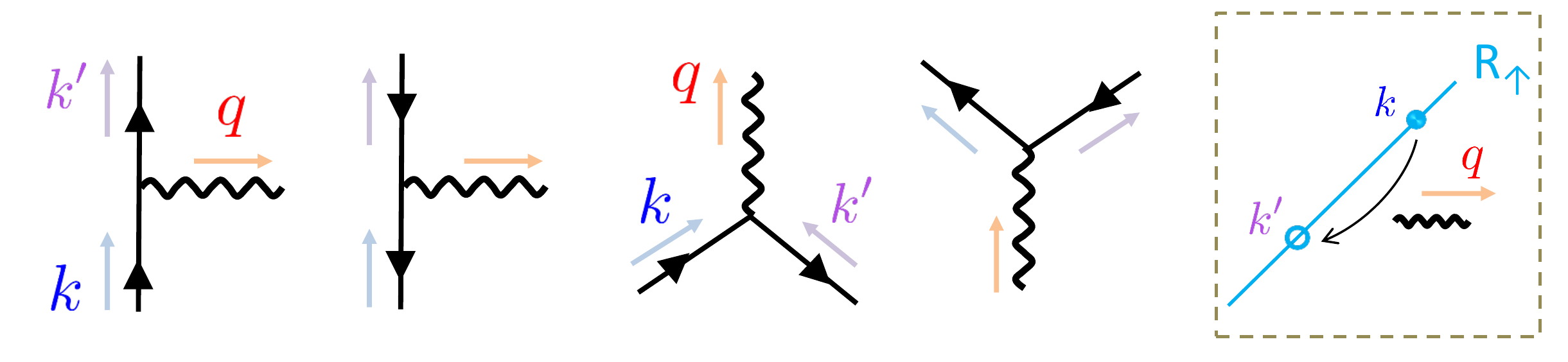}
		\end{center}
		\vspace{-0.5cm}
\caption{The 1-to-2 processes in a strong magnetic field. Time goes from bottom to top.
All these processes can be interpreted as a transition on the diagonal dispersion relation (rightmost).
The chirality conservation at the vertices gives crucial constraints on the relaxation dynamics. }
		\label{fig:diagrams}
\end{figure}

The kinematics in the presence of a magnetic field can be understood from 
the dispersion relation in the Landau quantization. 
The quantized energy levels are obtained from the Dirac equation in a constant magnetic field 
as 
\begin{eqnarray}
\label{eq:Landau-levels}
  \e_k  = \sqrt{ k_z^2 + m^2 + (2n+1 \pm 1 ) eB }
  \, ,
 \end{eqnarray} 
with $ n $ and $ k_z $ being the principal quantum number 
and the kinetic momentum along the magnetic field. 
The Zeeman energy shifts for spin-$1/2$ particles are included with the alternative signs.  
This dispersion relation has a (1+1)-dimensional form. On the other hand, photons have the normal (3+1)-dimensional dispersion relations without
being subject to the Landau-level discretization. 
Because of this mismatch in the dimensions (cf. Fig.~\ref{fig:conductivity}), 
the photons transverse momentum $ |\bq_\perp|^2 = q_0^2-q_z^2 $
can be regarded as an ``effective mass'' in the (1+1)-dimensional kinematics.
One may be then convinced that the kinematics of the 1-to-2 processes
with a ``massive gauge boson'' can be satisfied~\cite{Hattori:2016lqx, Hattori:2016cnt}.
This is a general property in a magnetic field that is valid not only in the LLL but also in the higher Landau levels (hLLs).

However, one needs to look into the kinematics more closely in case of the LLL.
Actually, we recognize prohibition of those 1-to-2 processes 
when the massless LLL fermions have the (1+1)-dimensional linear 
dispersion relations $ k^0 = \pm k_z $~\footnote{
Note that the correct dispersion relation that satisfies the Dirac equation is 
$ k^0 = \pm k_z $ instead of $ k^0 = \pm |k_z| $ in the massless case. 
}.
To understand the kinematics, we start with a particle scattering process in the leftmost diagram of Fig.~\ref{fig:diagrams}
which is obviously understood as a transition from one filled particle state to one of vacant states on the dispersion relation (see the rightmost panel).
One may regard an antiparticle as a hole of a negative-energy particle state, 
and include both positive and negative particle states on the same footing.
Then, all the 1-to-2 and 2-to-1 processes in Fig.~\ref{fig:diagrams}, 
including the pair creation/annihilation channels,
can be also interpreted just as a transition from one to another particle state (with either a positive or negative energy).
In the presence of the chiral symmetry, this transition has to occur on one diagonal line 
due to the absence of the chirality mixing. The origin of energy is not shown since it is not relevant here.

Now, notice that this diagonal transition does not change quantum numbers other than the energy and momentum 
since the initial and final particle states carry the same quantum numbers, especially the same chirality and spin.
This means that an emitted photon can only take away the energy and momentum
with the same amount, i.e., $ q_0^2 - q_z^2 =  \{ \pm(q_z - q_z^\prime) \}^2 - (q_z - q_z^\prime)^2 =0 $,
and is not allowed to carry a nonzero spin along the magnetic field.
%Without a chirality mixing in the massless limit,
%the energy difference is the same as the momentum difference up to
%an overall sign which does not make a difference after squared.
This kinematical constraint readily implies that a transverse photon 
cannot satisfy the kinematics of the 1-to-2 or 2-to-1 processes.
Indeed, the kinematics is only satisfied in the collinear limit $ (q \para k \para k^\prime )  $ 
with  $ |\bq_\perp| =0 $,
where the coupling between a physical transverse photon and fermions vanishes.
In other words, the angular momentum along the magnetic field is not conserved 
with the transverse photons~\footnote{
One may satisfy the angular momentum conservation
when the photon momentum is completely perpendicular to the magnetic field, i.e., $ q_z = 0 $.
However, the on-shell conditions for the fermions and photon cannot be satisfied simultaneously in this case.
}.

Following the above discussions, we conclude that the collision term vanishes in the massless limit
as a consequence of the chirality conservation and the linear dispersion relation in the LLL.
This is a expected from the fact that 
there is no transverse photon in a purely (1+1)-dimensional system. 
On the other hand, a finite fermion mass allows for chirality mixing at interaction vertices.
Then, the kinematics is no longer limited to the collinear configuration,
and we find a finite collision term which is proportional to the squared fermion mass $  m^2$.
It is worth mentioning that a similar mass dependence in the charged-pion decay rate 
is known as the {\it helicity suppression}~\cite{Donoghue:1992dd, Zyla:2020zbs}; 
This is the essential reason why the muonic channel dominates over the electronic channel
in spite of the smaller phase-space volume of the heavier particle. 
Technical details of the collisional effects are given in Refs.~\cite{Hattori:2016lqx,Hattori:2016cnt}.

We obtain the induced currents by inserting the solutions of
Eqs.~(\ref{eq:Boltzmann1}) and (\ref{eq:Boltzmann2}) into Eqs.~(\ref{eq:conductivity}) and (\ref{eq:bulk-viscosity}), respectively.
Finally, the electric conductivity and the bulk viscosity in the LLL
and at the leading-log accuracy with respect to the coupling constant
are obtained as~\cite{Hattori:2016lqx,Hattori:2016cnt,Hattori:2017qih}
\begin{subequations}
\begin{eqnarray}
&&
\sigma_\para = e^2 \frac{ eB }{2\pi} \frac{4T}{e^2  m^2 \ln(T/M)}
\label{eq:sol-conductivity}
\, ,
\\
&&
\zeta_\para = 3  \frac{ eB }{2\pi} \frac{4m^2}{e^2 \ln(T/M)}
\left[ \frac{1}{\pi^2} - \frac{14}{3} \times (0.0304 \cdots) \right]
\label{eq:sol-bulk-viscosity}
\, .
\end{eqnarray}
\end{subequations}
%where $ C_F = (N_c^2-1)/(2N_c) $.
The logarithmic factor originates from the collision integral
with the ultraviolet and infrared cutoff scales at $  T$ and
$M^2 = {\rm min} [ m^2, e^2/(2\pi) eB /(2\pi) ] $, respectively.
The latter is the Debye screening mass from the fermion loop. In addition to the above contributions from the fermion carriers,
there is a finite contribution from photon carriers to the bulk viscosity.
The latter contribution is, however, suppressed by a factor of $ 1/(eB) $
because the photon carriers are scattered by the abundant fermions
of which the phase space volume is enhanced by the Landau degeneracy factor.
Moreover, the phase space volume of the photon carriers is $\sim T^3 $ instead of $ \sim T eB$,
so that there is no enhancement by the factor of $eB $ as compared
to the fermion contribution included in the above result.
Therefore, the photon contribution is subleading to the fermion contribution in the strong-field limit.

We have discussed the fermion-mass suppression of the collision term
and have understood it as a consequence of the chirality conservation in the dimensionally reduced system.
This suppression is reflected in Eqs.~(\ref{eq:sol-conductivity}) and (\ref{eq:sol-bulk-viscosity})
as an enhancement by a factor of $ 1/m^2 $ when the fermion mass is small $  m^2 \lesssim eB  $.
Note that the bulk viscosity is suppressed by $  m^4$ when $ eB =0 $~\cite{Arnold:2006fz}
as a consequence of the conformal symmetry (or, more restrictively, the scale invariance)
as mentioned below Eq.~(\ref{eq:bulk-viscosity}).
The power dependence on the fermion mass is now reduced to $  m^2$ 
as a consequence of the competition between the scale invariance and the chirality conservation,
both of which govern the behaviors in the massless limit.
It is remarkable that the dependences on the fermion mass are important
even in the high temperature limit $ T \gg m $ as long as $ eB  \gg T^2$.

\subsection{Contributions of higher Landau levels}\label{sec:transport-coeff:hll}

As we decrease the ratio $ eB  / T^2$, 
contributions of the higher Landau levels become significant.
The fermion mass is less important for the kinematics in the higher Landau levels
because their dispersion relations are parabolic even for massless fermions.
In Refs.~\cite{Fukushima:2017lvb, Fukushima:2019ugr}, 
the authors have elaborated dependences of the electric conductivity
on the fermion mass and magnetic field strength,
and found a milder mass dependence after including the higher Landau levels~\footnote{
As in the above computation with the LLL, the 1-to-2 (2-to-1) scatterings,
the leading scattering channels in the coupling constant, are included in Ref.~\cite{Fukushima:2017lvb}.
}. Especially, a smooth massless limit was observed.
This result on the mass dependence implies that the LLL fermion carriers
acquire new scattering channels that open via transitions to the hLL and are at work even in the massless limit.
Indeed, thanks to the parabolic dispersion relation of the hLL,
the kinematics of the transition from the LLL to a hLL is satisfied
with a finite photon transverse momentum $ |\bq_\perp| > 0$ even in the massless limit,
and the hLL with a finite energy gap $ \sim |eB| $ has a smaller velocity than the LLL~\footnote{
Since the dynamics of the hLL, as well as of the LLL, is (1+1) dimensional in the momentum space,
relaxation of the longitudinal electric current occurs only when the velocities of fermion carriers are changed.
}. The transition to the hLL occurs only when assisted by absorption of thermal photons,
so that existence of the thermal photons, as well as of the hLL, are crucial for this relaxation dynamics.
As for scattering processes among the hLLs,
one of the important scattering channels may be back scatterings
which are allowed again thanks to the parabolic dispersion relations in the hLL,
but are prohibited for the massless fermions in the LLL.

In this subsection, we have discussed the perturbative computation of electric conductivities and viscosities.
In the strong field limit, they have a significant anisotropy and
their magnitudes depend on the magnetic field strength.
In addition, a special kinematics in the LLL gives rise to an interesting fermion mass dependence.
The hLL contributions become important in the weak to intermediate magnetic field strength.
Evaluation of the viscosities with the hLL contributions remains a challenging issue. Potential applications include relativistic heavy-ion collisions, Dirac/Weyl semimetals,
and neutron star physics.

\section{Chiral magnetohydrodynamics}\label{sec:CMHD}

The system composed of massless chiral fermions also supports the conserved axial charge density when there is no electromagnetic fields.
However, in the presence of the electromagnetic field, a conservation law of the axial charge is broken by the source term proportional to $E_\m B^\m$, which induces the coupling between the electromagnetism and dynamics of the axial charge~\cite{Fukuda-Miyamoto1949,Adler:1969gk, Bell:1969ts}. 
This anomalous violation of the axial charge conservation is a famous consequence of the chiral anomaly~(see, e.g., \cite{Bertlmann2000,Fujikawa-Suzuki2004} for a review).
When one treats the electromagnetic field as non-dynamical backgrounds, 
the effect of the chiral anomaly should be also present in the low-energy effective theory due to the 't Hooft anomaly matching condition~\cite{tHooft:1979rat,Frishman:1980dq}.

In the last decade, there is a significant progress on understanding  macroscopic manifestation of the underlying quantum anomaly in the hydrodynamic regime.
It has been clarified that the consistency to the chiral anomaly requires the novel transport phenomena, called the anomaly-induced transport (or anomalous transport), whose example includes the chiral magnetic effect (CME)~\cite{Vilenkin:1980fu, Nielsen:1983rb, Alekseev:1998ds, Kharzeev:2007jp, Fukushima:2008xe}. 
In Ref.~\cite{Son:2009tf}, Son and Surowka provides an elegant derivation of the anomaly-induced transport on the basis of the second law of local thermodynamics, which has been further investigated along this line 
\cite{Neiman:2010zi, Sadofyev:2010pr, Kharzeev:2011ds, Furusawa:2021vlw}.
On the other hand, there also appear the fruitful applications of the anomaly matching to the hydrodynamic effective action or thermodynamic functional~\cite{Lin:2011aa, Banerjee:2012iz,Jensen:2012jh,Jensen:2012jy,Jensen:2012kj,Jensen:2013vta,Jensen:2013kka,Jensen:2013rga,Haehl:2013hoa,Golkar:2015oxw,Chowdhury:2016cmh,Glorioso:2017lcn,Manes:2018llx,Manes:2019fyw} 
(see also, e.g., Refs.~\cite{Liao:2014ava, Kharzeev:2015znc, Huang:2015oca, Miransky:2015ava, Skokov:2016yrj, Hattori:2016emy,Landsteiner:2016led, Armitage:2017cjs,Hongo:2019rbd,Chernodub:2021nff} for reviews).

In this section, we generalize those frameworks to include dynamical magnetic fields. This can be also regarded as an extension of RMHD discussed in earlier sections with the chirality imbalance. This extension provides a hydrodynamic framework called the {\it chiral magnetohydrodynamics} (chiral MHD).
%(or, alternatively, anomalous magnetohydrodynamics).
In Sec.~\ref{sec:entropy-current-analysis-chiralMHD}, we formulate the chiral MHD and derive the CME on the basis of the entropy-current analysis~\cite{Hattori:2017usa}.
Then, after briefly reviewing the linear waves in RMHD in Sec.~\ref{sec:cmhd:waves}, we demonstrate the helical instability in the chiral MHD in Sec.~\ref{sec:cmhd:insta}.

\subsection{Entropy-current analysis with chiral anomaly}
\label{sec:entropy-current-analysis-chiralMHD}

Let us generalize the entropy-current analysis to derive the chiral MHD equations.
Having the massless Dirac fermions as microscopic constituents in mind, we consider the anomalous Ward-Takahashi identity for the axial current:
\begin{eqnarray}
\label{eq:jA}
 \partial_\mu J^\mu_{A} =  - C_A E^\mu  B_\mu
\, ,
\end{eqnarray}
where $J^\m_A$ and $C_A = e^2/(2\pi^2)$ denote the axial current and an anomaly coefficient for a single (colorless) Dirac fermion, respectively. 
This equation together with those in Eq.~\eqref{eq:EoMs} serves as the equation of motions.

It should be emphasized that the anomalous Ward-Takahashi identity~\eqref{eq:jA} gives a {\it non-conservation law} for the axial current due to the non-vanishing right-hand side. 
Thus, Eq.~\eqref{eq:jA} implies that the axial charge density is, in general, not a hydrodynamic variable, showing a transient dynamics towards the true hydrodynamic equilibrium.
Note, however, that the chiral anomaly is a quantum effect arising from the one-loop diagram, which is suppressed by the Planck constant. 
%Thus, we expect there is a regime where the symmetry-breaking term (\ref{eq:jA}) is smaller than the divergence of the zeroth-order terms. 
Motivated by this observation, we assume that the relaxation time of the axial charge density is longer than time scales of other non-hydrodynamic modes and treat it as a quasi-hydrodynamic quantity.
Based on this assumption and the second law of local thermodynamics, we formulate the chiral MHD as an effective model describing the transient coupled dynamics of strictly conserved quantities and non-conserved axial charge density. We keep track of the chiral anomaly effect with $C_A$ regarded as a small parameter.

Having identified the equations of motion, we then generalize the thermodynamic relation Eq.~\eqref{eq:1st-law-B} by including the axial charge $n_A \equiv u_\m J^\m_A$.
Introducing the axial chemical potential $ \mu_A $ as the thermodynamic conjugate of $ n_A $, we generalize the thermodynamic relation as 
\begin{eqnarray}
\label{eq:thermo-nA}
 \diff \epsilon = T \diff s  + H_\mu \diff B^\mu  + \mu_A \diff n_A 
\, .
\end{eqnarray}
In this section, we assume a neutrality in electric-charge density as before.
We parameterize the constitutive relation for the axial current as
\begin{eqnarray}
J_A^\mu = n_A u^\mu + J_{A\one}^\mu
\, ,
\end{eqnarray}
which enables us to rewrite Eq.~\eqref{eq:jA} as
\begin{eqnarray}
 - C_A E^\mu  B_\mu
 = \pd_\mu J_A^\mu = n_A \theta + D n_A + \pd_\mu   J_{A\one}^\mu
 \, .
\end{eqnarray}

Including the axial charge contribution, we compute the entropy production rate. As a result, the divergence of the entropy current (\ref{eq:entropy-current-MHD}) for RMHD is extend as 
\begin{eqnarray}
\label{eq:entropy-current-CMHD}
\pd_\mu s^\mu \= \beta (Ts - \epsilon - p_\perp +  B^\mu H_\mu + \mu_A n_A  )  \theta
- \beta [(p_\perp - p_\para) b^\mu b^\nu + B^\mu H^\nu ] \pd_\mu u_\nu 
\nnb
&&
+ T_\one^{\mu\nu} \pd_\mu ( \beta u_\nu )
+ \tilde F_\one^{\mu\nu} \pd_\mu ( \beta H_\nu )
-  J_{A\one}^\mu \pd_\mu ( \beta \mu_A) + \mu_A  C_A E_\one^\mu  B_\mu 
\nnb
&&
+  \pd_\mu ( \delta s^\mu - \beta u_\nu T_\one^{\mu\nu}  - \beta H_\nu \tilde F_\one^{\mu\nu} 
+  \beta  \mu_A  J_{A\one}^\mu  )
\nn
\, .
\end{eqnarray}
Note that the induced electric field is at most the first order in derivative; see Eq.~(\ref{eq:E-1st}). 
We now require that the above expression satisfy the second law of thermodynamics.  
At the leading-order in derivative and $C_A$, 
we find the constraints as in Eq.~(\ref{eq:constraints-LO})
but with the replacement of the thermodynamic relation by 
\begin{eqnarray}
Ts =  \epsilon + p_\perp -  B^\mu H_\mu - \mu_A n_A  
\, .
\end{eqnarray}
This is an extension with the contribution of the axial-charge density. 
The other constraints, the relation between $ p_\perp $ and $,  p_\perp $ 
and the absence of the zeroth-order electric field  $ E^\mu_\zero  =0 $, are intact.

In the correction terms, we focus on the regime 
where $ \pd \sim C_A$. 
Then, the semi-positive entropy production requires that \cite{Hattori:2017usa} 
\begin{subequations}
\begin{eqnarray}
&& s^\mu_\one =  \beta u_\nu  T_\one^{\mu\nu}  
+ \beta H_\nu  \tilde F_\one^{\mu\nu} 
-  \beta  \mu_A   J_{A\one}^\mu
\, ,
\\
 &&
 \label{eq:eq:semi-positive-E-chiral}
\tilde \R_E \equiv
- E_{(1)}^{\mu}  ( - \mu_A C_A B_\mu + J_{\one \mu}   )
\geq 0
\, ,
\\
&&
\R_A \equiv  J_{A(1)}^{\mu} \partial_\mu(- \beta \mu_A) \geq 0
\, .
\end{eqnarray}
\end{subequations} 
%The constraint on the electric field $ \tilde \R_E  $ contains the chiral anomaly effect. 
The corrections to the energy-momentum tensor are the same as in Eq.~(\ref{eq:semi-positive-u}), 
and there is no modification of the energy-momentum tensor by either $  C_A$ or $ \mu_A $ 
in the Landau frame. 
This is, however, a frame-dependent statement, and the the energy-momentum tensor in general can acquire 
anomalous contributions due to a redefinition of the flow vector (see, e.g., Ref.~\cite{Landsteiner:2016led}).

%%%%%%%%%

Let us focus on the chiral anomaly effect appearing in $ \tilde \R_E  $. 
Compared with the parity-even case (\ref{eq:E-bilinear2}), the electric current is shifted by the anomaly-induced term. 
The electric current is, therefore, given as 
\begin{eqnarray}
\label{eq:CME}
J^\mu_{(1)}
= C_A \mu_A B^\mu + 
 [ - \sigma_\para b^\mu b^\nu  + \sigma_\perp \varXi^{\mu\nu}
  + \sigma_\Hall  b_\star^{\mu\nu}] E_{\one\nu} 
\, .
\end{eqnarray}
The second term in the square brackets is 
the same as in Eq.~(\ref{eq:Ohmic}). 
What is new and remarkable is the first term that reproduces the CME \cite{Kharzeev:2007jp,Fukushima:2008xe}. 
This term does not depend on an undetermined coefficient like the Ohmic terms. 
The laws of thermodynamics uniquely determines the relation between the CME and the anomalous term in the equation of motion (\ref{eq:jA}). 
Notice that the anomalous term in $\tilde \R_E $ 
can take an arbitrary value in general hydrodynamic configurations. 
Therefore, unless this term is cancelled by the CME in Eq.~(\ref{eq:CME}), 
the entropy production is not necessarily positive semi-definite.
As first pointed out in Ref.~\cite{Son:2009tf}, 
the existence of the CME is demanded to ensure the semi-positive entropy production, 
and is more than allowed in the hydrodynamic frameworks. 
We have seen that the same logic works for the dynamical magnetic fields.

As for the axial current, one can make a bilinear form
\begin{eqnarray}
\R_A =   - J_{A(1)}^{\mu}  \rho^A_{\mu\nu}  J_{A(1)}^{\nu}
\, ,
\end{eqnarray}
with all the possible tensor structures
\begin{eqnarray}
\label{eq:rho-A}
\rho^A_{\mu\nu} =
\rho^A_\para (- b_\mu b_\nu ) +  \rho^A_\perp \varXi_{\mu\nu} - \rho^A_\Hall  b_\star^{\mu\nu}
\, .
\end{eqnarray}
The first two coefficients should be semi-positive quantities $ \rho_{\para,\perp}^A \geq0 $.
Similar to the Hall terms, the third term does not contribute to the entropy production,
so that $ \rho^A_\Hall  $ can take both positive and negative signs.
The constitutive relation of the axial current is obtained as
\begin{eqnarray}
J^\mu_{A\one} =  (\rho_A^{-1})^{\mu\nu}   \partial_\nu( \beta \mu_A)
= \kappa^A_\para \alpha^\mu_\para +  \kappa^A_\perp \alpha^\mu_\perp
+ \kappa^A_\Hall \alpha_\star^\mu
\, ,
\end{eqnarray}
where we defined
$ \alpha_\para^\mu = - b^\mu b^\nu \partial_\nu ( \beta \mu_A)$,
$ \alpha_\perp^\mu =   \varXi^{\mu\nu} \partial_\nu ( \beta \mu_A)$,
and $ \alpha_\star^\mu =  b_\star^{\mu\nu} \a_{\perp\nu}$. 
The longitudinal components are simply related with each other as $ \kappa^A_\para = 1/\rho^A_\para $,
while the transverse components are mixed with the Hall-like component,
$ \kappa^A_\perp = \rho^A_\perp/[ (\rho_\perp^{A })^2 + ( \rho_\Hall^{A})^ 2]$
and $ \kappa^A_\Hall = \rho^A_\Hall/[ (\rho_\perp^{A })^2 + ( \rho_\Hall^{A})^ 2]$. 
Three diffusion constants $ \kappa^A $ may take different values in a magnetic field. 
According to the sign constraints (\ref{eq:rho-A}), two of the diffusion constants 
should take semi-positive values, $ \kappa^A_{\perp,\para} \geq 0 $.

\subsection{Linear waves in relativistic MHD} \label{sec:cmhd:waves}

\begin{figure}
%\vspace{-1cm}
     \begin{center}
              \includegraphics[width=0.8\hsize]{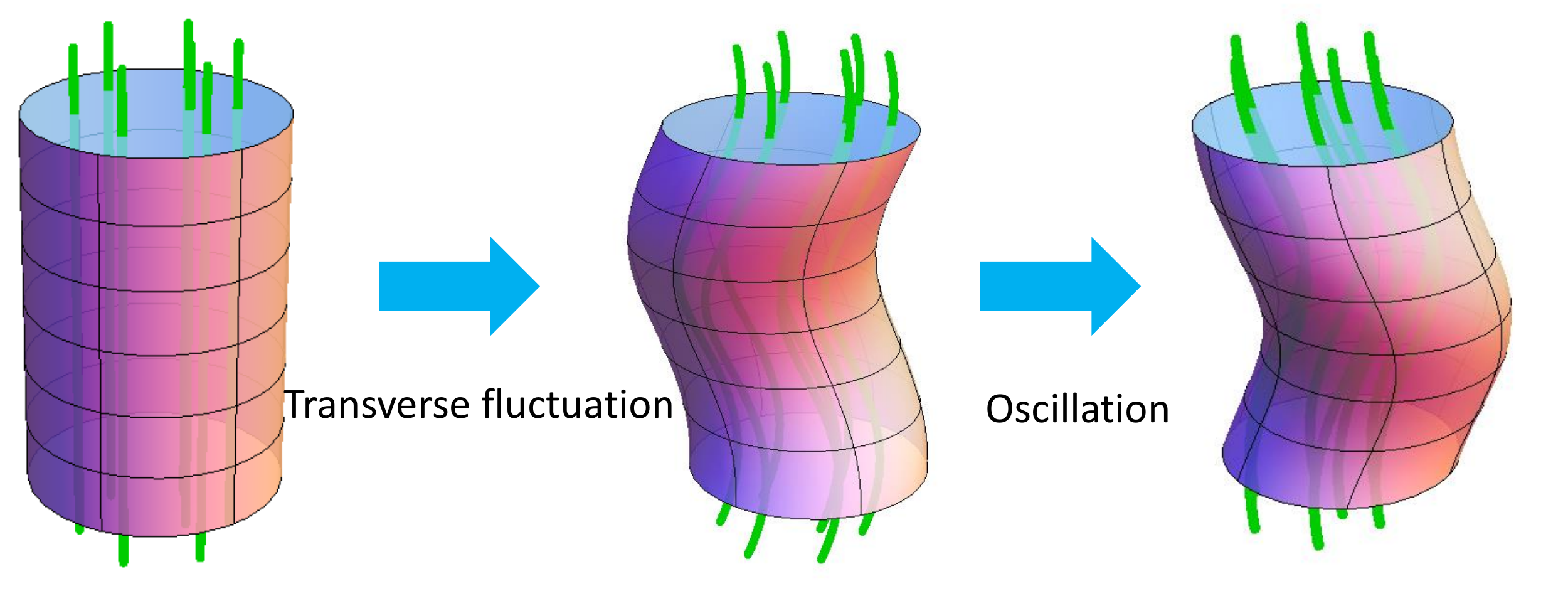}
     \end{center}
\vspace{-0.7cm}
\caption{Transverse Alfv\'en wave propagating along the magnetic field.
A resorting force is provided by the tension of the disturbed magnetic lines. }
\label{fig:Alfven}
\end{figure}

In general, it is far from a simple task to find solutions of hydrodynamic equations.
Nevertheless, one could find a solution near a stationary hydrodynamic configuration and study a linear wave describing propagation of a weak disturbance on top of the stationary state.
Here, before investigating the linear wave in the chiral MHD, 
we briefly describe the linear wave in RMHD. 
Starting from the stationary state $ \epsilon_0 $, $p_0  $, $ u^\mu = (1, {\bm 0}) $, and $ B^\mu = (0, \bB_0) $
with a constant $  \bB_0$,
we apply perturbations, $ \delta \epsilon $, $\delta p  $, $ \delta u^\mu  $, and $ \delta B^\mu  $.
Maintaining the linear terms with respect to those perturbations
in the MHD equations (\ref{eq:EoMs}) and (\ref{eq:TF_zeroth}),
we obtain the simple linearized wave equations.
The resulting linearized equation describe 
six modes, called the Alfv\'en wave and the fast and slow magnetosonic waves, which are three pairs of the waves propagating in opposite directions.
The number of modes coincide with that of the dynamical degrees of freedom counted.
%in Sec.~\ref{sec:entropy}. 
Those waves have been well-known in the non-relativistic theory~\cite{biskamp1997nonlinear, davidson2002introduction}
(see below and also Ref.~\cite{PhysRevE.47.4354} for relativistic cases).

One finds a simple physical mechanism that induces the Alfv\'en wave as follows.
%The Alfv\'en wave is induced by a simple mechanism.
If one applies perturbations to the flow velocity and the magnetic field
in perpendicular to the static field $ \bB_0 $,
the tension of the magnetic lines acts as a restoring force
that tends to bring the fluid volume back to the original stationary position (cf. Fig.~\ref{fig:Alfven}).
However, the energy density (or mass density in the non-relativistic case) provides an inertia
which prevents the fluid volume from stopping at the original position.
Consequently, the fluid volume and the penetrating magnetic lines start oscillating just like a string.
The Alfv\'en wave is, thus, a transverse wave.
Let us denote the perturbations $ \delta {\bm u}_\perp (t,z) $ and $ \delta \bB_\perp (t,z) $
which are perpendicular to $ \bB_0 $ and depend on time and the spatial coordinate along $ \bB_0 =(0,0,B_0) $.
Then, the linearized MHD equations read
\begin{subequations}
\label{eq:linearMHD}
\begin{eqnarray}
B_0 \partial_z \delta {\bm u}_\perp - \partial _t \delta \bB_\perp 
&=& 0
\, ,
\\
(\epsilon+p) \partial_t \delta {\bm u}_\perp - B_0 \partial_z \delta \bB_\perp &=& 0
\, ,
\end{eqnarray}
\end{subequations}
where $B_0 = \vert \bB_0 \vert  $ and we assumed that $ \mu_m = 1 $ for simplicity.
Eliminating $  \delta B_\perp$, we indeed find a wave equation
\begin{eqnarray}
\partial_t^2 \delta {\bm u}_\perp = \frac{B_0^2}{ \epsilon+p} \partial_z^2 \delta {\bm u}_\perp
\, .
\end{eqnarray}
This equation simply means that two transverse waves are propagating
in opposite directions along $  \bB_0$ with a velocity $ v_A^2 = B_0^2/( \epsilon+p)  $ 
called the Alfv\'en velocity.  This form is anticipated 
since the tension is proportional to the magnetic field strength. 
Remember that $ \epsilon $ and $ p $ are the total energy density and pressure 
including the magnetic-field contributions. 
When the magnetic field is so strong that its contribution dominates
in the energy density and pressure, the  Alfv\'en velocity approaches the speed of light from below.
Thus, the propagation of the Alfv\'en wave respects the causality.
Eliminating $  \delta {\bm u}_\perp$ in Eq.~(\ref{eq:linearMHD}),
one gets the same wave equation for $  \delta \bB_\perp$. 
This means that the magnetic field lines are ``frozen-in'' to the fluid volume. 
and their disturbances propagate together (cf. Fig.~\ref{fig:Alfven}).

The disturbance of the magnetic field lines also induces
the compression of the energy density $ \delta \epsilon \sim \delta \epsilon_{\rm matt} + \bB_0 \cdot \delta \bB $
and the pressure fluctuation $ \delta p  \sim \delta p_{\rm matt} + \bB_0 \cdot \delta \bB$. 
Therefore, the velocities of sound modes are modified in RMHD.
When the compression of the magnetic field lines is
in-phase (out-of-phase) to that of the fluid volume,
the restoring force becomes larger (smaller) as compared to the case without a magnetic field.
Consequently, the wave velocities are enhanced or reduced depending on the relative phases,
so that the sound modes split into the fast and slow magnetosonic waves~(see, e.g., Ref.~\cite{biskamp1997nonlinear} for more discussions).
Their dispersion relations can be also obtained from the linearized MHD equations in the same manner.
The full linearlized equation will be a $ 6\times6 $ matrix equation,
determining all of the six dispersion relations after the diagonalization.
The secular equation will be a cubic equation for the squared frequency $ \omega^2 $ for the three pairs of the waves.

\subsection{Helical instabilities in the chiral MHD}\label{sec:cmhd:insta}

\begin{figure}
%\vspace{-1cm}
     \begin{center}
              \includegraphics[width=0.5\hsize]{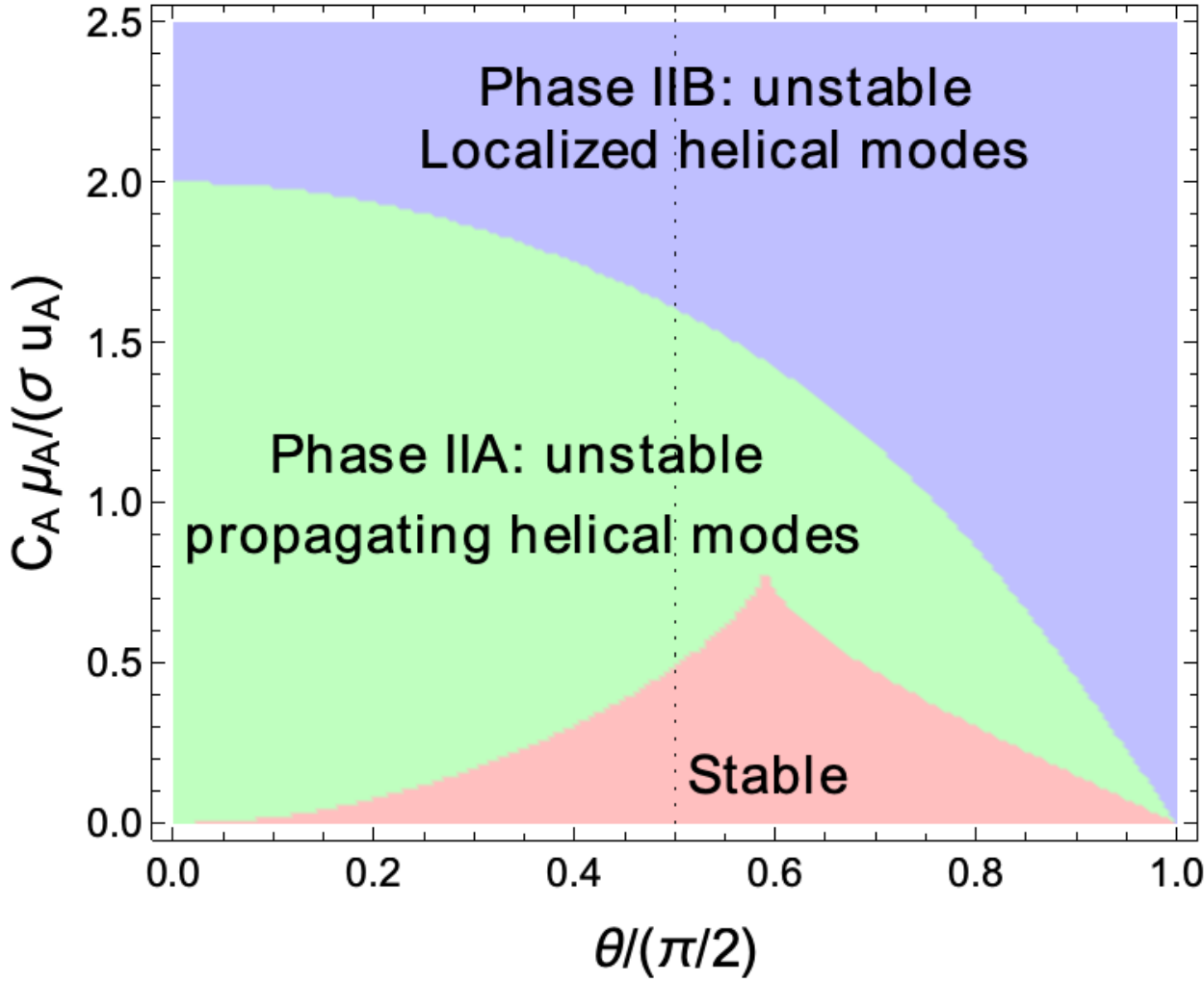}
     \end{center}
\vspace{-0.5cm}
\caption{``Phase diagram'' of the collective excitations in the chiral MHD
with respect to the axial chemical potential $ \mu_A $ and the angle $  \theta$
between the directions of the propagation and the magnetic field
(with $ \sigma $ and $ u_A $ being the Ohmic conductivity and the Alfv\'en velocity) \cite{Hattori:2017usa}.
}
\label{fig;diagram-modes}
\end{figure}

We found the six stable waves in RMHD in the above.
It is interesting to see how the CME affects those propagating modes. 
Each mode in RMHD is paired with another mode propagating 
in the opposite direction with the same velocity. 
This degeneracy is, however, resolved in the chiral MHD 
since the parity symmetry is broken by the axial chemical potential.
As a result, the dispersion relations could acquire different dispersion relations 
depending on the propagation directions. 
The dispersion relations can be obtained by diagonalizing the linearized equations 
in the same manner as above.

The ``phase diagram'' of the linear modes are drawn in Fig.~\ref{fig;diagram-modes}
with respect to the propagation direction $ \theta $ and the axial chemical potential $ \mu_A $
(with $ \sigma $ and $ u_A $ being the Ohmic conductivity and the Alfv\'en velocity). 
The phases are classified with respect to the imaginary parts of the dispersion relations 
that appears at the linear order in the momentum. 
As we increase $ \mu_A $ in the Phase IIA and IIB, one of the paired waves exponentially grows in time, while the other is damped out.
The exponentially growing mode indicates a hydrodynamic instability induced by the CME current.
Moreover, the growing and damping modes are found to be helicity eigenstates and carry {\it opposite} helicities~\cite{Hattori:2017usa}.
Remarkably, this means that a particular helicity mode is selectively excited in the chiral MHD.
As expected, the roles of the growing and damping helical modes are interchanged if we flip the sign of the axial chemical potential.
Namely, the excess of the R (L) fermionic chirality induces the exponential grows of the R (L) helical waves
as the mixture of the disturbances in the magnetic field and the fluid velocity.

This instability reminds us of the chiral plasma instability (CPI) 
that is an exponential amplification of the magnetic field 
in the presence of the CME current \cite{Akamatsu:2013pjd, Akamatsu:2014yza} 
(see also Refs.~\cite{Carroll:1989vb, Garretson:1992vt, 
Garretson:1992vt,Joyce:1997uy, Field:1998hi, Semikoz:2004rr, Laine:2005bt}). 
Here, the chiral anomaly is coupled to the fluid dynamics as well as the magnetic field,
so that the total helicity conversion can be extended 
with the inclusion of the ``fluid helicity'' \cite{Yamamoto:2015gzz}.
As in the CPI, one may expect the conversion of the fermionic helicity to the fluid and magnetic helicities as the topological origin of the instability.
Recently, the instability was also found in Ref.~\cite{Kojima:2019eed}~\footnote{
%The secular equation in Eq.~(8) of Ref.~\cite{Kojima:2019eed} agrees to that in Eq.~(14) of Ref.~\cite{Hattori:2017usa}.
%However, there is a critical sign error in Eq.~(61) of Ref.~\cite{Rogachevskii:2017uyc},
%where the overall sign of the anomalous term should be positive instead of minus.
There is a disagreement between the secular equations in Eq.~(61) of Ref.~\cite{Rogachevskii:2017uyc}
and in Eq.~(14) of Ref.~\cite{Hattori:2017usa}.
Seemingly, the overall sign of the anomalous term in Ref.~\cite{Rogachevskii:2017uyc} 
may be positive instead of minus.
This may be the reason why the authors did not find unstable modes.
}.
This new hydrodynamic instability deserve further studies beyond the linear-mode analysis.

\section{Summary and future prospects}\label{sec:futur}

We have reviewed the progresses in the theory of RMHD over the last decade. We do not go into details on the applications of RMHD; Instead, we provide the literature pertinent to the applications of RMHD to relativistic heavy-ion collisions, astrophysics, cosmology, and so on, so that the reader can follow the trails. We have structured the discussions in a pedagogical manner so that this review could benefit the readers who are not familiar with hydrodynamics. To derive the constitutive relations of RMHD up to the first order in derivative expansion, we first employ a phenomenological approach based on the entropy-current analysis (Sec.~\ref{sec:entropy}). 
In this construction, the Bianchi identity is regarded as a conservation law for the magnetic flux, and the magnetic field  is treated as a leading-order quantity in the derivative expansion. 
At both the ideal and dissipative levels, we show how the presence of such a leading-order magnetic field gives rise to anisotropic constitutive relations. Then, we discussed the nonequilibrium statistical operator technique  (Sec.~\ref{sec:mhd:qed}) that verifies the constitutive relations from the microscopic theory. 
In this method, the phenomenological parameters (transport coefficients introduced in the entropy-current analysis) are expressed as the low-energy limit of certain correlation functions, the Green-Kubo formulas, that connect the underlying quantum theories (or QED) to  macroscopic RMHD. In Sec.~\ref{sec:comparison}, we have discussed how this recent formulations presented in Sec.~\ref{sec:entropy} and Sec.~\ref{sec:mhd:qed} are related to the conventional formulation of RMHD in the literature. Based on this conventional view,
we then used the kinetic theory to re-derive the constitutive relations at the first order of derivatives by treating the EM fields as external fields and evaluate the transport coefficients in a relaxation time approximation~(Sec.~\ref{sec:kine}). By going beyond the relaxation time approximation, we also demonstrate how one can use the perturbative QCD or QED to compute these transport coefficients (Sec.~\ref{sec:transport-coeff}). 
Finally, we considered the situation in which the parity is allowed to be violated (Sec.~\ref{sec:CMHD}). 
This brings new, non-dissipative, transport terms, such as the chiral magnetic effect, into the constitutive relations, and leads to new types of wave modes and instabilities.

Certainly, the theory of RMHD has progressed significantly in the previous decade. Yet, there are many challenges to be addressed in the future works. Here are a couple of them that, in our opinion, require further investigation. 
 
{\bf (1) Spin hydrodynamics.}
One profound effect of the magnetic field is to polarize spin. This naturally addresses the question that in the situation where the spin can be a quasi-hydrodynamic mode, how the formulation of RMHD is modified by the spin degree of freedom. Quite recently, relativistic spin hydrodynamics has attracted intensive discussions. It is theoretically very interesting and potentially applicable to the study of spin transport phenomena in QGP or supernova matter. 
In fact, one of the main motivations that drive the study of relativistic spin hydrodynamics is the recent experimental breakthrough of the observation of the spin polarization in the vortical QGP produced in heavy-ion collisions (see Refs.~\cite{Liu:2020ymh,Gao:2020vbh,Huang:2020xyr,Becattini:2020ngo,Becattini:2022zvf} and references therein). 

In the absence of EM fields, the basic symmetries underlying spin hydrodynamics are the translational and Lorentz symmetries, which give the conservation laws of energy-momentum and angular momentum
\begin{align}
\label{spin:energy-momentum}
\pt_\m T^{\m\n} &= 0,\\
\label{spin:angular-momentum}
\pd_\m {\mathcal J}^{\m\rho\sigma} 
&=\pd_\m(\Sigma^{\m\rho\sigma}+x^\rho T^{\m\sigma}-x^\sigma T^{\m\rho})=0.
\end{align}
where ${\mathcal J}^{\m\r\s}$ is the angular momentum tensor and $\Sigma^{\m\r\s}$ is the spin current. 
Re-arranging Eq.~(\ref{spin:angular-momentum}) into the form
\begin{eqnarray}
\label{spin:spincurrent}
&\pd_\m \Sigma^{\m\rho\sigma}=-2 T^{[\r\s]},
\end{eqnarray}
shows that the spin current is not conserved but is sourced by the anti-symmetric part of the energy-momentum tensor,  representing the conversion of angular momentum between spin and orbital components. The non-conservation of spin current renders spin density not a strict hydrodynamic variable. However, when the spin-orbit coupling is weak (see Ref.~\cite{Hongo:2022izs} for a concrete analysis showing how the spin-orbit coupling can be parametrically weak), the spin relaxation time can be parametically large, making spin density a quasi-hydrodynamic mode~\cite{Hongo:2021ona}. The spin hydrodynamics is thus well formulated in this regime as a {\it quasi}-hydrodynamics~\cite{Grozdanov:2018fic} or Hydro+~\cite{Stephanov:2017ghc}. 
A striking consequence of the spin-orbit coupling is the pseudo-gauge ambiguity in defining the spin current, namely, a finite shift of $\Sigma^{\m\r\s}$
\begin{eqnarray}
\label{spin:pgspin}
&\Sigma^{\m\rho\sigma} \rightarrow \Sigma^{\m\rho\sigma} -\Phi^{\m\r\s},
\end{eqnarray}
is compensated by a corresponding change in $T^{\m\n}$
\begin{eqnarray}
\label{spin:pgtmn}
&T^{\m\n} \rightarrow T^{\m\n}+\frac{1}{2}\pt_{\l}(\Phi^{\l\m\n}-\Phi^{\m\l\n}-\Phi^{\n\l\m}),
\end{eqnarray}
leaving the conservation laws (\ref{spin:energy-momentum}) and (\ref{spin:spincurrent}) preserved. Noticeably, choosing $\Phi^{\m\r\s}=\Sigma^{\m\r\s}$ eliminates the spin current and results in a symmetric energy-momentum tensor, called Belinfante-Rosenfeld improved energy-momentum tensor~\cite{Belinfante1939,Belinfante1940,Rosenfeld1940}.

Such pseudo-gauge ambiguity brings freedom of choosing different structures for the spin current and energy-momentum tensor in constructing the spin hydrodynamics. Then, in a similar manner as we present in this article, one can construct the spin hydrodynamics in an order-by-order analysis in the derivative expansion (see, \eg, Refs.~\cite{Montenegro:2017rbu,Florkowski:2017ruc,Florkowski:2018fap,Montenegro:2018bcf,Hattori:2019lfp,Fukushima:2020ucl,Bhadury:2020puc,Li:2020eon,She:2021lhe,Gallegos:2021bzp,Peng:2021ago,Hongo:2021ona,Hu:2021pwh,Daher:2022xon,Gallegos:2022jow,Weickgenannt:2022zxs, Cao:2022aku}). Especially, anisotropic constitutive relations arise as in RMHD when there is a zeroth-order vector provided by a strong spin polarization \cite{Cao:2022aku}. 
When dynamical EM fields are present, it is worth looking into how we can unify the frameworks of spin hydrodynamics and RMHD and clarify the resulting new phenomena in such a unified formulation (see recent attempts in Refs.~\cite{Singh:2022ltu,Bhadury:2022qxd} and also in Ref.~\cite{Koide:2012kx} and references therein for non-relativistic cases). In this case, we need to choose a pseudo-gauge (\eg, a totally antisymmetric spin current) and couple Eqs.~(\ref{spin:energy-momentum})-(\ref{spin:spincurrent}) with the magnetic flux conservation 
\begin{eqnarray}
\label{eq:magnetic-flux-conservaton}
&\pt_\m\tilde F^{\m\n}=0,
\end{eqnarray}
and work out a derivative expansion with an appropriate power counting scheme, as we outlined in Sec.~\ref{sec:entropy} and \ref{sec:mhd:qed}.

{\bf (2) Kinetic theory and transport coefficient with the new formulation.} As we demonstrated in Secs.~\ref{sec:entropy} and \ref{sec:mhd:qed}, the formulation based on magnetic flux conservation provides a new view on RMHD based on the symmetries of the system. Noting that the hydrodynamics provides a universal description of low-energy behaviors of the system, we expect that this formulation of RMHD should be also derived based on the kinetic theory. Since the kinetic theory reviewed in Sec.~\ref{sec:kine} is based on the conventional approach with background magnetic fields, it is an interesting problem to reconstruct the kinetic description along the line of the recent formulation of RMHD with dynamical magnetic fields.

In establishing such a kinetic theory, it is again crucial to consider the Bianchi identity \eqref{eq:magnetic-flux-conservaton} as an additional equation of motion together with the energy-momentum conservation law. Thus, the reformulated kinetic theory should be equipped with the distribution function of photons as well as matters, and we need to clarify how we can extract information of $\tilF^{\m\n}$ based on the distribution functions. Expanding the off-equilibrium part of the distribution function by generalizing the approaches in Sec.~\ref{sec:kine}, one may obtain the dissipative corrections to the constitutive relations. Once this is established, one can extract values of all the transport coefficients by the use of the resulting kinetic theory.

As for the computation of transport coefficients, we have already presented the Green-Kubo formulas in terms of the correlation functions in the recent formulation of RMHD. Thus, one can also compute the Green-Kubo formulas \eqref{eq:GK} with field theoretical techniques. In Ref.~\cite{Grozdanov:2017kyl}, such a calculation at a certain strong-coupling regime is performed based on the holographic principle, but there is little discussion about the perturbative evaluation at the weak-coupling regime in Ref.~\cite{Grozdanov:2016tdf}. It is worth working out the perturbative evaluation of the Green-Kubo formulas \eqref{eq:GK} and clarifying their relations to those obtained from the conventional approach partly reviewed in Sec.~\ref{sec:transport-coeff}.

\vspace{1cm}

{\it Acknowledgments ---} We thank G. Denicol, K. Fukushima, Y. Hirono, M. Huang, T. Koide, S. Li, E. Molnar, G. M. Monteiro, H. Niemi, J. Noronha, D. Rischke, A. Sedrakian, D. Satow, Q. Wang, H.-U. Yee, Y. Yin for collaborations on topics in this article and Y. Hidaka, J. Liao, S. Lin, P. Kovtun, D. Kharzeev, S. Pu, M. Stephanov, H. Taya, N. Yamamoto, D.-L. Yang for useful discussions.
X.G.H. is supported by NSFC under Grant No.~12075061 and Shanghai NSF under Grant No.~20ZR1404100. 
This work is supported in part by JSPS KAKENHI under grant Nos.~20K03948 and 22H02316 and RIKEN iTHEMS Program (in particular iTHEMS Non-Equilibrium Working group and iTHEMS Mathematical Physics Working group).

 \appendix

\section{Electric charges in MHD}
\label{sec:time}

In neutral plasmas, the electric field $ \bE $ in the fluid rest frame and the electric charge-density fluctuation $ \delta J^0 $ are damped out in a finite time scale and are therefore not qualified as hydrodynamic variables~\cite{Hattori:2017usa}. This can be shown in the following way. An electric field $ \bE$, induced by the charge-density fluctuation, 
obeys the Gauss's law $ \nabla \cdot \bE  = \delta J^0/\epsilon_e$ with $ \epsilon_e$ being a dielectric constant.
Charges will be redistributed via an Ohmic current $ {\bm J} = \sigma \bE $.
Plugging the Gauss's law into the current conservation $  \partial_\mu J^\mu =0 $,
we find
\begin{eqnarray}
\frac{\partial \delta J^0 (t,\bx)}{\partial t}
= - \nabla \cdot {\bm J} (t,\bx)
=  - \frac{ \sigma } { \epsilon_e }\delta J^0 (t, \bx)
\, .
\end{eqnarray}
Thus, the charge-density fluctuation is damped out in time as
\begin{eqnarray}
\delta J^0 (t,\bx) = \delta J^0 (0,\bx) \exp \left( - \frac{ \sigma } { \epsilon_e } t \right)
\, .
\end{eqnarray}
Plugging this solution back to the Gauss's law, we find that
the electric field is also damped out in the same time scale. 

Therefore, neither of them are qualified to be included in a set of hydrodynamic variables in the formulation of RMHD. Nevertheless, they can be induced in the dynamics,and should be expressed as functionals of hydrodynamic variables, i.e., constitutive relations. On the other hand, the magnetic field persists in a plasma. This difference arises from the presence of the current term in the Maxwell equation, which breaks the duality between electric and magnetic fields.

\section{Matching and frame conditions in relativistic MHD}
\label{sec:T1}

For the systems invariant under the charge-conjugation and parity transformations, 
the most general tensor structures of the first-order dissipative corrections $ T_{(1)}^{\mu\nu}  $ 
and $\tilde F^{\mu\nu}_\one$ can be written as 
%Since the energy-momentum tensor is even under the parity and charge-conjugation transformations,
%a term proportional to $u^{(\mu} b^{\nu)}  $ requires its coefficient be odd under those transformations.
%Therefore, this term is not allowed in those systems where the parity and charge-conjugation symmetries are preserved.
%}
\begin{eqnarray}
&&
T_{(1)}^{\mu\nu} = \delta \epsilon u^\mu u^\nu
- \delta p_\perp \varXi^{\mu\nu} 
+ \delta p_\para b^\mu b^\nu
+ 2 h^{(\mu} u^{\nu)} + 2 f^{(\mu} b^{\nu)} + \pi_\perp^{\mu\nu}
\label{eq:T1_mn0}
\, ,
\\
&&
 \tilde F^{\mu\nu}_\one = 2 \delta B b^{[\mu} u^{\nu]}
  +  2 g^{[\mu} u^{\nu]} + 2 b^{[\mu} \ell^{\nu]}   + m^{\mu\nu}  
 \, .
\end{eqnarray}
We denote the derivative corrections as $ \{ \delta \epsilon , \delta p_\para , \delta p_\perp,  h^\mu, f^{\mu} , \pi_\perp^{\mu\nu} , 
 \delta B ,g^\mu,  \ell^\mu, m^{\mu\nu}  \} \sim {O} (\partial^1) $.  
The off-equilibrium pressure correction, which is written as $ \delta p \Delta^{\mu\nu} $
in the absence of a magnetic field, is split into two terms due to the preferred orientation in a magnetic field.
The fourth term in $T_{(1)}^{\mu\nu}$ is proportional to the familiar heat current $ h^\mu $. The second-rank tensors are 
symmetric $  \pi_\perp^{\mu\nu}  =  \pi_\perp^{\nu\mu}  $ 
and antisymmetric $ m^{\mu\nu}  = - m^{\nu\mu}  $ 
in their Lorentz indices. 
Those tensors are all transverse to both $  u^\mu$ and $b^\mu  $, i.e., 
$ u_\mu h^\mu =0=   b_\mu h^\mu $, $ u_\mu f^\mu = 0=  b_\mu f^\mu $, 
$   u_\mu \pi_\perp^{\mu\nu} = 0 = b_\mu \pi_\perp^{\mu\nu} $, 
$u_\mu  g^\mu = 0 = b_\mu g^\mu $,
$u_\mu  \ell^\mu = 0 = b_\mu \ell^\mu $, 
and $ u_\mu m^{\mu\nu} = 0 = b_\mu m^{\mu\nu} $. 
If some of them, \eg, the heat current $h$, has longitudinal components, we can always decompose it into the longitudinal and transverse parts
as $ h^\mu = (\varXi^{\mu\alpha} + u^\mu u^\alpha - b^\mu b^\alpha) h_\alpha
=\varXi ^{\mu\alpha} h_\alpha  + \{ \, (u^\alpha h_\alpha) u^\mu  +  (b^\alpha h_\alpha) b^\mu\, \} $.
Only the first term provides an independent structure to the energy-momentum tensor,
and the other terms between the braces can be absorbed by the terms existing in Eq.~(\ref{eq:T1_mn0}). 
%Similar decompositions work for $ f^\mu $ and $\tau^{\mu\nu}  $,
%and the independent tensors are transverse to both $ u^\mu $ and $ b^\mu $
%because of the projection operator $ \varXi^{\mu\nu} $.
%Therefore, we assume the transversality, instead of writing the projection operators explicitly.

%$ u_\mu h^\mu = b_\mu h^\mu =0 $, $ u_\mu f^\mu = b_\mu f^\mu =0 $, and
%$ u_\mu \tau^{\mu\nu} =u_\nu \tau^{\mu\nu} = b_\mu \tau^{\mu\nu} =b_\nu \tau^{\mu\nu}  =0$ in Eq.~(\ref{eq:T1_mn0}).

In an off-equilibrium state, the local thermodynamic quantities as well as the flow velocity and the magnetic field are not uniquely defined,
but are subject to ambiguities of the order of derivative~\cite{landau1959course:fluid,
Hiscock:1985zz, Bhattacharya:2011tra, Kovtun:2012rj, Minami:2012hs, Jensen:2012jh, Hernandez:2017mch}.
The derivative corrections vanish as the system approaches an equilibrium state,
and those quantities reduce to the same equilibrium values
which may be defined unambiguously by the expectation values of the microscopic current operators
in equilibrium statistical mechanics.
We shall, therefore, express the coefficients as
\begin{eqnarray}
&&
\epsilon + \delta \epsilon = u_\alpha u_\beta  T^{\alpha\beta}
, \quad
p_\perp  + \delta  p_\perp = - \frac{1}{2} \varXi_{\alpha\beta}   T^{\alpha\beta}
,\quad
p_\parallel + \delta  p_\parallel = + b_\alpha b_\beta  T^{\alpha\beta}
, 
\nn
\\
&&
h^\mu =   u_{(\alpha} \varXi^{\mu}_{ \beta)}    T^{\alpha\beta}
, \quad
f^\mu =  -   b_{(\alpha} \varXi^{\mu}_{ \beta)}  T^{\alpha\beta}
, \quad
\pi_\perp^{\mu\nu} = ( \varXi^{  (\mu} _\alpha \varXi^{\nu) }_\beta
 - \frac12 \varXi^{\mu\nu}\varXi_{\alpha\beta} )   T^{\alpha\beta}
 ,
 \\
 &&
 B + \delta B = -  b_\mu u_\nu \tilde F^{\mu\nu}  , \quad 
g^\nu = - \Xi^{\nu}_\a u _\mu \tilde F^{\mu\a} , \quad
\ell^\nu = - \Xi^\nu_\a b _\mu \tilde F^{\mu\a} , \quad
m^{\mu\nu} = 2 \varXi^{\mu [\a} \varXi^{\b] \nu} \tilde F_{\a\b}
.\nn
\end{eqnarray}

The conserved currents may be parametrized by either set of variables
before and after the redefinition 
$ T \to T' = T + \delta T$, 
$u^\mu \to u^{\prime \mu} = u^\mu + \delta u^\mu$, 
$H \to H' = H + \delta H$,
and $b^\mu \to b^{\prime \mu} = b^\mu + \delta b^\mu$
with $ \delta T, \, \delta u^\mu , \, \delta H, \, \delta b^\mu \sim O(\partial^1) $, 
where $H = \sqrt{- H_\mu H^\mu}$. 
We examine how those shifts change the constitutive relations. 
Under the redefinition of $T$, the scalar quantities are transformed as, e.g.,  
$ \epsilon(T) \to \epsilon(T') \sim \epsilon(T) + (\partial \epsilon/\partial T) \delta T $ and so on, 
while their derivative corrections as $ \delta \epsilon(T) \to \delta \epsilon(T')
\sim \delta \epsilon(T) + \order(\partial^2) $. A similar transformation holds under the redefinition of $H$. Therefore, up to the first order in derivatives, 
the redefinition induces a set of the shifts 
\begin{subequations}
\label{eq:redef-scalar}
\begin{eqnarray}
&&
\epsilon + \delta \epsilon \to
\epsilon^\prime :=
\epsilon  + \frac{\partial \epsilon}{\partial T} \delta T
 + \frac{\partial \epsilon}{\partial H} \delta H
+ \delta \epsilon + \order(\partial^2)
,
\\
&&
p_\parallel + \delta  p_\parallel \to
p_\para^\prime := 
p_\para + \frac{\partial p_\para}{\partial T} \delta T
+ \frac{\partial p_\para}{\partial H} \delta H
+ \delta p_\para+ \order(\partial^2)
,
\\
&&
p_\perp  + \delta  p_\perp \to
p_\perp^\prime :=
p_\perp + \frac{\partial p_\perp}{\partial T} \delta T 
+ \frac{\partial p_\perp}{\partial H} \delta H 
+ \delta p_\perp+ \order(\partial^2)
,
\\
&&
B + \delta B \to 
B^\prime := B + \frac{\pd B}{\pd T} \delta T
+ \frac{\pd B}{\pd H} \delta H 
+ \delta B + \order (\pd^2)
. 
\end{eqnarray}
\end{subequations}
We will neglect the second-order terms which are beyond the current working accuracy. 
By using the transforms of the scalar quantities (\ref{eq:redef-scalar}),
one can eliminate the off-equilibrium quantities $ \delta \epsilon $,  $ \delta p_\para $, or $\delta p_\perp $ 
in $T^{\mu\nu}_\one$ 
and $\delta B$ in $\tilde F_\one ^{\mu\nu}$.
It is customary to eliminate the off-equilibrium energy density and magnetic field so that the equilibrium energy density $ \epsilon $ and $B$ are maintained in the derivative expansion. Those demands can be achieved by adjusting $\delta T$ and $\delta H$ so that $\epsilon = \epsilon'$ and $B = B'$ 
up to the second-order corrections, 
while the pressure components remain to have the off-equilibrium contributions that correspond to the bulk viscosities. These are the matching conditions for $\e$ and $B$.

One can generalize this redefinition process to the case with multiple conserved charges. If there are $ n $ conserved scalar quantities, 
%such as an energy density and electric charge ($ n=2 $),
one can maintain their equilibrium quantities by the redefinitions of the $ n $ thermodynamic conjugate parameters, 
%the temperature and chemical potential, 
leaving the components of pressures subject to off-equilibrium corrections.

Similarly, one can eliminate some of the dissipative currents by the redefinition of zeroth-order vectors, $u^\mu \to u^{\prime \mu} = u^\mu + \delta u^\mu$ 
and $b^\mu \to b^{\prime \mu} = b^\mu + \delta b^\mu$
with $  \delta u^\mu , \, \delta b^\mu \sim \order(\partial^1) $. 
These shifts in the ideal constitutive relations  (\ref{eq:TF_zeroth}) 
induce shifts of the first-order derivative terms 
\begin{subequations}
\label{eq:redef-tensor}
\begin{eqnarray}
&&
h^\mu %\to h^\mu + 2\delta u^\mu u_\alpha u_\beta T^{\al\be}
\to h^{\prime\mu} =  h^\mu + (\epsilon+p_\perp)\delta u^\mu + \order (\partial^2)
,
\\
&&
f^\mu \to 
f^{\prime\mu} = 
f^\mu -  ( p_\perp - p_\para ) \delta b^\mu  + \order (\partial^2)
,
\\
&&  g^\mu \to
g^{\, \prime\mu} = g^\mu + B \delta b^\mu 
,
\\
&& \ell^\mu \to
\ell^{\, \prime\mu} = \ell^\mu + B \delta u^\mu 
.
\end{eqnarray}
\end{subequations}
Making a choice of $\delta u^\mu$ is simultaneously reflected in $h^\mu$ and $\ell^\mu$, and so is making a choice of $\delta b^\mu$ in $f^\mu$ and $g^\mu$. One traditional choice is the Landau-Lifshitz frame \cite{landau1959course:fluid, Minami:2012hs}
with $  \delta u^\mu = -h^\mu /(\epsilon+p_\perp) $, eliminating the heat current $  h^{\prime\mu}$. 
In addition, we choose $\delta b^\mu = - g^\mu/B$ to eliminate $g^{\prime \mu}$. 
This is the frame used in Sec.~\ref{sec:RMHD-1st}: 
\begin{subequations}
 \begin{eqnarray}
 T_{(1)} ^{\mu\nu} \= 
-  \delta p^\prime_\perp \varXi^{\mu\nu} 
+ \delta p^\prime_\para b^\mu b^\nu
+ 2 f^{\prime (\mu} b^{\nu)} + \pi_\perp^{\mu\nu} 
\, ,
\\
 \tilde F^{\mu\nu}_\one \=  
 2 b^{[\mu}  \ell^{\, \prime\nu]}  + m^{\mu\nu}
\, .
 \end{eqnarray}
 \end{subequations}
One can make another possible choice 
with $\delta u^\mu = - \ell^\mu/B$. 
This is an analogue of the Eckart frame \cite{Eckart:1940te}, leading to 
\begin{subequations}
 \begin{eqnarray}
 T_{(1)} ^{\mu\nu} \= 
-  \delta p^\prime_\perp \varXi^{\mu\nu} 
+ \delta p^\prime_\para b^\mu b^\nu
+ 2 h^{\prime (\mu} u^{\nu)}
+ 2 f^{\prime (\mu} b^{\nu)} + \pi_\perp^{\mu\nu} 
\, ,
\\
 \tilde F^{\mu\nu}_\one \=  m^{\mu\nu}
\, ,
 \end{eqnarray}
 \end{subequations} 
where we chose the same $\delta b^\mu$ as above. 
In this frame, the components of the first-order electric field in perpendicular to $b^\mu$ 
has been eliminated from $ \tilde F^{\mu\nu}_\one$ 
at the price of the presence of the heat current in $ T_{(1)} ^{\mu\nu} $. 
It is also possible to eliminate $f^{\prime \mu}$ by choosing $\delta b =  f^\mu/(p_\perp - p_\para)$. 
However, $f^\mu$ appears in $g^{\prime \mu} = g^\mu + B f^\mu/(p_\perp - p_\para)$ in $\tilde F_\one^{\mu\nu}$. 

In case there is also a conserved number current $N^\m=n u^\m+\n^\m$ in the absence of dynamical electromagnetic fields, 
the shift in $u^\m$ leads to a shift in $\n^\m$ by 
\begin{eqnarray}
\label{eq:redef-nu}
&&
\n^\mu %\to h^\mu + 2\delta u^\mu u_\alpha u_\beta T^{\al\be}
\to  \n^\mu +n \delta u^\mu + \order (\partial^2)
.
\end{eqnarray}
One can choose $ \delta u^\mu = - \n^\mu /n$ to eliminate $\n^\mu$. Such a choice is called the Eckart frame~\cite{Eckart:1940te}. It is worth noting that the following combination is frame-choice independent:
\begin{eqnarray}
\label{eq:redef-ind}
&&
\n^\mu-\frac{n}{\e+p}h^\m
\to  \n^\mu-\frac{n}{\e+p}h^\m + \order (\partial^2)
.
\end{eqnarray}
%The heat current is actually absent at the first order in gradient (see, e.g., Ref.~\cite{Hattori:2019lfp}). 
See more discussions in Refs.~\cite{Bhattacharya:2011tra, Kovtun:2012rj, Jensen:2012jh, Hernandez:2017mch}
for other possible frame choices.

%In the current case, this is trivial choice since there is no conserved currents
%other than the energy-momentum tensor.
%The heat current is actually absent at the first order in gradient (see, e.g., Ref.~\cite{Hattori:2019lfp}).
%The reader is referred to the discussion around (1.15) of Ref.~\cite{Kovtun:2012rj} for the case with a conserved charge,
%where one can eliminate either a charge or heat current.
 
\section{Derivation of Eqs.~\eqref{eq:Dbeta-ideal}-\eqref{eq:Du-ideal}}
\label{sec:solving-ideal-RMHD}

We here present a derivation of Eq.~\eqref{eq:eom-conjugate-parameters} by rewriting the ideal RMHD equation \eqref{eq:RMHD-ideal-eom}.
For that purpose, we first call our attention to the thermodynamic relation $\beta p_{\perp} = s - \beta \epsilon + \Hcal B$. By using the definition of the conjugate variables, this relation enables us to find the differential thermodynamic relation as 
\begin{equation}
 \label{eq:dp-perp}
 \begin{split}
  \beta \diff p_{\perp} 
  &= - ( \epsilon + p_{\perp} ) \diff \beta + B \diff \Hcal, 
 \end{split}
\end{equation}
from which we find a set of relations as
\begin{equation}
\label{eq:thermodynamic-relations}
 \epsilon + p_{\perp}
  = - \beta \frac{\partial p_{\perp}}{\partial \beta}, \quad 
  B = \beta \frac{\partial p_{\perp}}{\partial \Hcal} 
  \quad \mathrm {and} \quad 
  \nabla_\mu p_{\perp} 
  = \beta^{-1} 
  \big[ 
  -  (\epsilon + p_{\perp}) \nabla_\mu \beta +  B \nabla_\mu \Hcal
  \big].
\end{equation}
In addition, another thermodynamic relation 
$\diff s 
= \beta \diff \epsilon - \Hcal \diff B $ 
enables us to find the following Maxwell relation 
\begin{equation}
\label{eq:Maxwell-relation}
  \frac{\partial \beta}{\partial B} 
  = - \frac{\partial \Hcal}{\partial \epsilon} . 
 \end{equation}
Combining Eqs.~\eqref{eq:RMHD-ideal-eom} and \eqref{eq:dp-perp}, we also find the divergence of the magnetic flux as
\begin{equation}
\partial_\mu B^\mu
  = \beta^{-1} B^\nu \nabla_{\nu} \beta.
 \label{eq:divergence-B}
\end{equation}

Thanks to these equations, we can simplify the leading-order equations of motion for the thermodynamic parameters.
For instance, the time-derivaitve of the local inverse temperature $\beta$ reads
\begin{equation}
 \begin{split}
  D \beta 
  &= \frac{\partial \beta}{\partial \epsilon} D \epsilon
  + \frac{\partial \beta}{\partial B} D B
  \\
  &= \frac{\partial \beta}{\partial \epsilon} 
  \big[ 
  - (\epsilon + p_{\perp} ) \theta + HB \theta_{\para} 
  \big]
  - \frac{\partial \Hcal}{\partial \epsilon} (- B \theta_{\perp})
  \\
  &= \frac{\partial \beta}{\partial \epsilon} 
  \left(
  \beta \frac{\partial p_{\perp}}{\partial \beta}
  \theta + \beta H \frac{\partial p_{\perp}}{\partial \Hcal}
  \theta_{\para} 
  \right)
  + \beta \frac{\partial \Hcal}{\partial \epsilon}  
  \frac{\partial p_{\perp}}{\partial \Hcal}
  \theta_{\perp}
  \\
  &= \beta 
  \left( 
  \frac{\partial p_{\perp}}{\partial \epsilon}
  - B \frac{\partial H}{\partial \epsilon}  
  \right) 
  \theta_{\para}
  + \beta \frac{\partial p_{\perp}}{\partial \epsilon}
  \theta_{\perp},
 \end{split}
\end{equation}
where we used Eqs.~\eqref{eq:RMHD-ideal-eom}, \eqref{eq:thermodynamic-relations} and \eqref{eq:Maxwell-relation}, and simplified the result by using the chain rule.
This is Eq.~\eqref{eq:Dbeta-ideal} given in the main text.
Likewise, we can compute $D \Hcal$ to derive Eq.~\eqref{eq:DH-ideal}
as 
\begin{equation}
 \begin{split}
  D \Hcal
  &= \frac{\partial \Hcal}{\partial \epsilon} D \epsilon
  + \frac{\partial \Hcal}{\partial B} D B
  \\
  &= \frac{\partial \Hcal}{\partial \epsilon}
  \big[ 
  - (\epsilon + p_{\perp} ) \theta
  + HB \theta_{\para}
  \big]
  + \frac{\partial \Hcal}{\partial B} 
  (- B \theta_{\perp} )
  \\
  &= \frac{\partial \Hcal}{\partial \epsilon}
  \left(
  \beta \frac{\partial p_{\perp}}{\partial \beta} \theta
  + \beta H \frac{\partial p_{\perp}}{\partial \Hcal}
  \theta_{\para}
  \right)
  - \beta \frac{\partial \Hcal}{\partial B} 
  \frac{\partial p_{\perp}}{\partial \Hcal}
  \theta_{\perp} 
  \\
  &= - \beta \frac{\partial p_{\perp}}{\partial B} \theta_{\perp}
  - \beta 
  \left(
  \frac{\partial p_{\perp}}{\partial B}
  - B \frac{\partial H}{\partial B}
  \right)
  \theta_{\para}.
 \end{split}
\end{equation}
Furthermore, relying on the differential thermodynamic relation and Eq.~\eqref{eq:divergence-B}, we can also simplify the time-derivative of the fluid velocity as 
\begin{equation}
 \begin{split}
  D u_\nu 
  &= \frac{1}{\epsilon + p_{\perp}} 
  \left[
  - \frac{\epsilon + p_{\perp}}{\beta} \nabla_{\nu} \beta
  + \frac{1}{\beta} B^\mu \nabla_{\nu} \Hcal_\mu
  \right]
  - \frac{1}{\epsilon + p_{\perp}}
  ( \beta^{-1} H_\nu B^\mu \nabla_\mu \beta
  + \Delta_\nu^{\rho} B^\mu \partial_\mu H_\rho )
  \\
  &= - \beta^{-1} \nabla_{\nu} \beta
  + \frac{1}{\epsilon + p_{\perp}} 
  \left[
  \beta^{-1} B^\mu \nabla_{\nu} \Hcal_\mu
  - H_\nu \beta^{-1} B^\mu \nabla_{\mu} \beta
  - \Delta_\nu^{\rho} B^\mu \partial_\mu H_\rho 
  \right]
  \\
  &= - \beta^{-1} \nabla_{\nu} \beta
  - \frac{1}{\epsilon + p_{\perp}} 
  \left[
  2 \beta^{-1} B b^\mu \nabla_{[\mu} \Hcal_{\nu]} 
  + \theta_{\para} H B u_\nu 
  \right].
 \end{split}
\end{equation}
This gives Eq.~\eqref{eq:Du-ideal}.

\section{Viscous coefficients and inequalities}
\label{sec:MHD-viscosities}

The viscous tensor in magnetic fields has been investigated in classic works~\cite{hooyman1954coefficients, Landau10:kinetics},
and recently in Refs.~\cite{Huang:2011dc, Critelli:2014kra, Hernandez:2017mch, Grozdanov:2016tdf, Hongo:2020qpv} for relativistic systems.
Those authors obtained the same number of independent viscous coefficients,
so that their viscous tensors are expected to be equivalent. 
Here, we compare the viscous tensors in the literature 
and clarify discrepancies in the inequalities that should be satisfied by the viscous coefficients.

The construction of the independent tensor structures are elaborated
by Huang, Sedrakian, and Rischke (HSR) in Sec.~2.2 of Ref.~\cite{Huang:2011dc},
providing useful building blocks and a relativistic extension
of the nonrelativistic theory in a classic textbook~\cite{Landau10:kinetics}.
Therefore, the result by HSR serves as a good starting point for comparisons to other works.
Since the technical machinery has been already detailed there,
%the reader is referred to it for the construction of the independent tensors.
we shall start with the result given in Eq.~(39) of Ref.~\cite{Huang:2011dc}:
\begin{eqnarray}
\label{eq:T1-HSR}
T_\one^{\mu\nu}&=&
2\eta_0 ^{\rm HSR}  \left(w^{\mu\nu}-\frac{1}{3}\Delta^{\mu\nu}\theta\right)
- \eta_1^{\rm HSR}  \left( b^\mu b^\nu + \frac{1}{2}\varXi^{\mu\nu} \right)
\left( \theta_\para - \frac{1}{2}\theta_\perp\right)
\nnb 
&&
-4\Big( \,
\eta_2^{\rm HSR}   b^\alpha b^{(\mu} \varXi^{\nu) \beta}
- \eta_3^{\rm HSR}  \varXi^{\alpha (\mu}  b_\star^{\nu)\beta}
+\eta_4^{\rm HSR}  b^{\alpha}b^{(\mu} b_\star^{\nu)\beta}
\, \Bigr)  w_{\alpha\beta}
\nnb
&&
-3\zeta_\parallel^{\rm HSR} b^\mu b^\nu \theta_\para
+ \frac{3}{2}\zeta_\perp^{\rm HSR} \varXi^{\mu\nu} \theta_\perp
\, .
\end{eqnarray}
Here, we used our notations introduced in Sec.~\ref{sec:MHD}.
Especially, note that $ \theta = \theta_\perp + \theta_\para $.
Nevertheless, the above expression and definitions of the viscous coefficients
are exact duplicates of the result by HSR.

The tensors in the last two terms have nonvanishing traces in Eq.~(\ref{eq:T1-HSR}),
while the others are traceless.
Therefore, the two $\zeta^{\rm HSR} $'s and the five $\eta^{\rm HSR} $'s
are called the bulk and shear viscosities in HSR~\cite{Huang:2011dc}, respectively (see also Ref.~\cite{Landau10:kinetics}).
The term with $ \eta_1^{\rm HSR}  $ is proportional to 
the expansion/compression rates $ \theta_{\para, \perp} $.
However, this term is a traceless part of the viscous tensor,
so that the fluid volume does not change as the expansion in the parallel/perpendicular direction
to the magnetic field is compensated by the compression in the perpendicular/parallel direction.
In this sense, this term was not regarded as a bulk viscosity in HSR~\cite{Huang:2011dc}. 
We will see that this term generates the cross viscosity $ \zeta_\times $ discussed in Sec.~\ref{sec:MHD}.

Now, by using the identity (\ref{eq:w_mn}), 
one can arrange Eq.~(\ref{eq:T1-HSR}) into the form 
expressed with the tensor basis introduced in Eqs.~(\ref{eq:T-1-decomposed}) and (\ref{eq:Hall-viscous-tensor}). 
One can identify the correspondences between the viscous coefficients introduced
in Sec.~\ref{sec:MHD} and in HSR~\cite{Huang:2011dc} as
\begin{eqnarray}
\label{eq:correspondences}
&&
\eta_\perp = \eta_0 ^{\rm HSR}
, \quad
\eta_\para = 2(\eta_0^{\rm HSR}  + \eta_2^{\rm HSR} )
, \quad
\eta_{\Hall\para} = 2\eta_4^{\rm HSR}
, \quad
\eta_{\Hall\perp} =  2\eta_3^{\rm HSR}
,
\\
&&
\zeta_\para = 3\zeta_\para^{\rm HSR}  + \left( \frac{4}{3} \eta_0^{\rm HSR}  + \eta_1^{\rm HSR}   \right)
, \quad
\zeta_\perp = \frac{3}{2} \zeta_\perp ^{\rm HSR} + \frac{1}{4} \left( \frac{4}{3} \eta_0 ^{\rm HSR} + \eta_1 ^{\rm HSR}  \right)
, 
\nnb
&&
\zeta_\times = - \frac{1}{2} \left( \frac{4}{3} \eta_0 ^{\rm HSR} + \eta_1 ^{\rm HSR}  \right)
\nn
.
\end{eqnarray}
It is now clear that the cross viscosity $ \zeta_\times  $ was generated
from the traceless terms proportional to $ \eta_0 $ and $ \eta_1  $ in Eq.~(\ref{eq:T1-HSR}). 
The correspondences between the viscous coefficients in HSR~\cite{Huang:2011dc} and
Hernandez and Kovtun (HK)~\cite{Hernandez:2017mch} are available in Eq.~(B.1) of HK~\cite{Hernandez:2017mch}.
Putting Eq.~(\ref{eq:correspondences}) together, the list of correspondences is expanded as
\begin{eqnarray}
&&
\eta_\perp^{\rm HK} = \eta_0^{\rm HSR} =  \eta_\perp
 \, ,
 %\quad
\nn \\
&&
\tilde \eta_\perp^{\rm HK} = - 2 \eta_3^{\rm HSR} =  - \eta_{\Hall\perp}
 \, ,
 \nn \\
 &&
 \eta_\para^{\rm HK} =  \eta_0^{\rm HSR} +  \eta_2^{\rm HSR} = \frac{1}{2} \eta_\para
 \, ,
 %\quad
\nn \\
&&
\tilde \eta_\para^{\rm HK} = - \eta_4^{\rm HSR} = - \frac{1}{2}\eta_{\Hall\para}
 \, ,
\nn \\
 &&
 \eta_1^{\rm HK} = - \frac{1}{2} \eta_0 ^{\rm HSR} - \frac{3}{8} \eta_1^{\rm HSR}  - \frac{3}{4} \zeta_\perp ^{\rm HSR}
 = - \frac{1}{2} ( \zeta_\perp -  \zeta_\times)
\, ,
\\
 &&
\eta_2^{\rm HK} = \frac{3}{2} \eta_0 ^{\rm HSR} + \frac{9}{8} \eta_1^{\rm HSR}
+ \frac{3}{4} \zeta_\perp ^{\rm HSR} + \frac{3}{2} \zeta_\para ^{\rm HSR}
= \frac{1}{2} ( \zeta_\para - 2  \zeta_\times +\zeta_\perp )
 \, ,
 \nn \\
 &&
\zeta_1^{\rm HK} = \zeta_\perp ^{\rm HSR} = \frac{1}{3}(2 \zeta_\perp +  \zeta_\times)
 \, ,
 %\quad
\nn \\
&&
\zeta_2^{\rm HK} = \zeta_\para ^{\rm HSR} -\zeta_\perp ^{\rm HSR}
= \frac{1}{3} ( \zeta_\para  +  \ \zeta_\times - 2  \zeta_\perp)
\nn
\, .
\end{eqnarray}
We note that the viscous coefficients in the current paper agree with those
in Eqs.~(3.9), (3.10), (3.13), and (3.14) of Grozdanov, Hofman, and Iqbal (GHI)~\cite{Grozdanov:2016tdf}
up to conventions in the signs and factors. 
Further correspondences to the conventions in Ref~\cite{Critelli:2014kra} are discussed in
the appendix of HK~\cite{Hernandez:2017mch}. 
The viscosities were compute in an external weak magnetic fields by Li and Yee (LY) \cite{Li:2017tgi}. 
Compared with Eq.~(A5) therein, one finds the correspondences 
\begin{eqnarray}
&&
\eta^{\rm LY} = \eta_0 ^{\rm HSR}  + \eta_2^{\rm HSR} = \eta_\para
, \quad
\zeta^{\rm LY} = 3 \zeta_\para ^{\rm HSR} + \frac{4}{3} \eta_0 ^{\rm HSR} + \eta_1 ^{\rm HSR} = \zeta_\para
, 
\\
&&
\zeta^{\prime \, {\rm LY} }= - \frac{2}{3} \eta_0 ^{\rm HSR} - \frac{1}{2} \eta_1 ^{\rm HSR} = \zeta_\times
\nn
.
\end{eqnarray}
All three coefficients quantify the responses along the external magnetic field.

%\footnote{It would be also useful to simplify
%the equalities in Eq.~(B.1) of Ref.~\cite{Hernandez:2017mch} as
%$ \eta_1^{\rm HK} = -(\tilde \zeta_\perp - \tilde \zeta_\times)/2 $,
%$ \eta_2^{\rm HK} = (\tilde \zeta_\perp + \tilde \zeta_\para)/2 - \tilde \zeta_\times $,
%$ \zeta_1^{\rm HK} = (2\tilde \zeta_\perp + \tilde \zeta_\times)/3 $, and
%$ \zeta_2^{\rm HK} = (\tilde \zeta_\para - 2 \tilde \zeta_\perp +3  \tilde \zeta_\times)/3 $,
%where $ \eta_{1,2}^{\rm HK} $ and $\zeta_{1,2}^{\rm HK} $ denote the viscous coefficients
%in the conventions of Hernandez and Kovtun~\cite{Hernandez:2017mch}.
%With these identifications, one can directly verify
%that the inequalities in Eq.~(3.19) of Ref.~\cite{Hernandez:2017mch}
%results in those in Eq.~(\ref{eq:ineqalities}).}

By the use of the above identifications, the somewhat involved inequalities 
in Eq.~(B.2) of HK~\cite{Hernandez:2017mch} can be simplified as 
\begin{eqnarray}
\label{eq:inequalities-HK}
\eta_\perp \geq 0
\, , \ \ \
\eta_\para \geq 0
\, , \ \ \
\zeta_\perp \geq 0
\, , \ \ \
\zeta_\para + \zeta_\perp + 2 \zeta_\times \geq 0
\, , \ \ \
\zeta_\para \zeta_\perp - \zeta_\times^2 \geq 0
\, .
\end{eqnarray} 
As mentioned below Eq.~(\ref{eq:ineqalities}),
one may remove one inequality $  \zeta_\para \geq0$ from the list (\ref{eq:ineqalities}),
or putting it differently, one can show that $  \zeta_\para \geq0$
when the two inequalities $ \zeta_\perp \geq 0$ 
and $\zeta_\para \zeta_\perp - \zeta_\times^2 \geq 0 $ are satisfied.
Therefore, an essential difference from Eq.~(\ref{eq:ineqalities})  
is only the existence of the second-last inequality, $ \zeta_\para + \zeta_\perp + 2 \zeta_\times \geq 0 $.
This inequality can be, however, deduced from the others in Eq.~(\ref{eq:inequalities-HK}), as follows. 
We can immediately show that
$ ( \zeta_\para +  \zeta_\perp)^2 - (2  \zeta_\times)^2
=  (  \zeta_\para - \zeta_\perp)^2
+ 4( \zeta_\para   \zeta_\perp -  \zeta_\times^2) \geq 0 $.
Since we have $ \zeta_{\para,\perp} \geq0 $,
we find that $   \zeta_\para  + \zeta_\perp + 2  \zeta_\times \geq 0$, 
regardless of the sign of $   \zeta_\times$. 
This proof suggests that the fourth inequality in Eq.~(\ref{eq:inequalities-HK}),
and thus in Eq.~(B.2) of HK~\cite{Hernandez:2017mch}, is redundant.
The minimal list of the inequalities is given by 
\begin{eqnarray}
\label{eq:inequalities-HK-final}
\eta_\perp \geq 0
\, , \ \ \
\eta_\para \geq 0
\, , \ \ \
\zeta_\perp \geq 0
\, , \ \ \
\zeta_\para \zeta_\perp - \zeta_\times^2 \geq 0
\, .
\end{eqnarray}
The inequality $ \zeta_\perp \geq0 $ can be alternatively replaced by the other one, $ \zeta_\para \geq0  $.

\cout{
\section{Kubo formulas}\label{sec:kubo}

\com{This section will be removed.}

%Kubo formuals \cite{Kubo:doi:10.1143/JPSJ.12.570}

See Zubarev \cite{Zubarev1979}, Hosoya et al. \cite{Hosoya:1983id}, and Huang et al. \cite{Huang:2011dc},
Becattini et al. \cite{Becattini:2019dxo}.

We start with the density operator defined as
\begin{eqnarray}
\hat \rho [\hat Z] = \frac{ \exp\left( - \int d^3\bx \hat Z \right)}{ \tr \exp\left( - \int d^3\bx \hat Z \right)}
,
\quad
\hat Z(t_1,\bx) = \epsilon \int _{-\infty}^{t_1} dt e^{ \epsilon(t-t_1)} \hat \Phi^0 (t,\bx)
\end{eqnarray}
where
\begin{eqnarray}
\Phi^\mu (t,\bx) =  \beta_\nu (t,\bx)\hat  T^{\mu\nu} (t,\bx)
\end{eqnarray}
They are conserved currents
\begin{subequations}
\begin{eqnarray}
\partial_\mu \hat T^{\mu\nu} (t,\bx) =0
\, .
\end{eqnarray}
\end{subequations}
Integrating by parts,
we have
%\begin{eqnarray}
%\hat Z  (t_1,\bx)
%&=&
%\left[ e^{ \epsilon(t-t_1)} \hat  \Phi^0 (t,\bx) \right]^{t_1}_{-\infty}
%- \int _{-\infty}^{t_1} dt e^{ \epsilon(t-t_1)} \partial_t  \hat \Phi^0 (t,\bx)
%\nn
%\\
%&=& \hat \Phi^0 (t_1,\bx) -  \int _{-\infty}^{t_1} dt e^{ \epsilon(t-t_1)}  \partial_t \hat  \Phi^0 (t,\bx)
%\end{eqnarray}
%Furthermore, we get
\begin{eqnarray}
\int d^3 \bx \hat Z  (t_1,\bx)
%&=&
%\int d^3 \bx \hat  \Phi^0 (t_1,\bx) -    \int _{-\infty}^{t_1} dt  e^{ \epsilon(t-t_1)}
%\int d^3 \bx \left[ \partial_t \hat  \Phi^0 (t,\bx) -  \partial_i \hat \Phi^i (t,\bx) \right]
%\nn
%\\
%&=&
=
\int d^3 \bx \hat \Phi^0 (t_1,\bx) -   \int _{-\infty}^{t_1} dt  e^{ \epsilon(t-t_1)}
\int d^3 \bx \partial_\mu  \hat \Phi^\mu (t,\bx)
\end{eqnarray}
where we assumed that $ \hat \Phi^\mu (t,\bx)  \to 0 $ as $ |\bx| \to \infty $.

Accordingly, we define
\begin{subequations}
\begin{eqnarray}
 \hat Z  (t_1,\bx) &=& \hat Z_\eq (t_1,\bx) - \delta \hat Z (t_1,\bx)
\\
\hat Z _\eq (t_1,\bx)&\equiv& \hat  \Phi^0 (t_1,\bx)
\\
\delta \hat Z  (t_1,\bx)&\equiv& \int _{-\infty}^{t_1} dt  e^{ \epsilon(t-t_1)}  \partial_\mu  \hat \Phi^\mu (t,\bx)
\end{eqnarray}
\end{subequations}
Assuming that the perturbation is small, we expand the density operator as
\begin{eqnarray}
\hat \rho
%&=& \frac{ \exp\left[ - \int d^3\bx (\hat Z_\eq - \delta \hat Z) \right]}
%{ \tr \exp\left[ - \int d^3\bx (\hat Z_\eq - \delta \hat Z) \right]}
\sim
\left[ \, 1 + \int^1_0 d\tau  \int d^3\bx \left( e^{-  \int d^3\bx^\prime\hat Z_\eq \tau}
\delta \hat Z e^{ \int d^3\bx^\prime\hat Z_\eq \tau}  \right)
- \LR{ \delta \hat Z}_\eq  \,\right] \hat \rho_\eq
\end{eqnarray}
where
\begin{eqnarray}
\hat \rho_\eq = \hat \rho[\hat Z_\eq]
, \quad
\LR{\order}_\eq  = \tr[ \hat \rho_\eq \order ]
\end{eqnarray}
Inserting the above expansion, we get
\begin{eqnarray}
\LR{\order} &\sim& \LR{\order}_\eq + \delta \LR{\order}
\nn
\\
\delta \LR{\order}
%&=&
%\tr\sbra{
%\int^1_0 d\tau\left( e^{- \hat Z_\eq \tau} \delta \hat Z e^{\hat Z_\eq \tau} \hat \rho_\eq \order\right)
%- \LR{ \delta \hat Z}_\eq  \LR{\order}_\eq
%}
%\int^1_0 d\tau  \int d^3\bx \LR{
%\left( e^{-  \int d^3\bx^\prime\hat Z_\eq \tau} \delta \hat Z e^{ \int d^3\bx^\prime \hat Z_\eq \tau}
%-  \LR{ \delta \hat Z}_\eq  \right)  \order
%}_\eq
%\nn
%\\
&=&
\int_{-\infty}^{t_1} dt e^{ \epsilon(t-t_1)} \int d^3 \bx
\int^1_0 d\tau\LR{
\left( e^{-  \int d^3\bx^\prime \hat Z_\eq \tau} \hat f e^{ \int d^3\bx^\prime\hat Z_\eq \tau}
 -  \LR{ \hat f}_\eq  \right)  \order
}_\eq
\end{eqnarray}
where
\begin{eqnarray}
\hat f  \equiv \partial_\mu \hat  \Phi^\mu (t,\bx)
= \hat T^{\mu\nu} (t,\bx) \partial_\mu \beta_\nu (t,\bx)
\end{eqnarray}
\com{[Need to check the sign. Is it consistent to that in the entropy analysis?]}
One can express the (sightly) off-equilibrium component
of an operator expectation value in terms of the thermal retarded Green's function
\begin{eqnarray}
\delta \LR{\order} =
\int_{-\infty}^{t_1}dt e^{ \epsilon(t-t_1)} \int d^3 \bx \LR{\order, \hat f}
 = - T \lim_{\omega\to0} \lim_{\bk\to0} \frac{\partial \ }{\partial \omega} \Im G _R^{\order, \hat f}  (\omega,\bk)
\end{eqnarray}
where $ \Im G _R^{\order, \hat f}  $ is the imaginary part of $ G _R^{\order, \hat f}  $.

The remaining task is to match $ \delta \LR{\order}  $ to the constitutive equations.
Then, one can express the transport coefficients in terms of the equilibrium correlation functions.

The operator expectation value $  \LR{\order} $ is identified with a hydrodynamic constitutive equation.
Specifically, we need matching of $  \Theta^{\mu\nu} $ and $J^{\mu\a\b}  $.

The existence of conserved charges implies
\begin{eqnarray}
u_\mu u_\nu \LR{\hat  \Theta^{\mu\nu}} = u_\mu u_\nu \LR{\hat  \Theta^{\mu\nu}}_\eq
\end{eqnarray}
These relations implicitly define the thermodynamic conjugate quantities (Lagrange multipliers)
$ \beta $, $ \omega^{\mu\nu} $, etc in terms of the conserved charges.

\begin{eqnarray}
\hat T_{(1) {\rm dis}}^{\mu\nu} = \delta \hat p_\para (-b^\mu b^\nu)
+ \delta \hat p_\perp \varXi^{\mu\nu}  + \hat  f^{(\mu} b^{\nu)} + \hat \tau^{\mu\nu}
\label{eq:T1_mn-ope}
\, ,
\end{eqnarray}

\begin{eqnarray}
&&
%\hat e = u_\alpha u_\beta \hat  T^{\alpha\beta}
%, \quad
\delta \hat p_\parallel =(-b_\alpha b_\beta) \delta \hat T^{\alpha\beta}
, \quad
\delta \hat p_\perp = \frac{1}{2} \varXi_{\alpha\beta}  \delta \hat T^{\alpha\beta}
,
\\
&&
\hat f^\mu =  - b_{(\alpha} \varXi^{\mu}_{ \beta)}  \delta \hat T^{\alpha\beta}
, \quad
\hat \tau^{\mu\nu} =( \varXi^{  (\mu} _\alpha \varXi^{\nu) }_\beta - \varXi^{\mu\nu}\varXi_{\alpha\beta} ) \delta \hat T^{\alpha\beta}
\end{eqnarray}
where $  \delta \hat T^{\alpha\beta}  \equiv \hat T^{\alpha\beta} - \LR{\hat T^{\alpha\beta} }_\eq$

\begin{eqnarray}
\LR{ \hat T_{(1) {\rm dis}}^{\mu\nu} } =
\LR{ \hat T_{(1) {\rm dis}}^{\mu\nu} ,  \hat T_{(1) {\rm dis}}^{\alpha\beta} w_{\alpha\beta} }
\end{eqnarray}
The left-hand side is matched to the hydrodynamic constitutive equation ().

\begin{subequations}
\begin{eqnarray}
&&
\zeta_\para \theta_\para + \zeta_\times \theta_\perp
= \LR{ \delta \hat p_\para ,  \delta \hat p_\para \theta_\para +  \delta \hat p_\perp \theta_\perp}
\\
&&
\zeta_\times \theta_\para + \zeta_\perp \theta_\para
= \LR{ \delta \hat p_\perp ,  \delta \hat p_\para \theta_\para +  \delta \hat p_\perp \theta_\perp}
\\
&&
\eta_\para = - \frac{1}{2} \LR{ \hat f^\mu, \hat f_\mu}
\\
&&
\eta_\perp = \LR{ \hat \tau^{\mu\nu}, \hat \tau_{\mu\nu}}
\end{eqnarray}
\end{subequations}

\begin{subequations}
\begin{eqnarray}
&&
\zeta_\para
 = - T \lim_{\omega\to0} \lim_{\bk\to0} \frac{\partial \ }{\partial \omega}
 \Im G _R^{\delta \hat p_\para,\delta \hat p_\para}  (\omega,\bk)
\\
&&
\zeta_\perp
 = - T \lim_{\omega\to0} \lim_{\bk\to0} \frac{\partial \ }{\partial \omega}
 \Im G _R^{\delta \hat p_\perp,\delta \hat p_\perp}  (\omega,\bk)
\\
&&
\zeta_\para
 = - T \lim_{\omega\to0} \lim_{\bk\to0} \frac{\partial \ }{\partial \omega}
 \Im G _R^{\delta \hat p_\para,\delta \hat p_\perp}  (\omega,\bk)
 = - T \lim_{\omega\to0} \lim_{\bk\to0} \frac{\partial \ }{\partial \omega}
 \Im G _R^{\delta \hat p_\perp,\delta \hat p_\para}  (\omega,\bk)
\\
&&
\eta_\para
 = - T \lim_{\omega\to0} \lim_{\bk\to0} \frac{\partial \ }{\partial \omega}
 \Im G _R^{\hat f^\mu, \hat f_\mu}  (\omega,\bk)
\\
&&
\eta_\perp
 = - T \lim_{\omega\to0} \lim_{\bk\to0} \frac{\partial \ }{\partial \omega}
 \Im G _R^{\hat \tau^{\mu\nu} , \hat \tau_{\mu\nu}}  (\omega,\bk)
\\
\end{eqnarray}
\end{subequations}
where $ \Im G _R^{\order, \hat f}  $ is the imaginary part of $ G _R^{\order, \hat f}  $.

%\begin{eqnarray}
%U( \tau) \equiv \exp\left[ \int d^3\bx \hat Z_\eq \tau \right]
%\, , \quad
%\bar U( \tau)  \equiv \exp\left [-\int d^3\bx \hat Z_\eq \tau \right]
%\end{eqnarray}

\com{[Need to check signs.]}
The exponential factor can be regarded as the imaginary-time evolution operator,
which leads to relations
\bseq
\begin{eqnarray}
&&
e^{ -\int d^3\bx^\prime Z_\eq} \order(t,\bx) e^{  \int d^3\bx^\prime Z_\eq}  = \order(t + i\beta ,\bx)
\\
&&
\LR{A(t,\bx)} _\eq = \LR{ A(t+i\beta,\bx)}_\eq = \LR{ A(t - i \beta,\bx)}_\eq
\\
&&
\LR{A(t,\bx) B(t^\prime+i\beta,\bx)} _\eq = \LR{ B(t^\prime,\bx) A(t,\bx) } _\eq
\end{eqnarray}
\eseq
By using those relations, we have
\begin{eqnarray}
\LR{ A(t,\bx), B(t^\prime,\bx^\prime)}_\eq
&=& \LR{ A(t,\bx), B(t^\prime+i\beta  ,\bx^\prime)}_\eq
\nn
\\
&=&
\int_{-\infty}^{t^\prime} ds \frac{d \ }{ds} \LR{ A(t,\bx), B(s+i\beta  ,\bx^\prime)}_\eq
+ \lim_{s\to-\infty} \LR{ A(t,\bx), B(s+i\beta  ,\bx^\prime)}_\eq
\end{eqnarray}
The second term is assumed to vanish when the correlation vanishes in an infinite relative time separation, i.e.,
\begin{eqnarray}
\lim_{s\to-\infty} \LR{ A(t,\bx) B(s+i\beta  \bx^\prime)}_\eq
=\LR{ A(t,\bx)} \LR{ B(s+i\beta  ,\bx^\prime)}_\eq
\end{eqnarray}
Further simple arrangement leads to a relation to the thermal commutator

\begin{eqnarray}
\LR{ A(t,\bx), B(t^\prime,\bx^\prime)}_\eq
&=&
-i\beta^{-1} \int_{-\infty}^{t^\prime} ds \int_0^1 d\tau
\frac{d \ }{d\tau} \LR{ A(t,\bx) [ B(s+i\beta \tau ,\bx^\prime) - \LR{ B(s+i\beta \tau ,\bx^\prime)}_\eq ] }_\eq
\nn
\\
&=&
-i\beta^{-1} \int_{-\infty}^{t^\prime} ds
\LR{ A(t,\bx) [ B(s+i\beta,\bx^\prime) - \LR{ B(s+i\beta,\bx^\prime)}_\eq ] }_\eq
\nn
\\
&&
+ i\beta^{-1} \int_{-\infty}^{t^\prime} ds
\LR{ A(t,\bx) [ B(s,\bx^\prime) - \LR{ B(s,\bx^\prime)}_\eq ] }_\eq
\nn
\\
&=&
i\beta^{-1} \int_{-\infty}^{t^\prime} ds\LR{  [ A(t,\bx) , B(s,\bx^\prime) ]}_\eq
\end{eqnarray}
Now, we see the relation to the retarded Green's function
\begin{eqnarray}
G_R^{A,B} (t-s,\bx-\bx^\prime) = -i\theta(t-s) \LR{  [ A(t,\bx) , B(s,\bx^\prime) ]}_\eq
\end{eqnarray}
where we assume that the correlator in the thermal state
is a function only of the difference in the coordinate.
Then, one can arrange further to get a relation in the momentum space:
\begin{eqnarray}
\int_{-\infty}^{t} dt^\prime e^{ \epsilon(t^\prime-t)} \int d^3 \bx ^\prime
\LR{ A(t,\bx), B(t^\prime,\bx^\prime)}_\eq
&=&
- \beta^{-1}  \int_{-\infty}^{t} dt^\prime e^{ \epsilon(t^\prime-t)} \int_{-\infty}^{t^\prime} ds
\int d^3 \bx ^\prime  G_R^{A,B} (t-s,\bx-\bx^\prime)
\nn
\\
&=&
- \beta^{-1}  \int_{-\infty}^{0} du e^{ \epsilon u} \int_{-\infty}^{t+u} ds
\int d^3 \bx ^\prime  G_R^{A,B} (t-s,\bx-\bx^\prime)
\nn
\\
&=&
- \beta^{-1}  \int_{-\infty}^{0} du e^{ \epsilon u} \int_{-\infty}^{u} ds^\prime
\int d^3 \bx ^\prime  G_R^{A,B} (-s^\prime,\bx-\bx^\prime)
\nn
\\
&=&
- \beta^{-1}  \int_{-\infty}^{0} du e^{ \epsilon u} \int_{-\infty}^{u} ds^\prime
\lim_{\bk\to0} \int_{-\infty}^\infty \frac{d\omega}{2\pi} e^{ i \omega (-s^\prime)}
 G_R^{A,B} (\omega,\bk)
\nn
\\
\end{eqnarray}

\begin{eqnarray}
\int_{-\infty}^{0} du e^{ \epsilon u} \int_{-\infty}^{u} ds^\prime e^{ - i (\omega+i\delta) s^\prime}
&=&
\frac{i}{\omega} \int_{-\infty}^{0} du e^{ \epsilon u  - i (\omega+i\delta) u}
=
- \frac{1}{\omega(\omega+i\delta)}
\end{eqnarray}

\begin{eqnarray}
\int_{-\infty}^{t} dt^\prime e^{ \epsilon(t^\prime-t)} \int d^3 \bx ^\prime
\LR{ A(t,\bx), B(t^\prime,\bx^\prime)}_\eq
&=&
\beta^{-1} \lim_{\bk\to0} \int_{-\infty}^\infty \frac{d\omega}{2\pi}
\frac{1}{\omega(\omega+i\delta)}
 G_R^{A,B} (\omega,\bk)
\nn
\\
\end{eqnarray}
}

\bibliography{bib_review}

\end{document}